\documentclass[times,final]{elsarticle}
\usepackage{jcomp_arxiv}
\usepackage[labelfont=bf]{caption}
\usepackage{framed,multirow}

\usepackage{amssymb}
\usepackage{amsthm}
\usepackage{latexsym}
\usepackage{url}

\journal{Journal of Computational Physics}

\usepackage[T1]{fontenc}
\DeclareRobustCommand{\VAN}[3]{#2}
\let\VANthebibliography\thebibliography
\def\thebibliography{\DeclareRobustCommand{\VAN}[3]{##3}\VANthebibliography}

\usepackage{graphicx}	
\usepackage{amsmath}	
\usepackage{amssymb}	
\usepackage[dvipsnames,svgnames,table,xcdraw]{xcolor}
\definecolor{newcolor}{rgb}{.8,.349,.1}
\usepackage{physics}
\usepackage{xspace}
\usepackage{nicefrac}
\usepackage[normalem]{ulem}
\usepackage{newtxtext,newtxmath}
\usepackage{empheq}
\usepackage{tcolorbox}

\usepackage[colorlinks=true
,urlcolor=blue
,anchorcolor=blue
,citecolor=blue
,filecolor=blue
,linkcolor=red
,menucolor=blue
,linktocpage=true
,pdfproducer=medialab
,pdfa=true
]{hyperref}

\definecolor{mygreen}{RGB}{20,138,6}
\definecolor{myblue}{RGB}{52,180,230}
\definecolor{mygray}{RGB}{180,180,180}
\definecolor{mypurple}{RGB}{180, 1, 190}
\definecolor{myorange}{RGB}{209, 134, 0}

\theoremstyle{definition}
\newtheorem{remark}{Remark}
\newtheorem{definition}{Definition}
\newtheorem{example}{Example}
\newtheorem{counterexample}{Counterexample}
\theoremstyle{plain}
\newtheorem{proposition}{Proposition}

\AtBeginEnvironment{example}{%
  \pushQED{\qed}%
}
\AtEndEnvironment{example}{\popQED\endexample}
\AtBeginEnvironment{counterexample}{%
  \pushQED{\qed}%
}
\AtEndEnvironment{counterexample}{\popQED\endcounterexample}
\AtBeginEnvironment{remark}{%
  \pushQED{\qed}%
}
\AtEndEnvironment{remark}{\popQED\endremark}
\AtBeginEnvironment{definition}{%
  \pushQED{\qed}%
}
\AtEndEnvironment{definition}{\popQED\enddefinition}

\newcommand\like[1]{\begin{picture}(1,1)
\ifnum0=#1\put(.75,.5){\circle{1}}\else
\ifnum100=#1\put(.75,.5){\circle*{1}}\else
\put(.75,.5){\circle{1}}\put(.75,.5){\circle*{.#1}}
\fi\fi\end{picture}\hspace{0.5\unitlength}}

\newcommand{\vecb}[1]{\ensuremath{\boldsymbol{#1}}}
\newcommand{\mPsi}{{\mathit \Psi}}
\newcommand{\mPi}{{\mathit \Pi}}
\providecommand{\dd}{\ensuremath{{\rm d}}}
\newcommand{\bnabla}{\ensuremath{\boldsymbol{\nabla}}}

\begin{document}

\verso{Florian List \& Oliver Hahn}

\begin{frontmatter}
\title{Perturbation-theory informed integrators for cosmological simulations}

\author[1,2]{Florian List}
\ead{florian.list@univie.ac.at}

\author[1,2]{Oliver Hahn}
\ead{oliver.hahn@univie.ac.at}

\address[1]{Department of Astrophysics, University of Vienna, Türkenschanzstraße 17, 1180 Vienna, Austria}
\address[2]{Department of Mathematics, University of Vienna, Oskar-Morgenstern-Platz 1, 1090 Vienna, Austria}

\received{xxxxx}
\finalform{xxxxx}
\accepted{xxxxx}
\availableonline{xxxxx}
\communicated{xxxxx}

\begin{abstract}
Large-scale cosmological simulations are an indispensable tool for modern cosmology. To enable model-space exploration, fast and accurate predictions are critical. In this paper, we show that the performance of such simulations can be further improved with time-stepping schemes that use input from cosmological perturbation theory.
Specifically, we introduce a class of time-stepping schemes derived by matching the particle trajectories in a single leapfrog/Verlet drift-kick-drift step to those predicted by Lagrangian perturbation theory (LPT). As a corollary, these schemes exactly yield the analytic Zel'dovich solution in 1D in the pre-shell-crossing regime (i.e.\ before particle trajectories cross). One representative of this class is the popular `\textsc{FastPM}' scheme by \emph{Feng et al.\ 2016} \cite{Feng:2016}, which we take as our baseline. We then construct more powerful LPT-inspired integrators and show that they outperform \textsc{FastPM} and standard integrators in fast simulations in two and three dimensions with $\mathcal{O}(1 - 100)$ timesteps, requiring fewer steps to accurately reproduce the power spectrum and bispectrum of the density field.
Furthermore, we demonstrate analytically and numerically that, for \emph{any} integrator, convergence is limited in the post-shell-crossing regime (to order $\nicefrac{3}{2}$ for planar-wave collapse), owing to the lacking regularity of the acceleration field, which makes the use of high-order integrators in this regime futile. Also, we study the impact of the timestep spacing and of a decaying mode present in the initial conditions. Importantly, we find that symplecticity of the integrator plays a minor role for fast approximate simulations with a small number of timesteps.
\end{abstract}

\end{frontmatter}



\section{Introduction}
Large-scale cosmological $N$-body simulations have been the driving force behind most of our theoretical understanding of the formation of cosmic structure, such as galaxies and their distribution, from minute metric fluctuations seeded in the primordial universe (see e.g.\ \cite{AnguloHahn:2022} for a recent review). These simulations are still mostly collisionless, i.e.\ they model the purely (Newtonian) gravitational interaction of cold matter in an expanding universe. Neglecting collisions is justified since in the $\Lambda$CDM ($\Lambda$ cold dark matter) paradigm, most matter ($\sim$80 per cent) is in the form of cold dark matter. The remaining electromagnetically interacting baryonic matter is also cold, and therefore the cold collisionless limit is an excellent approximation to the entirety of all non-relativistic components in the universe on large-enough scales. Thus, the equations of motion are given by the Vlasov--Poisson system with an additional explicitly time-dependent term due to the expansion of the universe \cite{Peebles:1980}. Due to the Hamiltonian structure underlying the Vlasov equation, symplectic integrators are typically used to advance the particles \cite{Quinn:1997,Springel:2005}. 

In order to sample both the volume of the universe covered by the next generation of galaxy surveys (such as \emph{Euclid} \cite{Euclid}, the \emph{Legacy Survey of Space and Time} (LSST) \cite{LSST} conducted by the Vera C. Rubin Observatory, and the \emph{Nancy Grace Roman Space Telescope} \cite{WFIRST}) and achieve sufficient resolution in individual observable structures, the largest simulations to date employ $10^{12}-10^{13}$ particles, e.g.\ the \emph{Euclid Flagship simulation} \cite{potter2017pkdgrav3}, \emph{AbacusSummit} \cite{AbacusSummit}, \emph{OuterRim} \cite{OuterRim1, OuterRim2}, \emph{TianNu} \cite{TianNu}, and \emph{Uchuu} \cite{Uchuu}, and are run on the biggest supercomputers available. At the same time, covering the space of cosmological parameters and computing covariance matrices along with quantifying statistical uncertainties requires large suites of such simulations. In order to make optimal use of computational resources, it is therefore imperative to minimise the computational cost of any given simulation.

The dominant computational cost in collisionless cosmological simulations is due to the gravitational force calculation, and much progress has been made over the last decades through algorithmic improvements. The tree-PM method \cite{Bagla:2002} has been the most popular algorithmic approach until recently. It combines the Barnes~\&~Hut tree method \cite{Barnes:1986,Appel:1985} with a particle-mesh-Ewald (PME) method \cite{Darden:1993} to achieve $\mathscr{O}(N\log N)$ complexity for one full interaction calculation for $N$ particles. With ever increasing particle numbers, more recently, the Fast-Multipole method (FMM) \cite{Greengard:1987} has gained popularity, which achieves $\mathscr{O}(N)$ complexity. FMM has a significantly higher pre-factor at fixed accuracy, so that it has only more recently been adopted in most cosmological simulation codes. If less spatial resolution is sufficient, the much simpler particle-mesh (PM) methods can be used, which also have $\mathscr{O}(N)$ complexity w.r.t.\ the particle number, but typically are dominated by the $\mathscr{O}(N_c \log N_c)$ complexity when employing $N_c$ mesh cells (and an FFT or multigrid-based Poisson solver). Due to the coldness of CDM, it is usually acceptable to have $N_c\gtrsim N$, unlike in typical hot particle-in-cell simulations, where many particles per mesh cell are used (cf.~\cite{Hockney:1988}). At the same time, discreteness errors in this high-force-resolution at low-mass-resolution regime of $N$-body simulations do occur \cite{Hahn:2013,Michaux:2021,Colombi:2021}, and need to be taken into account when interpreting the simulation results.

We may thus assume a near optimal force calculation (in terms of computational speed) for a given set-up. This leaves only one possibility for further minimisation of the computational cost: \emph{to reduce the number of timesteps} at fixed accuracy. One may consider two approaches in this regard: (1) a \emph{global} reduction by optimising the time-stepping scheme, or (2) a \emph{local} reduction by introduction of a multi-stepping scheme where particles are assigned individual (quantised) optimal timesteps according to properties of their trajectories \cite{Hayli:1974,McMillan:1986}. 

Here, we focus on the possibility of a global reduction. We leave a possible extension of our results to block-stepping schemes to future work. Naively, one might expect that a global reduction at fixed accuracy is not possible, since in time integration schemes the global integration error is generally dictated by the number of timesteps. This is not so, however, in the case of \emph{cosmological} $N$-body simulations. It is perhaps one of the most famous results of large-scale structure cosmology due to Zel'dovich \cite{Zeldovich:1970} that in suitable time coordinates, the leading-order dynamics at early times (specifically prior to the crossing of particle trajectories) is simply given by free inertial motion despite the presence of gravitational interactions. This is due to a leading-order cancellation between acceleration due to gravity and deceleration due to the expansion of space. This cancellation is, however, not reflected in the Hamiltonian, so that many timesteps are needed to reproduce the relatively simple dynamics in this regime. This observation has motivated the recent so-called \textsc{FastPM} scheme \cite{Feng:2016}, which makes an ad-hoc modification to the drift and kick operators of a second-order kick-drift-kick (KDK) integrator to reproduce these leading-order dynamics exactly. Although `PM' in the name \textsc{FastPM} stands for particle mesh, the time integration scheme presented in that work is in fact independent of the employed force-computation method, and all references to \textsc{FastPM} herein exclusively concern the time integration.

Ref.~\cite{Feng:2016} demonstrate the performance of this approach in terms of the effect on summary statistics; however, we are not aware of a formal analysis or proofs of properties of this integrator in the literature. At the same time, it is clear that such `perturbation-theory informed' integrators might be optimal for a highly efficient and accurate integration of the cosmological Vlasov--Poisson system. Given the discreteness errors mentioned above (see further \cite{Michaux:2021,Marcos:2006,Joyce:2009} arising particularly during this early phase of integration), it might be preferable to perform as few timesteps as possible during this phase. 

Recently, fast approximate simulations have also garnered much interest as forward models in the context of simulation-based inference (e.g.\ \cite{Jasche2019, Cranmer2020}). In this setting, simulations need to be run many times for different parameter values, for which reason efficiency is imperative. A particularly interesting avenue is the development of differentiable cosmological simulations such as the \textsc{FlowPM} \cite{Modi2021} offspring of \textsc{FastPM}, or using the adjoint method as recently proposed by Ref.~\cite{Li2022}. These differentiable simulations allow for gradient-based optimisation and can be utilised for various inference tasks such as the estimation of cosmological parameters and for reconstructing the initial conditions of the universe \cite{Modi2021}. The new integrators we will present herein are ideally suited for the use in differentiable simulations.

In this article, we perform a thorough analysis of perturbation-theory informed integration schemes for cosmological simulations, along with an analytical characterisation of their properties. Importantly, we demonstrate that whereas \textsc{FastPM} incorporates the dynamics described by the growing-mode solution of first-order Lagrangian perturbation theory (LPT), i.e.\ the Zel'dovich approximation \cite{Zeldovich:1970}, similar second-order Strang splitting schemes with the same computational cost can be devised incorporating knowledge of \emph{second-order} LPT (2LPT), which leads to more accurate results in the pre-shell-crossing regime, i.e.\ at early times / on large scales. We will explicitly construct such 2LPT-inspired integrators and demonstrate their effectiveness in numerical experiments in 1D, 2D, and 3D. 

Recall that in 1D prior to shell crossing, the Zel'dovich `approximation' is, in fact, the exact solution. Therefore, although our integrators are universally defined regardless of the spatial dimensionality, we will enforce that they exactly reproduce the Zel'dovich trajectories before shell-crossing in 1D. We show numerically that this property for cosmological time integrators is also highly beneficial for realistic 3D simulations of cosmological structure formation, where it leads to a significant reduction in the number of timesteps required to obtain accurate results on mildly non-linear scales at late times.

In Section \ref{sec:theory}, we review the cosmological Vlasov--Poisson system and the resulting equations of motions for an $N$-body system; moreover, we provide a brief overview of LPT, which provides a perturbative solution for the Vlasov--Poisson system in the pre-shell-crossing regime. Then, we consider general drift-kick-drift (DKD) schemes that evolve the canonical position and momentum variables (w.r.t.\ an arbitrary time coordinate) and prove the following results regarding the symplecticity and accuracy of the time integrators in Section~\ref{sec:general_integrators}.
\begin{itemize}    
\item \emph{All} DKD schemes for the cosmological Vlasov--Poisson equations that are defined in terms of the canonical variables (see Definition~\ref{def:canonical_dkd_integrator}) are symplectic ($\to$~\textbf{Proposition~\ref{prop:symplectic}}).
\item Provided that the acceleration field is sufficiently smooth, globally 2$^\text{nd}$-order-in-time accuracy is achieved \emph{irrespective of the chosen time coordinate} such as cosmic time, superconformal time, scale-factor time, etc. ($\to$~\textbf{Proposition~\ref{prop:convergence}}). However, the commutator between the drift and kick operators introduces an additional globally $2^\text{nd}$-order-in-time error, which adds to the `usual' leapfrog approximation error. In order to construct higher-order schemes for cosmological simulations by sandwiching multiple drift and kick steps, these commutator terms need to be eliminated.

\item The shell-crossing of particles leads to a singularity in the derivative of the acceleration field, which formally limits the global order of convergence to $\nicefrac{3}{2}$ for collapse along a single spatial axis in the cold limit ($\to$~\textbf{Proposition~\textbf{\ref{prop:convergence_loss}}}).
\end{itemize}

After having studied general DKD integrators for cosmological simulations, we turn towards the second major subject of this work in Section~\ref{sec:lpt_integrators}, namely integrators inspired by LPT. We proceed as follows:

\begin{itemize}
    \item We define a class of integrators for which the kick operation is formulated in terms of a modified momentum variable $\vecb{\mPi}$, which describes the rate of change of the coordinates w.r.t.\ the growth factor $D^{+}$ ($\to$~\textbf{Definition~\ref{def:pi_integrator}}).

    \item We derive a simple necessary and sufficient condition for such integrators to be `Zel'dovich consistent', by which we mean that the exact solution in 1D prior to shell-crossing (given by the Zel'dovich solution) is correctly reproduced ($\to$ \textbf{Proposition~\ref{prop:characterisation_of_zeldovich_consistency}}).

    \item We revisit (the DKD variant of) the \textsc{FastPM} integrator introduced by Ref.~\cite{Feng:2016} and show that it is the only integrator in this class that is both Zel'dovich consistent and symplectic ($\to$ \textbf{Proposition}~\ref{prop:fastpm}).

    \item Then, we explicitly construct several new Zel'dovich-consistent schemes and discuss their properties. The schemes that will be most interesting for practical applications are \textsc{LPTFrog} ($\to$~\textbf{Example~\ref{example:lptfrog}}, inspired by 2LPT and rooted in contact geometry) and \textsc{TsafPM} ($\to$~\textbf{Example~\ref{example:tsafpm}}), which can be directly used as drop-in replacements of \textsc{FastPM}, and our most powerful integrator \textsc{PowerFrog} ($\to$~\textbf{Example~\ref{example:powerfrog}}, which asymptotes to 2LPT as $a \to 0$).     
\end{itemize}

In Section~\ref{sec:results_1D}, we present our numerical results in one dimension. We evaluate different integrators in an idealised scenario before and after shell-crossing occurs. Also, we consider the presence of a decaying mode in the initial conditions, and we analyse how well the cosmic energy balance as given by the Layzer--Irvine equation is satisfied for different integrators. In Section~\ref{sec:results_2D}, we compute a single timestep in two dimensions with a tree-PM code and compare the results of different integrators. Finally, we consider the \textsc{Quijote} \cite{quijote} and \textsc{Camels} \cite{Villaescusa-Navarro2020} $N$-body simulation suites as benchmarks for the realistic 3D case with $\Lambda$CDM cosmology in Section~\ref{sec:results_3D}. Starting from redshift $z = 127$, we perform PM simulations where we evolve the particles until $z = 0$ in a few steps using different integrators. We find that our new LPT-inspired integrators reduce the number of timesteps required to recover the statistics of the density field at a given accuracy. We conclude this work in Section~\ref{sec:conclusions}.


\section{Theory}
\label{sec:theory}
In this section, we present the cosmological Vlasov--Poisson equations, which describe the motion of collisionless self-gravitating matter in an expanding universe, before discussing how these equations are discretised in cosmological $N$-body simulations. Then, we summarise perturbation-theory solutions of these equations given cold initial data, which will be the basis for the development of our new numerical integrators in the main part of this work.

\subsection{Cosmological equations of motion for cold collisionless fluids}
Non-relativistic large-scale cosmological simulations \cite{AnguloHahn:2022} focus on evolving the Vlasov--Poisson system of equations \cite{Peebles:1980,Rampf:2021}, which describes the collisionless evolution of the phase-space distribution function $f(\vecb{x},\vecb{p},t):\;\mathscr{C}\times\mathbb{R}^3\times\mathbb{R}\to\mathbb{R}_0^+$  (of dark matter, galaxies, stars,...) in an expanding universe. Coordinates $\vecb{x}\in\mathscr{C}\subseteq\mathbb{R}^3$ live in a configuration space co-moving with the expansion of the universe, where typically for simulations a flat 3-torus topology is assumed on $\mathscr{C}=[0,1)^3$. Moreover, $\vecb{p}$ denotes the momentum coordinate, and $t$ is cosmic time. The Vlasov--Poisson system reads
\begin{subequations}
\label{eq:Vlasov-Poisson}
\begin{align}
    \dd_t f = \partial_t f + \frac{\vecb{p}}{a^2}\cdot\bnabla_{\vecb{x}} f + \frac{\vecb{g}}{a}\cdot\bnabla_{\vecb{p}} f = 0, \qquad\textrm{where}\qquad \vecb{g}:= -4\pi G M \, \bnabla_{\vecb{x}} \nabla_{\vecb{x}}^{-2} (n-1)  = - \frac{3 \Omega_\mathrm{m}}{2} \bnabla_{\vecb{x}} \nabla_{\vecb{x}}^{-2} \delta \label{eq:Vlasov-Poisson_a}
\end{align}
is the gravitational acceleration given through Poisson's equation sourced by the density $n$, which is the momentum-space marginal of the distribution function $f$
\begin{align}
    n \phantom{:}= \int_{\mathbb{R}^3}\dd^3p \;f,\qquad\textrm{with normalisation}\qquad\int_{\mathscr{C}}\dd^3x\;n = 1.
\end{align}
Here, $G$ is the gravitational constant, $M$ is the total mass contained in the simulation box $\mathscr{C}$, and we defined the density contrast $\delta = n - 1$.
Throughout this paper, we use time units relative to the Hubble time $1 / H_0$, for which reason $H_0$ does not appear in Eq.~\eqref{eq:Vlasov-Poisson}. The last identity for the acceleration in Eq.~\eqref{eq:Vlasov-Poisson_a} follows from the definition of the critical density of the universe $\rho_{\text{crit}} := 3 /(8 \pi G)$ and $\Omega_\mathrm{m} := M / \rho_{\text{crit}}$ in our unit system.
The cosmological scale factor $a(t)$ that relates co-moving coordinates $\vecb{x}$ and physical coordinates $\vecb{r}$ via $\vecb{r} = a \vecb{x}$ follows Friedmann's equation. For the purposes of this paper, we focus on Friedmann equations of the type
\begin{align}
    H:=\frac{\dot{a}}{a} = \sqrt{\Omega_\mathrm{m} a^{-3} + \Omega_\Lambda},\qquad\textrm{with}\qquad\Omega_\mathrm{m}+\Omega_\Lambda=1,
\end{align}
\end{subequations}
where $\Omega_\mathrm{m}$ and $\Omega_\Lambda$ are the present-day density parameters of non-relativistic matter and of the cosmological constant (i.e.\ the simplest description of dark energy), respectively. We will often consider the Einstein--de~Sitter (EdS) limit with $\Omega_\Lambda\to0$, where $a(t)\propto t^{2/3}$. We ignore other forms of energy here, but our results are straightforwardly extended. 

In the $\Lambda$CDM paradigm, we can focus on cold dark matter (CDM), which means that we can assume the cold limit for the distribution function $f$, specifically
\begin{align}
    f(\vecb{x},\vecb{p},t) =  \int_{\mathscr{Q}} \dd^3 q\;\delta_\textrm{D}(\vecb{x}-\vecb{x}(\vecb{q},t))\,\delta_\textrm{D}(\vecb{p}-\vecb{p}(\vecb{q},t)),
\end{align}
where $\delta_\textrm{D}$ denotes the Dirac delta distribution, $\mathscr{Q} = [0, 1)^3$ (equipped with a flat 3-torus topology) is the Lagrangian space, and $\vecb{q} \in \mathscr{Q}$ is the (3-dimensional) Lagrangian coordinate. The positions and momenta lie on a single 3-dimensional Lagrangian submanifold \cite{Arnold:1989,Abel:2012} of 3+3-dimensional phase space, i.e.\ we have the 1-parameter family $(\vecb{q};t)\mapsto(\vecb{x}(\vecb{q},t),\vecb{p}(\vecb{q},t))$.

\subsection{Cosmological equations of motion}
The equations of motion~\eqref{eq:Vlasov-Poisson} can be discretised in terms of the $N$-body method through a finite number of $N$ characteristics (cf.~\cite{AnguloHahn:2022}) by selecting a discrete subset of Lagrangian coordinates $\{\vecb{q}_i\}_{i=1\dots N}$. This implies approximating $f$ with the $N$-body distribution function
\begin{align}
    f(\vecb{x},\vecb{p},t) = \frac{1}{N} \sum_{i=1}^N \delta_\textrm{D}(\vecb{x}-\vecb{X}_i(t)) \, \delta_\textrm{D}(\vecb{p}-\vecb{P}_i(t))\;.
\end{align}
The corresponding system of characteristics then obeys a separable but explicitly time-dependent Hamiltonian
 \begin{align}
\mathscr{H} = \sum_{i=1}^N \frac{\|\vecb{P}_i\|^2}{2a^2} + \frac{1}{2a} \sum_{i=1}^N \sum_{j\neq i} I(\vecb{X}_i,\vecb{X}_j)  \label{eq:Hamiltonian_t}
\end{align}
in cosmic time $t$. We assume $N$ particles of identical mass, and we will write $\vecb{\xi}_i(t):=(\vecb{X}_i(t),\vecb{P}_i(t))$, $ \mathbb{R}\to\mathbb{R}^{3+3}$ for the 1-parameter family of conjugate comoving-coordinate and momentum trajectory associated with particle $i$. The gravitational pair-interaction term has the form
\begin{align}
\sum_{j \neq i} I(\vecb{X}_i,\vecb{X}_j) =: \varphi_N(\vecb{X}_i), 
\end{align}
where the (discretised) gravitational potential $\varphi_N(\vecb{X}_i)$ satisfies the Poisson equation
\begin{equation}
    \nabla_{\vecb{x}}^2 \varphi_N = 4 \pi G M (n - 1) = \frac{3}{2} \Omega_{\mathrm{m}} \delta.
    \label{eq:poisson}
\end{equation}
The canonical equations of motion are then given in cosmic time by the pair \citep{Peebles:1980}
\begin{equation}
    \dd_t \vecb{X}_i = \bnabla_{\vecb{P}_i} \mathscr{H} = a^{-2}\vecb{P}_i \qquad \text{and} \qquad
    \dd_t \vecb{P}_i = -\bnabla_{\vecb{X}_i} \mathscr{H} = - a^{-1}\bnabla_{\vecb{x}} \varphi_N( \vecb{X}_i ) =: a^{-1}\vecb{A}(\vecb{X}_i),
    \label{eq:eqs_of_motion_cosmic_time}
\end{equation}
where $\vecb{A}(\vecb{X}_i)$ is the acceleration of particle $i$.

There is exactly one time-coordinate, in which the kinetic energy in the Hamiltonian in Eq.~\eqref{eq:Hamiltonian_t} is not explicitly time-dependent. This is the well-known superconformal time defined by $\dd\tilde{t} := a^{-2} \, \dd t$ \citep{Doroshkevich:1973,Martel:1998}, for which the Hamiltonian becomes
\begin{align}
\tilde{\mathscr{H}} = \sum_{i=1}^N \frac{\|\vecb{P}_i\|^2}{2} + \frac{a}{2} \sum_{i=1}^N \sum_{j\neq i} I(\vecb{X}_i,\vecb{X}_j).  \label{eq:Hamiltonian_ttilde}
\end{align}
This leads to a natural grouping of the time dependency with the coordinates in the potential part, allowing an embedding in an extended phase space \citep[cf.][]{AnguloHahn:2022}. The equations of motion in superconformal time are given by
\begin{equation}
    \dd_{\tilde{t}} \vecb{X}_i = \bnabla_{\vecb{P}_i} \tilde{\mathscr{H}} =\vecb{P}_i \qquad \text{and} \qquad
    \dd_{\tilde{t}} \vecb{P}_i = -\bnabla_{\vecb{X}_i} \tilde{\mathscr{H}} = - a\,\bnabla_{\vecb{x}} \varphi_N( \vecb{X}_i ).
\label{eq:eqs_of_motion}
\end{equation}

\subsection{Perturbation theory results}
We will compare our numerical results against exact results and also make use of various standard results of LPT \citep[e.g.][]{Zeldovich:1970,Buchert:1993,Bouchet:1995,Rampf:2012} later in this paper; therefore, we will briefly review some key concepts here. Standard LPT provides a perturbative solution applicable for cold initial data and before the moment of `shell-crossing', i.e.\ the moment when multi-kinetic solutions appear (e.g.\ \cite{AnguloHahn:2022, Rampf:2021a}). Specifically, in LPT the Vlasov--Poisson system is recast into a fluid-like system of equations, which is then solved in Lagrangian coordinates using standard perturbative techniques. Most importantly, LPT at first order is \emph{exact} for cold one-dimensional initial data until shell-crossing. Second-order LPT predicts the evolution of particle trajectories as products of temporal parts $D(a)$ and spatial parts $\vecb{\psi}_{\vecb{q}}$ as
\begin{subequations}
\begin{align}
    \vecb{X}_{\vecb{q}}(a) &= \vecb{q} \, + \,  \vecb{\mPsi}_{\vecb{q}}(a) = \vecb{q} \;+\; D^{+}(a) \, \vecb{\psi}^{+}_{\vecb{q}} \;+\; D^{-}(a) \, \vecb{\psi}^{-}_{\vecb{q}}\; + \; D^{++}(a)\,\vecb{\psi}^{++}_{\vecb{q}} \;+\; D^{+-}(a)\,\vecb{\psi}^{+-}_{\vecb{q}} \;+\;D^{--}(a)\,\vecb{\psi}^{--}_{\vecb{q}} \;+\; \textrm{h.o.t.,} \label{eq:X_lpt} \\
    \vecb{P}_{\vecb{q}}(a) &= F^{+}(a) \, \vecb{\psi}^{+}_{\vecb{q}} \;+\; F^{-}(a)\, \vecb{\psi}^{-}_{\vecb{q}}\;+\; F^{++}(a)\, \vecb{\psi}^{++}_{\vecb{q}}\;+\; F^{+-}(a)\, \vecb{\psi}^{+-}_{\vecb{q}}\;+\; F^{--}(a)\, \vecb{\psi}^{--}_{\vecb{q}} \;+\; \textrm{h.o.t.,}
\end{align}
\label{eq:XP_lpt}
\end{subequations}
where $\vecb{\mPsi}_{\vecb{q}}$ is the total displacement of the particle with Lagrangian coordinate $\vecb{q}$ and $F^\pm(a) := a^3 H \,\dd_a D^\pm$ since $\vecb{P_q}=a^3H \,\dd_a \vecb{X_q}$. The $^{\pm}$ superscripts stand for growth ($^+$) and decay ($^-$). Note that for our purposes it is good enough to approximate the second-order terms as $D^{++} \simeq (D^+)^2$, $D^{+-} \simeq D^+D^-$, and $D^{--} \simeq (D^-)^2$ \cite{Bouchet:1995,Rampf:2022}. The spatial parts to all orders are fully determined by the first-order pieces $\vecb{\psi}^+$ and $\vecb{\psi}^-$; explicit expressions for the higher-order spatial terms can be found e.g.\ in Ref.~\cite{Bouchet:1995}. It is common in studies of the large-scale structure to focus only on the fastest growing modes (since decaying modes are typically small in the regime of interest) so that one sets $\vecb{\psi}^-\equiv0$ and as a consequence also has $\vecb{\psi}^{+-}=\vecb{\psi}^{--}=0$. In this \emph{fastest growing limit}, one has at second order
\begin{subequations}
\begin{align}
    \vecb{X}_{\vecb{q}}(a) &\asymp \vecb{q} \;+\; D^{+}(a) \, \vecb{\psi}^{+}_{\vecb{q}} \;+\;\left[D^{+}(a)\right]^2 \, \vecb{\psi}^{++}_{\vecb{q}} \;+\;\textrm{h.o.t.,} \label{eq:2lpt_fastest_growing_X} \\
    \vecb{P}_{\vecb{q}}(a) &\asymp F^{+}(a) \, \vecb{\psi}^{+}_{\vecb{q}} \;+\; 2 D^{+}(a)\, F^{+}(a)\, \vecb{\psi}^{++}_{\vecb{q}} \;+\; \textrm{h.o.t.,} 
\end{align}
\label{eq:2lpt_fastest_growing}
\end{subequations}
where
\begin{subequations}
\begin{align}
    \vecb{\psi}^{+}_{\vecb{q}} &= - \bnabla_{\vecb{q}} \varphi^{(1)}, & \text{with} \  \varphi^{(1)} &= \varphi_{\mathrm{ini}}, \\    
    \qquad \vecb{\psi}^{++}_{\vecb{q}} &= -\frac{3}{7} \bnabla_{\vecb{q}} \varphi^{(2)}, & \text{with} \ \varphi^{(2)} &= \frac{1}{2} \nabla^{-2}_{\vecb{q}} \left[\varphi^{(1)}_{,ii} \varphi^{(1)}_{,jj} - \varphi^{(1)}_{,ij} \varphi^{(1)}_{,ij} \right].    
\end{align}
\label{eq:lpt_displacements}
\end{subequations}
Here, we used Einstein's sum convention and defined $\varphi_{,ij} := \partial_{{q}_i} \partial_{{q}_j} \varphi$; further, $\varphi_{\mathrm{ini}}$ is the gravitational potential for $a \to 0$, and $\varphi^{(1)}$ and $\varphi^{(2)}$ are the potentials sourcing the 1LPT and 2LPT contributions to the displacement field, respectively.
The \emph{Zel'dovich approximation} consists in neglecting second and higher-order contributions in $D^{+}(a)$ in Eqs.~\eqref{eq:2lpt_fastest_growing}.
For $\Lambda$CDM cosmologies, the first-order time-dependent functions can be given in closed form as \citep{Chernin:2003,Demianski:2005}
\begin{subequations}
\begin{alignat}{-1}
     D^+(a) &= a\; {}_2{\rm F}_1 \! \left(\nicefrac{1}{3}, 1, \nicefrac{11}{6}, - \lambda_0\, a^3\right) &&\stackrel{\textrm{EdS}}{\asymp} a, \label{eq:D+} \\
D^-(a)  &= a^{-\nicefrac{3}{2}} \;\sqrt{1+ \lambda_0\, a^3}&&\stackrel{{\textrm{EdS}}}{\asymp} a^{-\nicefrac{3}{2}} , \\
F^+(a) &=\frac{3}{2}\frac{a^3 H}{1+a^3\lambda_0} \left(\frac{5}{3}-\frac{D^+}{a}\right) && \stackrel{{\textrm{EdS}}}{\asymp} a^{\nicefrac{3}{2}},\\
F^-(a) &= -\frac{3}{2} \frac{H}{a D^-(a)} && \stackrel{{\textrm{EdS}}}{\asymp} -\frac{3}{2a},
\end{alignat}
\label{eq:Ds_and_Fs}
\end{subequations}
where $\lambda_0:= \Omega_\Lambda/\Omega_{\rm m}$, ${}_2{\rm F}_1$ is Gauss' hypergeometric function, and the  EdS asymptotics is for $\Omega_{\rm m}=1,\Omega_{\Lambda}=0$. Note that we do not adopt here the usual normalisation that $D^+(a=1)=1$ so that the functions for EdS and $\Lambda$CDM agree for $a\ll1$.

\section{General results on numerical integrators}
\label{sec:general_integrators}
In this section, we will present some general results regarding the symplecticity and convergence of numerical integrators that are defined in terms of the canonical variables. For convenience, we collect the three canonical coordinates $(x, y, z)$ of the $N$ particles in the vectors $\vecb{X}, \vecb{P} \in \mathbb{R}^{3N}$ defined as
\begin{subequations}
\begin{align}
    \vecb{X} = \left(X_{1,x}, X_{1,y}, X_{1,z}, \ldots, X_{N,x}, X_{N,y}, X_{N,z}\right), \\
    \vecb{P} = \left(P_{1,x}, P_{1,y}, P_{1,z}, \ldots, P_{N,x}, P_{N,y}, P_{N,z}\right).
\end{align}
\end{subequations}

Let us note that regardless of the time coordinate (e.g.\ cosmic time $t$ or superconformal time $\tilde{t}$), the Hamiltonian of the cosmological Vlasov--Poisson system is \emph{separable}. Whenever the time coordinate is left unspecified, we will denote the Hamiltonian as $\mathcal{H}$ as opposed to $\mathscr{H}$. We take $\tau$ to denote a general time variable (\emph{not} the conformal time, as is common in the cosmology literature). Then, the $N$-body Hamiltonian can be written in the form
\begin{equation}
    \mathcal{H} = \mu(\tau) T(\vecb{P}) + \nu(\tau) U(\vecb{X}),
    \label{eq:Hamiltonian_general}
\end{equation}
where $\mu(\tau)$ and $\nu(\tau)$ are functions of the time variable $\tau$, and $T$ and $U$ are the kinetic and potential energy, respectively. In particular, this implies that the update of the positions (momenta) in the drift (kick) operator only depends on the momenta (positions). In fact, for the cosmological $N$-body problem, $\mu(\tau)$ and $\nu(\tau)$ are related by $\nu(\tau) = \mu(\tau) a(\tau)$ for any choice of $\tau$ (when defining the potential $\varphi_N$ in such a way that the Poisson equation \eqref{eq:poisson} is $a$-independent).

We start with the definition of a canonical three-step (leapfrog/Verlet) scheme, i.e.\ a time integration scheme that evolves the canonical position and momentum variables. 

\begin{definition}[Canonical DKD integrator]
\label{def:canonical_dkd_integrator}
We define a \emph{canonical DKD integrator} as a numerical integrator that updates the particle positions and momenta according to
\begin{subequations}
\begin{align}
\vecb{X}^{n+\nicefrac{1}{2}}_i &= \vecb{X}^n_i + \alpha(\tau_n,  \tau_{n+1}) \vecb{P}^n_i, \label{eq:dkd_drift_1} \\
\vecb{P}^{n+1}_i &= \vecb{P}^n_i + \beta(\tau_n, \tau_{n+1}) \vecb{A}\left(\vecb{X}^{n+\nicefrac{1}{2}}_i\right), \label{eq:dkd_kick} \\
\vecb{X}^{n+1}_i &= \vecb{X}^{n+\nicefrac{1}{2}}_i + \gamma(\tau_n, \tau_{n+1}) \vecb{P}^{n+1}_i, \label{eq:dkd_drift_2}
\end{align}
\label{eq:leapfrog_scheme}
\end{subequations}
with arbitrary functions $\alpha, \beta, \gamma$ of the initial time $\tau_n$ and end time $\tau_{n+1}$ of the $n$-th step. The superscript ${}^{n + \nicefrac{1}{2}}$ indicates the intermediate time, which splits the drift into two parts.
\end{definition}
We almost exclusively consider the DKD case in this work, but our theoretical results readily carry over to KDK schemes, see also \ref{sec:KDK}. While the position and momentum updates are modulated by the factors $\alpha$, $\beta$, and $\gamma$, we fix the factors in front of the previous positions $\vecb{X}_i^n$ and $\vecb{P}_i^n$ to unity here as they are intimately linked to the interpretation of the drift and kick as the infinitesimal generators of the Lie group of symplectic maps, see Eq.~\eqref{eq:drift_and_kick_as_generators} in \ref{sec:proof_proposition2}. In Section~\ref{sec:lpt_integrators}, we will relax this requirement for the kick and introduce a multiplicative factor that will allow us to match trajectories to perturbation theory results (at the cost of giving up exact symplecticity).

Due to the underlying Hamiltonian nature, it is desirable to preserve the symplectic form $\omega_i = \sum_{\ell=1}^3 \dd X_{i,\ell}\wedge \dd P_{i,\ell}$ on the conjugate variables along the flow $\vecb{\xi}_i(\tau)$.
This is achieved by \emph{symplectic integrators}, for which the mapping $(\vecb{X}_i^0, \vecb{P}_i^0) \mapsto (\vecb{X}_i^n, \vecb{P}_i^n)$ from the initial coordinates of a particle $i$ to those at timestep $n$ is a canonical transformation. Although the time-discrete system will not exactly follow the dynamics as governed by the true Hamiltonian, it can be shown that the time-discrete system follows the dynamics described by a perturbed so-called `shadow' Hamiltonian (e.g.\
\cite{skeel2001practical, hairer2006geometric}). Importantly, this implies that for autonomous Hamiltonians (such as in the case of a bound gravitating system not subject to the cosmological expansion), the energy error of symplectic integrators has no secular term that would cause spurious energy production or dissipation as time progresses (see e.g.\ Ref.~\cite{Yoshida:1990}). While rigorously analysing the stability of numerical integrators is challenging in the non-linear case, and multiple notions of stability exist (e.g.\ \cite{mclachlan2004nonlinear}), symplecticity endows numerical integrators with certain stability properties \cite{shang1999kam}, as can be shown e.g.\ with KAM theory \cite{kolmogorov1954conservation, moser1962invariant, Arnold_1963}.

As it turns out, symplecticity is no exceptional property among integrators of the type in Eqs.~\eqref{eq:leapfrog_scheme}, as the following proposition shows.

\begin{proposition}\label{prop:symplectic}
Let the acceleration field $\vecb{A} = \vecb{A}(\vecb{X})$ be a differentiable function of the spatial coordinates. Then, any canonical DKD integrator is symplectic.

\begin{proof}
We will suppress the dependence of $\alpha$, $\beta$, $\gamma$ on time in the notation for simplicity. To show symplecticity, we write Eqs.~\eqref{eq:leapfrog_scheme} as
\begin{subequations}
\begin{align}
\vecb{X}^{n+1}_i &= \vecb{X}^n_i + \alpha \vecb{P}^n_i + \gamma  \left( \vecb{P}_i^n + \beta \,\vecb{A}(\vecb{X}_i^n + \alpha \vecb{P}_i^n)\right), \\
\vecb{P}^{n+1}_i &= \vecb{P}^n_i + \beta \,\vecb{A}(\vecb{X}^n_i + \alpha \vecb{P}^n_i).
\end{align}
\label{eq:canonical_integrator_in_symplecticity_proof}
\end{subequations}
Let us define the standard symplectic matrix
\begin{equation}
    \mathbf{J} = 
    \begin{pmatrix}
    0 & \mathbb{I}_3 \\
    -\mathbb{I}_3 & 0
    \end{pmatrix},
\end{equation}
which defines the canonical structure. Here, $\mathbb{I}_3$ is the $3$-dimensional identity matrix. Further, we define a function $\mathcal{F}_i^n: \vecb{\xi}_i^n = (\vecb{X}_i^n, \vecb{P}_i^n) \mapsto (\vecb{X}_i^{n+1}, \vecb{P}_i^{n+1}) = \vecb{\xi}_i^{n+1}$ for the mapping of particle $i$ from timestep $n$ to timestep $n+1$ via Eqs.~\eqref{eq:canonical_integrator_in_symplecticity_proof}. The Jacobian matrix of this mapping can be computed in a straightforward manner as
\begin{equation}
    \bnabla_{\vecb{\xi}_i^n} \mathcal{F}_i^n = 
        \begin{pmatrix}
        \mathbb{I}_3 + \beta\gamma \bnabla_{\vecb{x}} \vecb{A}(\vecb{X}^n_i + \alpha \vecb{P}^n_i) & (\alpha+\gamma) \mathbb{I}_3 + \alpha\beta\gamma \bnabla_{\vecb{x}} \vecb{A}(\vecb{X}^n_i + \alpha \vecb{P}^n_i) \\
        \beta \bnabla_{\vecb{x}} \vecb{A}(\vecb{X}^n_i + \alpha \vecb{P}^n_i) & \mathbb{I}_3 + \alpha\beta \bnabla_{\vecb{x}} \vecb{A}(\vecb{X}^n_i + \alpha \vecb{P}^n_i).
        \end{pmatrix},
\end{equation}
From this, it is easily confirmed that 
\begin{equation}
(\bnabla_{\vecb{\xi}_i^n} \mathcal{F}_i^n)^\top \, \mathbf{J} \, (\bnabla_{\vecb{\xi}_i^n} \mathcal{F}_i^n) = \mathbf{J},
\end{equation}
implying that the mapping $\mathcal{F}$ is a canonical transformation, i.e.\ the symplectic form $\omega_i = \sum_{\ell=1}^3 \dd X_{i,\ell}^{n+1} \, \wedge \, \dd P_{i,\ell}^{n+1} = \sum_{\ell=1}^3 \dd X_{i,\ell}^{n} \, \wedge \, \dd P_{i,\ell}^{n}$ is conserved, and canonical DKD integrators are therefore symplectic.
\end{proof}
\end{proposition}
In particular, this means that the symplecticity of the DKD integrator is obtained irrespective of the chosen time coordinate, which only affects the coefficients $\alpha, \beta$, and $\gamma$.
Next, we discuss the order of accuracy of the canonical DKD integrator. While global second-order accuracy of the DKD leapfrog integrator for time-independent Hamiltonians can be shown in a straightforward manner based on Taylor expansions in time of the canonical variables, the situation is less clear in the time-dependent case of the cosmological $N$-body Hamiltonian where commutators between the functions $\mu(\tau)$ and $\nu(\tau)$ need to be taken into account. Fortunately, these terms can be shown to be of third order (for each timestep) such that the global accuracy is still second order, provided that some consistency conditions are satisfied, as the following proposition shows.

\begin{proposition} \label{prop:convergence}
For a canonical DKD integrator, let the following three properties hold for all $\tau$ and $\Delta \tau$:
\begin{subequations}
\begin{align}
\left(\alpha + \gamma\right)(\tau, \tau + \Delta \tau) &= C_\mu(\tau, \tau + \Delta \tau) + \mathscr{O}((\Delta \tau)^3), \\
\beta(\tau, \tau + \Delta \tau) &= C_\nu(\tau, \tau + \Delta \tau) + \mathscr{O}((\Delta \tau)^3), \\
\left((\alpha - \gamma) \beta\right)(\tau, \tau + \Delta \tau) &= \mathscr{O}((\Delta \tau)^3), \label{eq:general_dkd_consistency_conditions_3}
\end{align}
\label{eq:general_dkd_consistency_conditions}
\end{subequations}
where $\mu$ and $\nu$ express the time dependence of the Hamiltonian $\mathcal{H}$ in Eq.~\eqref{eq:Hamiltonian_general} and we defined for any function $f$
\begin{equation}
      C_f(\tau, \tau+\Delta \tau) := \int_\tau^{\tau+\Delta \tau} \dd \tau_1\, f(\tau_1).
\end{equation}
Then the DKD integrator is locally third-order (and globally second-order) accurate.
\begin{proof}
Due to its length, the proof is provided in \ref{sec:proof_proposition2}.
\end{proof}
\end{proposition}
The proof of Proposition~\ref{prop:convergence}, which is based on a Magnus expansion \cite{Magnus1954, Blanes2009} for the DKD operator, shows that even when choosing $\alpha, \beta, \gamma$ such that $\alpha + \gamma = C_\mu$ and $\beta = C_\nu$, a third-order error arises from commutators between the drift and kick operators, where the leading-order term is given by
\begin{equation}
    C_{[\mu, \nu]}(\tau, \tau + \Delta \tau) = \frac{1}{2} \int_{\tau}^{\tau + \Delta \tau} \dd \tau_1 \int_{\tau}^{\tau_1} \dd \tau_2 \, \left(\mu(\tau_1) \nu(\tau_2) - \mu(\tau_2) \nu(\tau_1)\right) = \mathscr{O}((\Delta \tau)^3).
\label{eq:C_mu_nu_commutator}
\end{equation}
Note that this term arises from the time dependence of the Hamiltonian in Eq.~\eqref{eq:Hamiltonian_general} and vanishes for the time-independent case $\mu \equiv \nu \equiv 1$. We will explicitly compute the terms $C_\mu$ and $C_\nu$, as well as the leading-order error term $C_{[\mu, \nu]}$ for cosmic time $\tau = t$ and superconformal time $\tau = \tilde{t}$ below. 

Let us also briefly inspect the `symmetry condition' in Eq.~\eqref{eq:general_dkd_consistency_conditions_3}. For an equal split of the drift, i.e.\ $\alpha = \gamma = C_\mu / 2$, this condition is trivially satisfied, and one hence obtains global second-order accuracy. On the other hand, for a fixed unequal split, i.e.\ $\alpha = \theta C_\mu$ and $\gamma = (1 - \theta) C_\mu$ for $\theta \in (0, 1) \setminus \{\nicefrac{1}{2}\}$, one finds that $(\alpha - \gamma) \beta = \mathscr{O}((\Delta \tau)^2)$ (assuming $\beta = C_\nu$), and the global accuracy is therefore only first order. Another possible option for achieving second-order accuracy is given by a midpoint approximation $\alpha = \int_{\tau}^{\tau_*} \dd \tau_1\, \mu(\tau_1)$ and $\gamma = \int_{\tau_*}^{\tau + \Delta \tau} \dd \tau_1\, \mu(\tau_1)$, where $\tau_*(\tau, \tau + \Delta \tau)$ can be any averaging function within the class of Kolmogorov's generalised $f$-means \cite{kolmogorov1991mean}, for example the arithmetic mean $\tau_*(\tau, \tau + \Delta \tau) = \tau + \Delta \tau / 2$ or the geometric mean $\tau_*(\tau, \tau + \Delta \tau) = (\tau (\tau + \Delta \tau))^{\nicefrac{1}{2}}$.

\begin{example}[Superconformal time]    
\label{example:superconft}
For the Hamiltonian in superconformal time in Eq.~\eqref{eq:Hamiltonian_ttilde}, i.e.\ $\tau = \tilde{t}$, one has
\begin{equation}
    \mu(\tilde{t}) \equiv 1 \qquad \text{and} \qquad \nu(\tilde{t}) = a(\tilde{t}).
\end{equation}
For a timestep $\Delta \tilde{t}$ in superconformal time, this yields 
\begin{subequations}
\begin{align}
    C_{\mu}(\tilde{t}, \tilde{t} + \Delta \tilde{t}) &= \Delta \tilde{t}, \\
    C_{\nu}(\tilde{t}, \tilde{t} + \Delta \tilde{t}) &= \int_{\tilde{t}}^{\tilde{t} + \Delta \tilde{t}} \dd \tilde{t}_1 \, a(\tilde{t}_1) = \int_{a(\tilde{t})}^{a(\tilde{t} + \Delta \tilde{t})} \frac{\dd a_1}{H(a_1) a_1^2} \stackrel{\mathrm{EdS}}{\asymp} 2 \left(\sqrt{a(\tilde{t} + \Delta \tilde{t})} - \sqrt{a(\tilde{t})}  \right).
\end{align}
\end{subequations}    
The integral for the kick can either be evaluated exactly or by a numerical quadrature formula that is at least second-order accurate, such as the midpoint rule $\beta = \Delta \tilde{t} \, a(\tilde{t} + \Delta \tilde{t} / 2) + \mathscr{O}((\Delta \tilde{t})^3)$. 

Note that although the explicit time dependence of the Hamiltonian has been shifted entirely to $\nu$ (and therefore to the potential energy) when using superconformal time and $\mu$ is constant, the $\mathscr{O}((\Delta \tilde{t})^3)$ error terms on the right-hand side of Eqs.~\eqref{eq:general_dkd_consistency_conditions}, which arise in the Magnus expansion from commutators between the drift and kick operators, do not vanish. For the leading-order term, one obtains
\begin{equation}
\begin{aligned}
    C_{[\mu, \nu]}(\tilde{t}, \tilde{t}+\Delta \tilde{t}) &\stackrel{\phantom{\text{EdS}}}{=} \frac{1}{2} \int_{\tilde{t}}^{\tilde{t}+\Delta \tilde{t}} \dd \tilde{t}_1 \int_{\tilde{t}}^{\tilde{t}_1} \dd \tilde{t}_2 \, \left(\mu(\tilde{t}_1) \nu(\tilde{t}_2) - \mu(\tilde{t}_2) \nu(\tilde{t}_1)\right) \\
    &\stackrel{\phantom{\text{EdS}}}{=} \frac{1}{2} \int_{a(\tilde{t})}^{a(\tilde{t} + \Delta \tilde{t})} \dd a_1 \int_{a(\tilde{t})}^{a_1} \dd a_2 \, \frac{a_2 - a_1}{H(a_1) a_1^3 H(a_2) a_2^3} \\
    &\stackrel{\mathrm{EdS}}{\asymp} 2 \left(\log \left(\frac{a(\tilde{t}+\Delta \tilde{t})}{a(\tilde{t})}\right) + \sqrt{\frac{a(\tilde{t})}{a(\tilde{t}+\Delta \tilde{t})}} - \sqrt{\frac{a(\tilde{t}+\Delta \tilde{t})}{a(\tilde{t})}}\right) \\
 &\stackrel{\phantom{\text{EdS}}}{=} - \frac{(\dd_{\tilde{t}} a)^3}{12 a(\tilde{t})^3} (\Delta \tilde{t})^3 + \mathscr{O}((\Delta \tilde{t})^4),
 \label{eq:C_mu_nu_ttilde}
\end{aligned}    
\end{equation}
which is indeed a third-order contribution, where $\dd_{\tilde{t}} a = H(a) a^3$.
Superconformal time is used for example by the cosmological $N$-body code \textsc{Ramses} \cite{Teyssier2002}.

\end{example}
\begin{example}[Cosmic time]
    When instead considering the Hamiltonian in cosmic time given by Eq.~\eqref{eq:Hamiltonian_t}, i.e.\ $\tau = t$, one has
    \begin{equation}
        \mu(t) = a(t)^{-2} \qquad \text{and} \qquad \nu(t) = a(t)^{-1}.
    \end{equation}
    Written in terms of cosmic time variables, the terms on the right-hand side of Eqs.~\eqref{eq:general_dkd_consistency_conditions} read as    
    \begin{subequations}
    \begin{alignat}{-1}
        C_{\mu}(t, t + \Delta t) &= \int_t^{t+\Delta t} \frac{\dd t_1}{a(t_1)^{2}} &{}= \int_{a(t)}^{a(t+\Delta t)} \frac{\dd a_1}{H(a_1) a_1^{3}} &\stackrel{\mathrm{EdS}}{\asymp} \frac{2}{\sqrt{a(t)}} - \frac{2}{\sqrt{a(t + \Delta t)}}, \\        
        C_{\nu}(t, t + \Delta t) &= \int_t^{t+\Delta t} \frac{\dd t_1}{a(t_1)} &{}= \int_{a(t)}^{a(t+\Delta t)} \frac{\dd a_1}{H(a_1) a_1^{2}} &\stackrel{\mathrm{EdS}}{\asymp} 2 \left(\sqrt{a(t + \Delta t)} - \sqrt{a(t)} \right).
    \end{alignat}
    \end{subequations}
    Written in terms of $a$, these are the same expressions as in Example~\ref{example:superconft}. Clearly, changing the coordinates also leaves the error terms unaffected, and Eqs.~\eqref{eq:C_mu_nu_ttilde} for $C_{[\mu, \nu]}$ still apply when replacing $(\tilde{t}, \Delta \tilde{t})$ by $(t, \Delta t)$.
\end{example}

\subsection{Higher-order symplectic integrators}
\label{sec:higher_order}
Higher-order integrators can be constructed by sandwiching lower-order operators and then requiring terms in the Baker--Campbell--Hausdorff \cite{campbell1897, baker1905, hausdorff1906} vs. Magnus expansions to match at increasingly higher order as done in the proof of Proposition~\ref{prop:convergence} -- see Ref.~\cite{Yoshida:1990} for a derivation of operators up to 8$^\text{th}$ order that involve, however, positive and negative time coefficients. Also, alternative symplectic formulations with purely positive coefficients are possible, see Ref.~\cite{Chin:2001} for a 4$^\text{th}$-order method. 

To our knowledge, integrators with order $> 2$ have not yet been employed in cosmological simulations, and it is an interesting question whether (and if so, for which degree of desired accuracy) higher-order integrators are able to achieve a given accuracy level with lower computational cost than the standard second-order leapfrog integrator.

In this work, we consider the well-known Forest--Ruth integrator \cite{Yoshida:1990,ForestRuth:1990,CandyRozmus:1991}, which simply consists of a threefold application of the canonical DKD integrator with the symmetric choice $\alpha = \gamma$, where two free parameters determine the length of the inner ($x_0$) and the two outer ($x_1$) DKD applications. These parameters $x_0$ and $x_1$ are chosen in such a way that the locally 3$^\text{rd}$ and $4^\text{th}$-order errors vanish. In superconformal time, this method can be readily extended to the cosmological $N$-body setting, yielding the following symplectic 4$^\text{th}$-order scheme:
\begin{equation}
\begin{aligned}
\vecb{X}^{n \like{25}}_i &= \vecb{X}^n_i + \frac{x_1}{2} \Delta \tilde{t} \, \vecb{P}^n_i, \\[0.25cm]
\vecb{X}^{n \like{50}}_i &= \vecb{X}^{n \like{25}}_i + \frac{x_0 + x_1}{2} \Delta \tilde{t} \, \vecb{P}^{n \like{33}}_i, \\[0.25cm]
\vecb{X}^{n \like{75}}_i &= \vecb{X}^{n \like{50}}_i + \frac{x_0 + x_1}{2} \Delta \tilde{t} \, \vecb{P}^{n \like{67}}_i, \\[0.25cm]
\vecb{X}^{n+1}_i &= \vecb{X}^n_i + \frac{x_1}{2} \Delta \tilde{t} \, \vecb{P}^{n+1}_i,
\end{aligned}
\qquad
\begin{aligned}
\vecb{P}^{n \like{33}}_i &= \vecb{P}^n_i + \vecb{A}(\vecb{X}^{n \like{25}}_i) \int_{a(\tilde{t}_n)}^{a(\tilde{t}_n + x_1 \Delta \tilde{t})} \frac{\dd a_1}{H(a_1) a_1^2}, \\
\vecb{P}^{n \like{67}}_i &= \vecb{P}^{n \like{33}}_i + \vecb{A}(\vecb{X}^{n \like{50}}_i) \int_{a(\tilde{t}_n + x_1 \Delta \tilde{t})}^{a(\tilde{t}_n + (x_0 + x_1) \Delta \tilde{t})} \frac{\dd a_1}{H(a_1) a_1^2}, \\
\vecb{P}^{n+1}_i &= \vecb{P}^{n \like{66}}_i + \vecb{A}(\vecb{X}^{n \like{75}}_i) \int_{a(\tilde{t}_n + (x_0 + x_1) \Delta \tilde{t})}^{a(\tilde{t}_{n+1})} \frac{\dd a_1}{H(a_1) a_1^2}, \\
\end{aligned}
\end{equation}
where the numerical factors $x_0$ and $x_1$ are given by \cite{Yoshida:1990}
\begin{equation}
    x_0 = - \frac{2^{\nicefrac{1}{3}}}{2 - 2^{\nicefrac{1}{3}}}, \qquad x_1 = \frac{1}{2 - 2^{\nicefrac{1}{3}}},
\end{equation}
and the partially filled circles indicate the progressive evolution from time $\tilde{t}_n$ to $\tilde{t}_{n+1}$. The choice of superconformal time turns out to be convenient here as it allows us to carry over the Forest--Ruth integrator for time-independent Hamiltonians to the cosmological $N$-body case without difficulty: since the kinetic part of the Hamiltonian has no explicit time dependence, the drifts for the cosmological $N$-body Hamiltonian are the same as for the time-independent case, and the usual multiplication of the kick factors with $x_0$ and $x_1$ now becomes a change of the integration boundaries. Another way to see this would be to switch to an extended phase space, where $a$ is appended to $\vecb{X}$ such that $(\vecb{X}, a)$ is the extended position variable. Then, the Hamiltonian becomes time-independent in terms of this extended set of variables. The same idea could be used to construct integrators for cosmological $N$-body simulations with even higher order, e.g.\ using the coefficients as derived in Ref.~\cite{Yoshida:1990}. 

So far, we have not discussed a potentially limiting factor for achieving convergence at high order, however, which is the lack of sufficient regularity of the Lagrangian acceleration field $\dd^2_a \vecb{X}_{\vecb{q}}(a) = \dd^2_a \left(\vecb{q} + \vecb{\mPsi}_{\vecb{q}}(a)\right) = \dd^2_a \vecb{\mPsi}_{\vecb{q}}(a)$, where $\vecb{\mPsi}_{\vecb{q}}(a)$ is the displacement field. In particular, the following proposition shows that even if the conditions in Eqs.~\eqref{eq:general_dkd_consistency_conditions} are satisfied, a singularity in the derivative of the acceleration field will affect the order of convergence:

\begin{proposition}\label{prop:convergence_loss}
If the derivative of the (Lagrangian) acceleration field $\dd^2_a \vecb{X}_{\vecb{q}}(a)$ w.r.t.\ the scale factor $a$ has a singularity of type
\begin{equation}
    \dd^3_a  \vecb{X}_{\vecb{q}}(a) = C (a - a_*)^{-\theta}
\label{eq:singularity}
\end{equation}
for some $C \in \mathbb{R} \setminus \{0\}$, $\theta \in (0, 1)$ and $a > a_*$, the local order of convergence of the particle positions $\vecb{X}$ and momenta $\vecb{P}$ in the regime $a > a_*$ is limited to $3 - \theta$.
\begin{proof}
Let us assume that $\alpha, \beta, \gamma$ are chosen in such a way that Eqs.~\eqref{eq:general_dkd_consistency_conditions} are satisfied, i.e.\ the error terms arising from the Magnus expansion up to (locally) second order vanish. Then, the leading-order error of the canonical DKD integrator is locally of third order and has two contributions: 1) the error from the unequal time commutator in the Magnus expansion expressed by Eq.~\eqref{eq:C_mu_nu_commutator} and 2) the approximation error arising from the fact that the DKD scheme is a (globally) second-order method. Note that the former term vanishes for time-independent Hamiltonians whereas the latter does not.
We will translate the arbitrary time coordinate $\tau$ and the timestep $\Delta \tau$ used by the DKD integrator to their counterparts in terms of the scale factor, i.e.\ $a$ and $\Delta a$. In view of Taylor's formula with integral remainder, the approximation error for the position of a particle $i$ of the DKD scheme for a timestep from $a_n$ to $a_{n+1}$ can be written as
\begin{equation}
    \mathcal{E}(\vecb{X}_i^n, \Delta a) = \Big{|}\int_{a_n}^{a_{n+1}} \dd \eta \, \frac{(\dd_a^3 \vecb{X}_i)(\eta)}{2} \, (a_{n+1} - \eta)^2 \Big{|}.
\end{equation}
Let us select the time index $n$ for which $a_* \in [a_n, a_{n+1})$ and consider the part of the integral from the scale factor of the singularity $a_*$ to $a_{n+1}$.
This yields
\begin{equation}
\begin{aligned}
\Big{|} \int_{a_*}^{a_{n+1}} \dd \eta \,  \frac{(\dd_a^3 \vecb{X}_i)(\eta)}{2} \, (a_{n+1} - \eta)^2 \Big{|} &=
     \frac{|C|}{2} \int_{a_*}^{a_{n+1}} \dd \eta \, (\eta - a_*)^{-\theta} \, (a_{n+1} - \eta)^2 \\         
     &\geq \frac{|C|}{2} \int_{a_*}^{a_{n+1}} \dd \eta \, (a_{n+1} - a_*)^{-\theta} \, (a_{n+1} - \eta)^2 \\ &= \frac{|C|}{6} (a_{n+1} - a_*)^{3 - \theta},
\end{aligned}
\end{equation}
where we used the fact that $\theta \in (0, 1)$. If we assume the scale factor of the singularity $a_*$ falls randomly in the interval $[a_n, a_{n+1})$, the expected value of $a_{n+1} - a_*$ is $(a_{n+1} - a_n) / 2 = \Delta a / 2$. Since we have $\theta > 0$, this is the dominant error contribution, which limits the local convergence order of numerical integrators to $3 - \theta$ for $\vecb{X}$. A similar argument can be made for the momentum $\vecb{P}$.
\end{proof}
\end{proposition}

\begin{remark}[Loss of regularity due to shell-crossing]
\label{remark:convergence_loss}
The relevance of Proposition~\ref{prop:convergence_loss} in the context of cosmological simulations is the following. At early times, the displacement of particles is small, and particle trajectories have not yet crossed. In this regime, the trajectory at each Eulerian coordinate $\vecb{X}_{\vecb{q}}(a) = \vecb{q} + \vecb{\mPsi}_{\vecb{q}}(a)$ can be uniquely mapped back to the (Lagrangian) initial position $\vecb{q}$ where the particle originated, meaning the fluid is `single stream'. In view of mass conservation, the Eulerian density contrast in the single-stream regime can be written as
\begin{equation}
    \delta(\vecb{X}_{\vecb{q}}(a)) + 1 = \int_{\mathscr{Q}} \dd^3 q_1 \, \delta_\textrm{D}(\vecb{X}_{\vecb{q}}(a) - \vecb{X}_{\vecb{q}_1}(a)) = \left(\det (\bnabla_{\vecb{q}} \vecb{X}_{\vecb{q}}(a)) \right)^{-1}.
\end{equation}
We remind the reader that $\delta(\vecb{X}_{\vecb{q}}(a))$ is the density contrast at the Eulerian position associated with the Lagrangian position $\vecb{q}$ evaluated at scale-factor time $a$, whereas $\delta_\textrm{D}$ denotes the Dirac delta distribution.
However, as time progresses, particle trajectories will eventually cross. At this critical transition, which is known as `shell-crossing', $\det (\bnabla_{\vecb{q}} \vecb{X}_{\vecb{q}})$ vanishes, and the density hence (formally) diverges. While the theoretical analysis of the particle trajectories after the first shell-crossing is challenging (see e.g.\ \cite{Taruya2017}) and, in particular, LPT is no longer applicable, Ref.~\cite{Rampf:2021} showed that for the planar collapse of a density perturbation, the displacement field $\vecb{\mPsi}_{\vecb{q}}$ is non-analytic at the moment of shell-crossing, with leading-order dynamics given by
\begin{equation}
     \vecb{\mPsi}_{\vecb{q}}(a) \propto (a - a_*(\vecb{q}))^{\nicefrac{5}{2}},
\label{eq:psi_singularity_behaviour}
\end{equation}
after the first shell-crossing, where $a_*(\vecb{q})$ is the space-dependent scale factor when the shell-crossing singularity reaches the Langrangian coordinate $\vecb{q}$.\footnote{Our $a_*$ is denoted as $\tau_1$ in Ref.~\cite{Rampf:2021}, see their Eq.~(15). The authors of that work also analyse another type of singularity (occurring at time $\tau_2$ in their Eq.~(15)); however, the smallest singularity exponent for $\vecb{\mPsi}_{\vecb{q}}$, which is given by $\nicefrac{5}{2}$, will limit the order of convergence after shell-crossing. Although Ref.~\cite{Rampf:2021} consider an EdS universe for simplicity, a cosmological constant will leave the singularity exponent unaffected.} This implies that the derivative of the acceleration has a singularity given by  $\dd_a^3 \vecb{\mPsi}_{\vecb{q}}(a) \propto (a - a_*(\vecb{q}))^{-\nicefrac{1}{2}}$. In view of Proposition~\ref{prop:convergence_loss} (with $\theta = \nicefrac{1}{2}$), we therefore expect that the (global) convergence order should be limited to $\nicefrac{3}{2}$ after a particle has undergone the first shell crossing. We will experimentally show this in Section~\ref{sec:results_1D} for planar-wave collapse.

Since density perturbations typically collapse along a single dimension first also in the three-dimensional case and form so-called Zel'dovich pancakes (e.g.\ \cite{Zeldovich:1970, Melott1989}), the loss of higher-order convergence after shell-crossing is also relevant for the three-dimensional case; however, the convergence behaviour can be expected to be more complex due to the occurrence of (quasi-)spherical collapse and collapse across two dimensions, in addition to the effect of gravitational softening, which we do not address herein. In our numerical experiments in 3D below, we will find that, as expected, the order of convergence in 3D depends on the range of considered scales: when simulating a small volume at high resolution in the presence of density fluctuations all the way down to the Nyquist scale, particle trajectories will cross early, limiting the regularity of the displacement field and consequently the convergence.

Another interesting avenue is the development of hybrid schemes, which would adapt the integrator depending on the regime (possibly locally). For instance, one could use higher-order schemes only in the pre-shell-crossing regime and eventually switch to the standard symplectic leapfrog integrator. We leave a more detailed study on the post-shell-crossing convergence of numerical integrators in three dimensions and the exploration of hybrid schemes for future work. 
\end{remark}

\section{Perturbation-theory informed integration schemes}
\label{sec:lpt_integrators}
First-order LPT is exact for one-dimensional initial data up to the shell-crossing scale factor $a_{\text{cross}}$, i.e.\ the exact solution can be obtained in a \emph{single} timestep until $a_{\text{cross}}$. However, standard symplectic integrators converge to this solution only at the respective order of the scheme, which implies that many timesteps may be necessary to reach a given error bound. While in more than one dimension, the exact solution cannot be reached in a single leapfrog step in general, an LPT-consistent step is nonetheless a reasonable guess. Below, we will discuss the \textsc{FastPM} scheme \citep{Feng:2016}, which was to our knowledge the first (and so far only) PT-informed $N$-body time-stepping scheme in the literature. We will show that \textsc{FastPM} is a representative of a more general class of `Zel'dovich-consistent' integrators, which we will define below. In particular, we will introduce more representatives of this class, for example \textsc{LPTFrog} and \textsc{PowerFrog}, which are 2LPT-informed and further improve on the performance of \textsc{FastPM}. 

An alternative approach for informing cosmological integrators with PT is to explicitly compute the LPT solution and use an $N$-body simulation to solve for a \emph{correction} to the LPT displacement field. This idea is known as the COLA (COmoving Lagrangian Acceleration) approach \cite{Tassev2013}, which has been widely used and led to various optimised implementations, see Refs.~\cite{Howlett2015, Tassev2015, Izard2016, Koda2016}. Since COLA -- just like \textsc{FastPM} -- is LPT-informed, it also achieves the correct growth of structures on large scales with few timesteps. However, COLA entails an additional computational overhead because the LPT solution needs to be explicitly computed and stored; further, COLA possesses a hyperparameter $n_\text{LPT}$, which requires tuning depending on the cosmology and number of timesteps. Although harnessing our new integrators within the COLA framework could also be an interesting avenue, we leave a study in this direction to future work. Instead, we focus on building LPT directly \emph{into} the numerical integrator here (rather than writing the total displacement as a sum of LPT and a correction term), leading to second-order integrators that are accurate on large scales with few timesteps, without the need for tuning any hyperparameters.

\begin{figure*}
\centering
  \noindent
   \resizebox{1\textwidth}{!}{
    \includegraphics{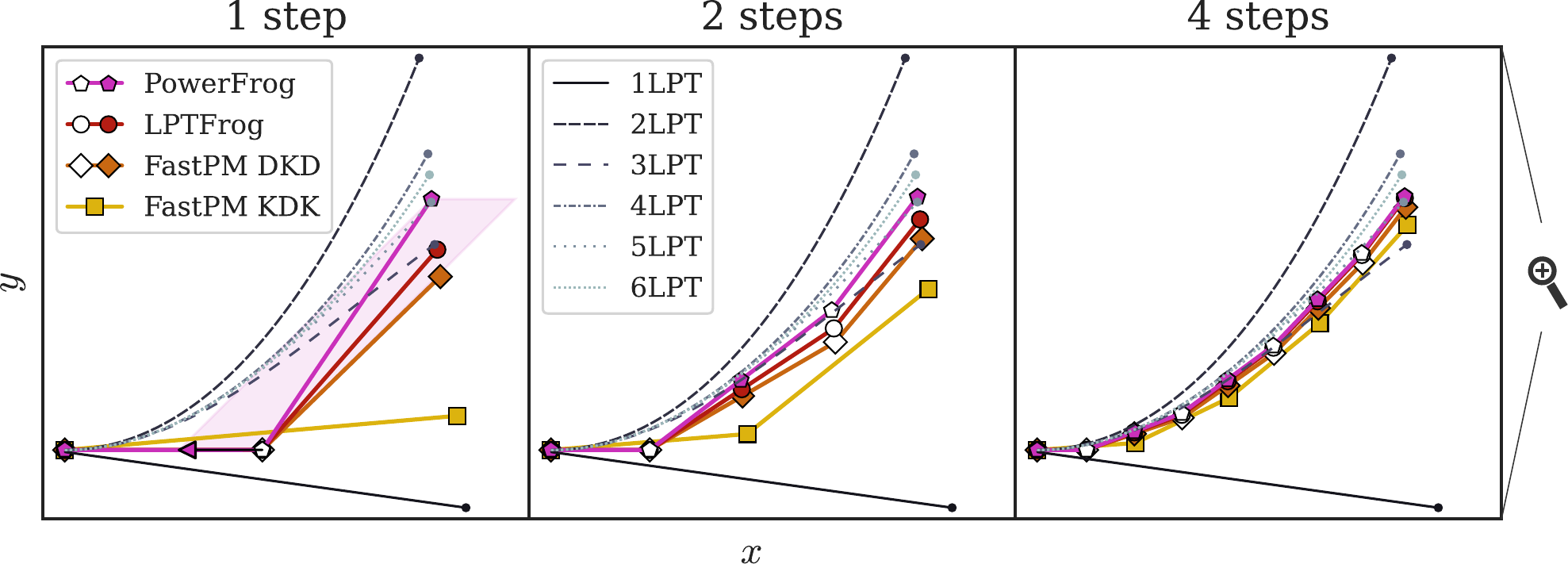}
    }
    \caption{Comparison between the trajectory of an individual particle extracted from a two-dimensional simulation using the $\mPi$-integrators \textsc{PowerFrog}, \textsc{LPTFrog} and \textsc{FastPM}, for 1, 2, and 4 integration steps to the end time. The initial particle position (marker in the lower left corner) and momentum are computed with 2LPT. For DKD schemes, hollow markers indicate the position of the particle after the first drift of each DKD step, whereas filled markers show the position at the end of each DKD (or KDK) step. The edges of the faint pink parallelogram for a single step decompose the second part of \textsc{PowerFrog}'s drift into its component parallel to the old momentum and another component parallel to the acceleration evaluated at the midpoint. Importantly, a \emph{negative} (!) factor for the old momentum ($p(\Delta D, D_n)$ in Eq.~\eqref{eq:kick_ansatz_pi}, in the figure indicated by the backwards-pointing arrow) is necessary to adjust the direction of the velocity sufficiently in order for the particle to move towards the higher-order LPT solutions when the timestep is large.
    The grey curves depict the trajectories as predicted by different orders of LPT, which can be seen to converge in an alternating manner in this case. To highlight the differences between the methods, we zoom into the $y$-direction such that the vertical deviation of the higher LPT trajectories from 1LPT appears enlarged. For better comparability, we take the arithmetic mean $a_{n+\nicefrac{1}{2}} = (a_{n} + a_{n+1}) / 2$ between the start and end point of each timestep also for the intermediate step of \textsc{FastPM} DKD here (rather than the geometric mean $a_{n+\nicefrac{1}{2}} = (a_{n} a_{n+1})^{\nicefrac{1}{2}}$ used in the original \textsc{FastPM} scheme).}    
    \label{fig:stepper_illustration}
\end{figure*}

\subsection{Introduction}
The initial conditions of cosmological simulations are typically computed with (some variant of) LPT; however, since standard numerical integrators are unaware of LPT, the particles will not remain on their LPT trajectory once the cosmological simulation starts. This is despite the fact that the LPT trajectories provide an excellent approximation (and even the exact solution when going to infinite order, i.e.\ $n \to \infty$) at early times before shell-crossing \cite{Rampf2021b}. Let us therefore sketch what is required for numerical integrators to follow LPT trajectories. During a single DKD step, each particle moves on a straight line to some intermediate point, receives a velocity update, and continues its trajectory in a different direction, again along a straight line (see the left panel in Fig.~\ref{fig:stepper_illustration}). Hence, the highest-order polynomial to which we can match the endpoint of two subsequent drifts is given by a parabola w.r.t.\ the time variable. The optimal parabolic trajectory as a function of the growth factor $D^{+}$ is given by the fastest growing limit of second-order LPT, see Eq.~\eqref{eq:2lpt_fastest_growing}. We will take this insight as the basis when constructing two of our integrators, namely \textsc{LPTFrog} and \textsc{PowerFrog}.

We will re-parametrise all quantities such as $\vecb{X}, \vecb{P}$, the Hubble parameter $H$, the momentum growth factor $F^{+}$, and the scale factor $a$ as functions of the growth factor $D^{+}$.\footnote{Since $H$ and $F$ are easily computed in terms of $a$, but not in terms of $D^{+}$ for $\Lambda$CDM cosmology (as the inverse of $D^{+}(a)$ in Eq.~\eqref{eq:Ds_and_Fs} cannot be expressed in terms of basic functions), it is in practice more convenient to start with a given scale factor $a$ and compute $D^{+}(a)$, $H(a)$, and $F^{+}(a)$ as functions of $a$. Nonetheless, our notation emphasises that we take the growth factor $D^{+}$ as the time variable for our integrators. Recall that derivatives transform as $\dd_D a = (\dd_a D)^{-1}$, $\dd_D^2 a = - \dd_a^2 D / (\dd_a D)^3$, etc.} Note that in an EdS universe, the growth factor reduces to the scale factor, i.e.\ $D^+(a) = a$.  Further, since we will only consider the growing-mode solution from here on, we will simplify the notation and write $D = D^{+}$, $F = F^{+}$, as well as $\vecb{\psi}_i^{(1)} = \vecb{\psi}_{\vecb{q}_i}^+$ and $\vecb{\psi}_i^{(2)} = \vecb{\psi}_{\vecb{q}_i}^{++}$ for the linear and quadratic displacements experienced by the particle $i$, respectively.

Herein, we limit ourselves to $\Lambda$CDM cosmologies (which include the EdS limit case of $\Omega_{\mathrm{m}} \to 1$); however, extending our results to other cosmologies with e.g. an evolving equation-of-state parameter $w$ for dark energy or incorporating massive neutrinos (see also Ref.~\cite{bayer2021fast}) should be possible without major difficulties -- as long as the growth-factor $D = D(a)$ defines a suitable time variable (in particular it should be a strictly monotonically increasing function of $a$ so that there is a one-to-one mapping between $a$ and $D$).
First, let us compute the velocity and acceleration of a particle in $D$-time:
\begin{subequations}
\begin{align}
\dd_D \vecb{X}_i &= \frac{\vecb{P}_i \, \dd_D a}{H \, a^3} = \frac{\vecb{P}_i}{F} =: \vecb{\mPi}_i \stackrel{\mathrm{EdS}}{\asymp} \frac{\vecb{P}_i}{a^{\nicefrac{3}{2}}}, \label{eq:dD_X_LCDM_computed} \\
\dd^2_D \vecb{X}_i &= \dd_D \vecb{\mPi}_i = - \vecb{P}_i \frac{\dd_D F}{F^2} + \frac{\dd_D \vecb{P}_i}{F} =: -\vecb{\mPi}_i \mathfrak{f} + \mathfrak{g} \vecb{A}(\vecb{X}_i), \label{eq:d2D_X_LCDM_computed}
\end{align}
where we defined 
\begin{equation}
\mathfrak{f}(D) := \frac{\dd_D F}{F} \stackrel{\mathrm{EdS}}{\asymp} \frac{3}{2a}, \qquad \mathfrak{g}(D) := \frac{a}{F^2} \stackrel{\mathrm{EdS}}{\asymp} \frac{1}{a^2}.
\label{eq:f_and_g}
\end{equation}
\end{subequations}

By leveraging the ordinary differential equation that defines the linear growth factor $D$, see e.g.\ Ref.~\cite[Eq.~(A15)]{Brenier2003}, one finds the simple relation
\begin{equation}
    \frac{\mathfrak{f}(D)}{\mathfrak{g}(D)} = \frac{F(D) \, (\dd_D F)(D)}{a(D)} =: G(D) = \frac{3}{2} \Omega_\mathrm{m} D.
    \label{eq:G_identity}
\end{equation}

We take Eq.~\eqref{eq:dD_X_LCDM_computed} as the motivation to consider schemes of the following form, which we call $\mPi$-integrators in view of the modified momentum variable $\vecb{\mPi} = \dd_D \vecb{X}$:

\begin{definition}[$\mPi$-integrator]
\label{def:pi_integrator}
We call an integrator of the form
\begin{subequations}
\begin{align}
    \vecb{X}_i^{n+\nicefrac{1}{2}} &= \vecb{X}_i^{n} + \frac{\Delta D}{2} \vecb{\mPi}^{n}_i, \label{eq:drift_ansatz_pi_1}\\
     \vecb{\mPi}_i^{n+1} &= p(\Delta D, D_n) \vecb{\mPi}_i^{n} + q(\Delta D, D_n) \vecb{A}(\vecb{X}_i^{n+\nicefrac{1}{2}}), \label{eq:kick_ansatz_pi} \\
    \vecb{X}_i^{n+1} &= \vecb{X}_i^{n+\nicefrac{1}{2}} + \frac{\Delta D}{2} \vecb{\mPi}^{n+1}_i, \label{eq:drift_ansatz_pi_2}
\end{align}
\label{eq:pi_integrator}
\end{subequations}
a $\mPi$-integrator. Here, $p(\Delta D, D_n)$ and $q(\Delta D, D_n)$ are functions that may explicitly depend on the growth factor $D_n$ at the beginning of the timestep and the timestep size $\Delta D$.
\end{definition}

As a straightforward extension, one could consider a more general intermediate point rather than $D_{n+\nicefrac{1}{2}} = D_n + \Delta D / 2$, but this will not change the fundamental properties of the methods.
While the definition of $\mPi$-integrators is independent of the number of spatial dimensions, it is worthwhile to study their behaviour in the one-dimensional case. In that case, a basic desideratum (which is however not satisfied for the standard symplectic leapfrog integrator, see below) is that they exactly follow the Zel'dovich solution before shell-crossing (when it is exact). 
We emphasise again that in the multidimensional case, the Zel'dovich trajectory is only a first-order approximation, and the exact solution prior to shell-crossing is described by infinitely many terms in the LPT series expansion. However, also then, one would expect that an integrator that correctly captures planar-wave collapse (i.e.\ the collapse along a single axis) as described by the Zel'dovich approximation improves the performance on large scales.

Motivated by this reasoning, we fixed the form of the drift in Eqs.~\eqref{eq:pi_integrator} in order to ensure that a particle lying on the Zel'dovich trajectory at $D_n$ will continue moving on the straight Zel'dovich trajectory to $D_{n+\nicefrac{1}{2}}$. For the kick, however, we introduced two degrees of freedom. 
It is now an interesting question how the functions $p$ and $q$ need to be chosen to this aim.
In order to answer this question, we define the following property of $\mPi$-integrators:
\begin{definition}[Zel'dovich consistency]
\label{def:zeldovich_consistency}
We call a $\mPi$-integrator \emph{Zel'dovich consistent} if it exactly produces the analytic solution of Eq.~\eqref{eq:eqs_of_motion} (supplemented with Poisson's equation \eqref{eq:poisson}) in 1D prior to shell-crossing, independent of the number of timesteps (and in particular with a single timestep).
\end{definition}
In the multidimensional case, the trajectories of $\mPi$-integrators do not exactly align with LPT at any order, as infinitely many LPT terms are excited in the computation of the Eulerian acceleration. Therefore, the requirement of Zel’dovich consistency is based entirely on the one-dimensional case, but all $\mPi$-integrators can be used in any number of dimensions without modifications.
It turns out that there is a simple relation between $p$ and $q$ that needs to be satisfied for Zel'dovich consistency:

\begin{proposition}[Characterisation of Zel'dovich consistency]
\label{prop:characterisation_of_zeldovich_consistency}
A $\mPi$-integrator is Zel'dovich consistent if and only if $p(\Delta D, D_n)$ and $q(\Delta D, D_n)$ satisfy the following relation:
\begin{equation}        
        \frac{1 - p(\Delta D, D_n)}{q(\Delta D, D_n)} = \frac{3}{2} \Omega_{\mathrm{m}} D_{n+\nicefrac{1}{2}} \stackrel{\mathrm{EdS}}{\asymp} \frac{3}{2} a_{n+\nicefrac{1}{2}}\;.
        \label{eq:zeldovich_consistency_condition}
\end{equation}
\begin{proof}
    First, note that a particle that moves on a Zel'dovich trajectory prior to shell-crossing remains on the Zel'dovich trajectory during the drift in 1D by construction of the $\mPi$-integrator, so that we only need to ensure the kick operator does not change the particle's velocity. In 1D prior to shell-crossing, the gravitational force can be computed explicitly as (e.g.\ \cite{mcquinn2016cosmological})
    \begin{equation}    
    A(X(D)) = - \dd_x \varphi_N(X(D)) = \frac{3}{2} \Omega_{\mathrm{m}} D \psi^{(1)},
    \end{equation}
    where $\psi^{(1)} = \psi^{+}$ is the Zel'dovich displacement in Eq.~\eqref{eq:2lpt_fastest_growing}. We dropped the bold vector notation here, as the 1D case is being considered. For a particle starting on the Zel'dovich trajectory, we have $\mPi_i^{n} = \psi_i^{(1)}$, and in order for the particle to remain on this trajectory, one therefore requires that
    \begin{equation}    
    \mPi_i^{n+1} \stackrel{!}{=} \psi_i^{(1)} + \frac{3}{2} \Omega_{\mathrm{m}} D_{n+\nicefrac{1}{2}} \psi_i^{(1)} q(\Delta D, D_n).
    \end{equation}
    This is equivalent to the relation between $p(\Delta D, D_n)$ and $q(\Delta D, D_n)$ stated in Eq.~\eqref{eq:zeldovich_consistency_condition}.
\end{proof}
\end{proposition}

This implies that each Zel'dovich consistent $\mPi$-integrator is completely specified by the choice of the function $p(\Delta D, D_n)$. How does $p(\Delta D, D_n)$ need to be chosen in order for the $\mPi$-integrator to be \textit{symplectic}? Computing the determinant of the Jacobian matrix $\bnabla_{\vecb{\xi}^{n}_i} \vecb{\xi}^{n+1}_i$, we find

\begin{equation}
    \det \bnabla_{\vecb{\xi}^{n}_i} \vecb{\xi}^{n+1}_i = p(\Delta D, D_n) \frac{F(D_{n+1})}{F(D_n)} \stackrel{\mathrm{EdS}}{\asymp} p(\Delta a, a_n) \left(\frac{a_{n+1}}{a_n}\right)^{\nicefrac{3}{2}} .
    \label{eq:symplecticity_of_pi_integrators}
\end{equation}
Since this Jacobian determinant must be unity for symplecticity, we immediately obtain the condition
\begin{equation}
    p(\Delta D, D_n) = \frac{F(D_n)}{F(D_{n+1})}  \stackrel{\mathrm{EdS}}{\asymp} \left(\frac{a_{n}}{a_{n+1}}\right)^{\nicefrac{3}{2}}.
\label{eq:symplecticity_of_pi_integrators_rewritten}
\end{equation}
Note that when re-writing the kick of the $\mPi$-integrator in terms of $\vecb{P}$, Eq.~\eqref{eq:symplecticity_of_pi_integrators_rewritten} is equivalent to the requirement that the factor $c$ in $\vecb{P}^{n+1}_i = c \vecb{P}^{n}_i + \beta \vecb{A}(\vecb{X}^{n+\nicefrac{1}{2}}_i)$ be unity. This is the only choice of $p$ for a $\mPi$-integrator for which the scheme belongs to the class of canonical DKD integrators as defined in Definition~\ref{def:canonical_dkd_integrator} and hence, we already know from Proposition~\eqref{prop:symplectic} that the scheme is symplectic in that case. It turns out that this choice of $p$ for a $\mPi$-integrator corresponds to the \textsc{FastPM} scheme, which we will revisit in what follows.

\subsection{\textsc{FastPM} vs. standard symplectic integrator}
\begin{example}[\textsc{FastPM}]
The {\sc FastPM} method was proposed by Ref.~\cite{Feng:2016} and is (to our knowledge) the only Zel'dovich-consistent $\mPi$-integrator that has been presented in the literature. While the original \textsc{FastPM} scheme is a KDK scheme, its DKD counterpart can be expressed as 

\begin{subequations}
\begin{alignat}{-1}
    \vecb{X}^{n+\nicefrac{1}{2}}_i &= \vecb{X}^n_i + \frac{D_{n+\nicefrac{1}{2}}-D_n}{F(D_n)}\; \vecb{P}^n_i &{} \stackrel{\mathrm{EdS}}{\asymp}& \ \ \vecb{X}^n_i + \frac{a_{n+\nicefrac{1}{2}} - a_n}{a_n^{\nicefrac{3}{2}}} \; \vecb{P}^n_i \label{eq:fastpm_step_first}, \\
\vecb{P}^{n+1}_i &= \vecb{P}^n_i + \frac{F(D_{n+1})-F(D_n)}{G(D_{n+\nicefrac{1}{2}})} \;\vecb{A}(\vecb{X}^{n+\nicefrac{1}{2}}_i) &{} \stackrel{\mathrm{EdS}}{\asymp}& \ \ \vecb{P}^n_i + \frac{2}{3} \frac{a_{n+1}^{\nicefrac{3}{2}} - a_{n}^{\nicefrac{3}{2}}}{a_{n+\nicefrac{1}{2}}} \vecb{A}(\vecb{X}^{n+\nicefrac{1}{2}}_i), \\
\vecb{X}^{n+1}_i &= \vecb{X}^{n+\nicefrac{1}{2}}_i + \frac{D_{n+1}-D_{n+\nicefrac{1}{2}}}{F(D_{n+1})}\; \vecb{P}^{n+1}_i &{} \stackrel{\mathrm{EdS}}{\asymp}& \ \ \vecb{X}^{n+\nicefrac{1}{2}}_i + \frac{a_{n+1}-a_{n+\nicefrac{1}{2}}}{a_{n+1}^{\nicefrac{3}{2}}} \; \vecb{P}^{n+1}_i,  \label{eq:fastpm_step_last}
\end{alignat}
\label{eq:fastpm}
\end{subequations}
where $G(D)$ is defined in Eq.~\eqref{eq:G_identity}.
\textsc{FastPM} clearly has the form of a canonical DKD integrator as defined in Definition~\ref{def:canonical_dkd_integrator}.
In order to obtain the equivalent formulation of \textsc{FastPM} as a $\mPi$-integrator, we rewrite the kick equation in terms of the modified momentum variable $\vecb{\mPi}$, which leads to the coefficient functions $p(\Delta D, D_n)$ and $q(\Delta D, D_n)$
\begin{equation}
    p_\text{\textsc{FastPM}}(\Delta D, D_n) = \frac{F(D_n)}{F(D_{n+1})} \stackrel{\mathrm{EdS}}{\asymp} \left(\frac{a_{n}}{a_{n+1}}\right)^{\nicefrac{3}{2}}, \qquad
    q_\text{\textsc{FastPM}}(\Delta D, D_n) = \frac{F(D_{n+1}) - F(D_n)}{F(D_{n+1}) \, G(D_{n+\nicefrac{1}{2}})} \stackrel{\mathrm{EdS}}{\asymp} \frac{2}{3} \frac{1 - \left(\frac{a_n}{a_{n+1}}\right)^{\nicefrac{3}{2}}}{a_{n+\nicefrac{1}{2}}}.
\end{equation}
One easily confirms that $p_\text{\textsc{FastPM}}$ and $q_\text{\textsc{FastPM}}$ satisfy Eq.~\eqref{eq:zeldovich_consistency_condition}, and \textsc{FastPM} is therefore Zel'dovich consistent.
\end{example}

The following characterisation of \textsc{FastPM} follows immediately from Eq.~\eqref{eq:symplecticity_of_pi_integrators}:
\begin{proposition}
\label{prop:fastpm}
    \textsc{FastPM} is the unique $\mPi$-integrator that is both Zel'dovich consistent and symplectic.
\end{proposition}

Before proceeding to introduce new $\mPi$-integrators, let us revisit the standard symplectic DKD integrator, which is \emph{not} Zel'dovich consistent and hence does not produce the exact pre-shell-crossing solution in 1D.

\begin{counterexample}[Symplectic~2 integrator]
First, note that the drift of the standard second-order symplectic integrator (`Symplectic~2') is \emph{not} of the form of Eqs.~\eqref{eq:drift_ansatz_pi_1} \& \eqref{eq:drift_ansatz_pi_2}, for which reason the Symplectic~2 integrator is not a $\mPi$-integrator. Rather, the first half of the drift of the Symplectic~2 integrator is given by
\begin{equation}
    \vecb{X}^{n+\nicefrac{1}{2}}_i = \vecb{X}^{n}_i + \vecb{P}^{n}_i \int_{a(D_n)}^{a(D_{n+\nicefrac{1}{2}})} \frac{\dd a}{H(a) a^3},
\end{equation}
and similarly for the second half of the drift.
Since the Symplectic~2 integrator does not make use of the growth factor $D$, we parametrise the Hubble parameter in terms of the scale factor $a$ here rather than rewriting the integral in terms of $D$.
The kick is given by
\begin{align}
    \vecb{P}^{n+1}_i = \vecb{P}^{n}_i + \vecb{A}(\vecb{X}^{n+\nicefrac{1}{2}}_i) \int_{a(D_n)}^{a(D_{n+1})}  \frac{\dd a}{H(a) a^2} \stackrel{\mathrm{EdS}}{\asymp} \vecb{P}^{n}_i + 2 \vecb{A}(\vecb{X}^{n+\nicefrac{1}{2}}_i) \left(\sqrt{a_{n+1}} - \sqrt{a_n}\right),
\end{align}
from which one finds (by rewriting $\vecb{P}$ in terms of $\vecb{\mPi}$)
\begin{subequations}
\begin{align}
    p_\text{{Symplectic~2}}(\Delta D, D_n) &= \frac{F(D_n)}{F(D_{n+1})} \stackrel{\mathrm{EdS}}{\asymp} \left(\frac{a_n}{a_{n+1}}\right)^{\nicefrac{3}{2}}, \\
    q_\text{{Symplectic~2}}(\Delta D, D_n) &= \frac{\int_{a(D_n)}^{a(D_{n+1})} \frac{\dd a}{H(a) a^2}}{F(D_{n+1})} \stackrel{\mathrm{EdS}}{\asymp}\frac{2 \left(\sqrt{a_{n+1}} - \sqrt{a_n}\right)}{a_{n+1}^{\nicefrac{3}{2}}}.
\end{align}
\end{subequations}
The standard Symplectic~2 integrator and \textsc{FastPM} thus have the same momentum factor $p$; however, \textsc{FastPM} adjusts the choice of the factor $q$, for which reason the kick of the Symplectic~2 integrator does not satisfy the Zel'dovich consistency condition in Eq.~\eqref{eq:zeldovich_consistency_condition}, in addition to the fact that the drift does not have the correct form in order to follow the Zel'dovich trajectories.
\end{counterexample}

\subsection{New Zel'dovich-consistent integrators}
Proposition~\ref{prop:characterisation_of_zeldovich_consistency} allows us to easily construct new Zel'dovich-consistent schemes (at the price of giving up exact symplecticity), to which we will turn our attention now. A requirement that we clearly need to satisfy when doing so is to ensure the second-order convergence of the new integrators to the correct solution. Therefore, let us expand $p_\text{{Symplectic~2}}$ and $q_\text{{Symplectic~2}}$ in a Taylor series w.r.t.\ $\Delta D$, which yields
\begin{subequations}
\begin{equation}
\begin{aligned}
        p_\text{{Symplectic~2}}(\Delta D, D) &\stackrel{\phantom{\text{EdS}}}{=} 1 - \frac{\dd_D F}{F} \Delta D + \frac{2 (\dd_D F)^2 - F \dd_D^2 F}{2 F^2} \left(\Delta D\right)^2 + \mathscr{O}\left((\Delta D)^3\right) \\
        &\stackrel{\mathrm{EdS}}{\asymp} 1 - \frac{3}{2} \frac{\Delta a}{a} + \frac{15}{8} \left(\frac{\Delta a}{a}\right)^2 - \frac{35}{16} \left(\frac{\Delta a}{a}\right)^3 + \frac{315}{128} \left(\frac{\Delta a}{a}\right)^4 + \mathscr{O}\left((\Delta a)^5\right),
\end{aligned}
\label{eq:p_consistency_with_symplectic_integrator}
\end{equation}
\begin{equation}
\begin{aligned}
      q_\text{Symplectic~2}(\Delta D, D) &\stackrel{\phantom{\text{EdS}}}{=} \frac{a}{F^2} \Delta D + \frac{F \dd_D a - 3 a \dd_D F}{2 F^3} \left(\Delta D\right)^2 + \mathscr{O}\left((\Delta D)^3\right) \\     
      &\stackrel{\mathrm{EdS}}{\asymp}
    \frac{1}{a} \frac{\Delta a}{a} - \frac{7}{4a} \left(\frac{\Delta a}{a}\right)^2 + \frac{19}{8a} \left(\frac{\Delta a}{a}\right)^3 - \frac{187}{64 a} \left(\frac{\Delta a}{a}\right)^4 + \mathscr{O}\left((\Delta a)^5\right),
\end{aligned}
\label{eq:q_consistency_with_symplectic_integrator}
\end{equation}
\end{subequations}
where $F$ and its derivatives are evaluated at $D$ (which we omitted here in order to avoid excessive notation). As it turns out, in order to construct second-order convergent Zel'dovich-consistent schemes, it is sufficient to consider $p(\Delta D, D)$ and to define $q(\Delta D, D)$ according to Eq.~\eqref{eq:zeldovich_consistency_condition}, as the following proposition shows:

\begin{proposition}
\label{prop:second_order_convergence}
    For a $\mPi$-integrator, let the following two conditions hold:
    \begin{enumerate}
        \item The associated coefficient function $p = p(\Delta D, D_n)$ agrees with $p_\mathrm{Symplectic~2}$ as given in Eq.~\eqref{eq:p_consistency_with_symplectic_integrator} up to second order w.r.t.\ $\Delta D$.
        \item The coefficient function $q = q(\Delta D, D_n)$ is chosen according to the Zel'dovich consistency condition in Eq.~\eqref{eq:zeldovich_consistency_condition}.
    \end{enumerate}
    Then, the $\mPi$-integrator is second-order accurate.
\begin{proof}
    The proof follows from a straightforward computation, which is provided in \ref{sec:proof_second_order}.
\end{proof}
\end{proposition}

The \textsc{FastPM} scheme is derived from the requirement that the drift and kick operations be consistent with the Zel'dovich approximation of the growing mode in the sense of Definition~\ref{def:zeldovich_consistency}. However, when plugging  Eqs.~\eqref{eq:drift_ansatz_pi_1} and \eqref{eq:kick_ansatz_pi} into Eq.~\eqref{eq:drift_ansatz_pi_2}, it becomes apparent that the trajectory of $\mPi$-integrators (just as for the standard DKD leapfrog scheme) contains both a linear and a quadratic term w.r.t.\ the timestep size $\Delta D$, suggesting that it should be possible to construct an integrator whose quadratic contribution matches the (quadratic) 2LPT correction to the Zel'dovich trajectory. We will derive a first 2LPT-inspired scheme in what follows, which we call \textsc{LPTFrog}. We remark that this scheme (just like the other schemes we present below) could in principle be extended to higher LPT orders by combining multiple drifts and kicks in a single timestep in order to match the endpoint of a higher-order polynomial; however, the practical gain from going to higher orders beyond 2LPT is unclear (also because the increased order of convergence that one might hope for as an additional side effect is limited to the pre-shell-crossing regime, see Remark~\ref{remark:convergence_loss}, Section~\ref{sec:1D_results_post}, and in particular Section~\ref{sec:3D_convergence_study} for the realistic 3D case.

\begin{example}[\textsc{LPTFrog}]
\label{example:lptfrog}
Let us consider particle $i$ moving from growth-factor time $D_n$ to $D_{n+1}$ on a trajectory that is quadratic w.r.t.\ $D - D_n$. Also, let us define the position and momentum of the particle at the beginning of the timestep $D_n$ as $\vecb{X}^n_i$ and $\vecb{P}^n_i$, respectively. This yields for $D \in [D_n, D_{n+1}]$:
\begin{subequations}
\begin{align}
\vecb{X}_i(D) &= \vecb{X}^{n}_i + (D - D_n) \vecb{\psi}_i^{n, (1)} + (D - D_n)^2 \vecb{\psi}_i^{n, (2)}, \label{eq:X_LCDM_lptfrog} \\
\dd_D \vecb{X}_i(D) &= \vecb{\psi}_i^{n, (1)} + 2 (D - D_n) \vecb{\psi}_i^{n, (2)}, \label{eq:dD_X_LCDM_lptfrog} \\
\dd^2_D \vecb{X}_i(D) &= 2 \vecb{\psi}_i^{n, (2)} = \text{const.} \label{eq:d2D_X_LCDM_lptfrog}
\end{align}
\label{eq:LCDM_lptfrog}
\end{subequations}
The superscript ${}^n$ for the displacements $\vecb{\psi}_i^{n, (1)}$ and $\vecb{\psi}_i^{n, (2)}$ indicates that they are defined \emph{locally} for the timestep starting from $D_n$, unlike the displacements in standard LPT, which are relative to $D = 0$.

By equating Eqs.~\eqref{eq:d2D_X_LCDM_lptfrog} and \eqref{eq:d2D_X_LCDM_computed} for the acceleration and using Eq.~\eqref{eq:dD_X_LCDM_lptfrog} for the momentum, we obtain
\begin{equation}
\dd^2_D \vecb{X}_i(D) = 2 \vecb{\psi}_i^{n, (2)} \stackrel{!}{=} -\left(\vecb{\psi}_i^{n, (1)} + 2 (D - D_n) \vecb{\psi}_i^{n, (2)}\right) \mathfrak{f}(D) + \mathfrak{g}(D) \vecb{A}(\vecb{X}_i(D)).
\end{equation}
With a DKD scheme in mind, we can solve this equation for $\vecb{\psi}_i^{n, (2)}$ at $D_{n+\nicefrac{1}{2}} = (D_n + D_{n+1})/2$:
\begin{equation}
\vecb{\psi}_i^{n, (2)} = \frac{-\vecb{\mPi}_i^n \, \mathfrak{f}(D_{n+\nicefrac{1}{2}}) + \mathfrak{g}(D_{n+\nicefrac{1}{2}}) \vecb{A}(\vecb{X}_i(D_{n+\nicefrac{1}{2}}))}{2 + \Delta D \mathfrak{f}(D_{n+\nicefrac{1}{2}})},
\end{equation}
where we have used $\vecb{\psi}_i^{n, (1)} = (\dd_D \vecb{X}_i)(D_n) = \vecb{\mPi}_i^n$ and $\Delta D := D_{n+1} - D_n$. Since $\vecb{\mPi}_i(D)$ is a linear function in $D$, we also have $\dd^2_D \vecb{X}_i(D) = (\vecb{\mPi}_i^{n+1} - \vecb{\mPi}_i^{n}) / (\Delta D)$,
which yields the following equation for the momentum update:
\begin{equation}
 \vecb{\mPi}_i^{n+1} = \frac{\vecb{\mPi}_i^n \left(1 - \frac{\Delta D}{2} \mathfrak{f}(D_{n+\nicefrac{1}{2}})\right) + \Delta D \mathfrak{g}(D_{n+\nicefrac{1}{2}}) \vecb{A}(\vecb{X}_i(D_{n+\nicefrac{1}{2}}))}{1 + \frac{\Delta D}{2} \mathfrak{f}(D_{n+\nicefrac{1}{2}})}.
\label{eq:kick_LPTFrog}
\end{equation}

From Eq.~\eqref{eq:kick_LPTFrog}, we can read off the functions $p$ and $q$, which are given by

\begin{subequations}
\begin{align}
    p_{\textsc{LPTFrog}}(\Delta D, D_n) &= \frac{1 - \frac{\Delta D}{2} \mathfrak{f}(D_{n+\nicefrac{1}{2}})}{1 + \frac{\Delta D}{2} \mathfrak{f}(D_{n+\nicefrac{1}{2}})} \stackrel{\mathrm{EdS}}{\asymp} \frac{4 a_{n+\nicefrac{1}{2}} - 3 \Delta a}{4 a_{n+\nicefrac{1}{2}} + 3 \Delta a}, \\
    q_{\textsc{LPTFrog}}(\Delta D, D_n) &= \frac{\Delta D \, \mathfrak{g}(D_{n+\nicefrac{1}{2}})}{1 + \frac{\Delta D}{2} \mathfrak{f}(D_{n+\nicefrac{1}{2}})} \stackrel{\mathrm{EdS}}{\asymp} \frac{4 \Delta a}{a_{n+\nicefrac{1}{2}} (4 a_{n+\nicefrac{1}{2}} + 3 \Delta a)}.
\end{align}
\end{subequations}
Using Eq.~\eqref{eq:G_identity}, one directly confirms that \textsc{LPTFrog} is Zel'dovich consistent in the sense of Eq.~\eqref{eq:zeldovich_consistency_condition}.
From Proposition~\ref{prop:symplectic}, it is clear that the \textsc{LPTFrog} integrator is not symplectic. Specifically, we find 
\begin{equation}
    \det \bnabla_{\vecb{\xi}^{n}_i} \vecb{\xi}^{n+1}_i = \frac{F(D_{n+1})}{F(D_n)} \left(\frac{2 F(D_{n+\nicefrac{1}{2}})}{F(D_{n+\nicefrac{1}{2}}) + \frac{\Delta D}{2} \, (\dd_D F)(D_{n+\nicefrac{1}{2}})} - 1 \right) = 1 + \mathscr{O}((\Delta D)^3).    
\end{equation}
For instance, one obtains for EdS cosmology
\begin{align}
    \det \bnabla_{\vecb{\xi}^{n}_i} \vecb{\xi}^{n+1}_i = 1 - \frac{5}{32} \left(\frac{\Delta a}{a_n}\right)^3 + \frac{15}{64} \left(\frac{\Delta a}{a_n}\right)^4 + \mathscr{O}((\Delta a)^5).
\end{align}
While \textsc{LPTFrog} does not conform with symplectic geometry, it is connected to the related framework of `contact geometry' (see~\ref{sec:contact}).
\end{example}

For the following example, let us consider the EdS case. As it turns out, \textsc{FastPM} and \textsc{LPTFrog} are representatives of a larger class of integrators, which we name $\epsilon$-integrators:
\begin{example}[$\epsilon$-integrators for EdS cosmology]
\label{example:epsilon_integrator}
Let us start with the ansatz
\begin{equation}
    p(\Delta a, a) = \left(\frac{\alpha a + \beta \Delta a}{\gamma a + \delta \Delta a}\right)^{\epsilon},
    \label{eq:epsilon_integrator_ansatz}
\end{equation}
with coefficients $\alpha, \beta, \gamma, \delta, \epsilon \in \mathbb{R}$. For the trivial case $\epsilon = 0$, one has $p(\Delta a, a) \equiv 1$ and therefore $q(\Delta a, a) = 0$ for a Zel'dovich-consistent integrator, which implies that the modified momentum variable $\vecb{\mPi}$ remains constant and is unaffected by the gravitational force. Thus, when initialised with Zel'dovich positions and momenta, this trivial $\mPi$-integrator will indefinitely move the particles along the Zel'dovich trajectories and not converge to the true solution.

Therefore, we assume $\epsilon \neq 0$ in what follows. In view of the symmetry of the numerator and denominator, we can assume $\epsilon > 0$ without loss of generality. Expanding this ansatz in a series yields
\begin{equation}
p(\Delta a, a) = \left(\frac{\alpha}{\gamma}\right)^{\epsilon} + \frac{\epsilon  \left(\frac{\alpha }{\gamma }\right)^{\epsilon - 1} (\beta  \gamma -\alpha  \delta )}{\gamma^2} \frac{\Delta a}{a}+ \frac{\epsilon  \left(\frac{\alpha }{\gamma }\right)^{\epsilon } (\alpha  \delta -\beta  \gamma ) (\alpha  \delta  \epsilon +\alpha  \delta -\beta  \gamma  \epsilon +\beta  \gamma )}{2 \alpha ^2 \gamma ^2} \left(\frac{\Delta a}{a}\right)^2 + \mathscr{O}\left((\Delta a)^3\right).
\end{equation}
Enforcing consistency with the standard symplectic integrator to second order leads to the following conditions (where we assume $\gamma \neq 0$)
\begin{equation}
    \gamma = \alpha \quad \text{($0^\text{th}$ order)}, \qquad
    \beta = \delta - \frac{3}{2}\frac{\alpha}{\epsilon} \quad \text{($1^\text{st}$ order)}, \qquad
    \delta = \frac{\alpha  (2 \epsilon + 3)}{4 \epsilon} \quad \text{($2^\text{nd}$ order)}.    
\end{equation}
Since $\beta$, $\gamma$, and $\delta$ are all multiples of $\alpha$, $\alpha$ cancels out in Eq.~\eqref{eq:epsilon_integrator_ansatz}, leaving us with the following class of $\epsilon$-integrators with a single free parameter $\epsilon$: 
\begin{equation}
    p_\epsilon(\Delta a, a) = \left(\frac{2 a \epsilon + \Delta a (\epsilon - \frac{3}{2})}{2 a \epsilon + \Delta a (\epsilon + \frac{3}{2})}\right)^{\epsilon},
    \label{eq:epsilon_integrator}
\end{equation}
where the associated $q_\epsilon(\Delta a, a)$ follows from the Zel'dovich consistency condition in Eq.~\eqref{eq:zeldovich_consistency_condition}.

Both \textsc{FastPM} and \textsc{LPTFrog} are included in this class of models; specifically, for the choice of $\epsilon = 1$ one recovers \textsc{LPTFrog}, whereas $\epsilon = \nicefrac{3}{2}$ is the \textsc{FastPM} integrator. Note that for $\epsilon < \nicefrac{3}{2}$, the term that is raised to the power of $\epsilon$ in Eq.~\eqref{eq:epsilon_integrator} can become negative, in which case $p(\Delta a, a)$ becomes complex for $\epsilon \notin \mathbb{Z}$. Although complex-time integrators have been successfully employed, see e.g.\ \cite{Corliss1980} (even symplectic ones \cite{Chambers2003}), we enforce $p(\Delta a, a)$ and $q(\Delta a, a)$ to be real in this work. This yields the following $\epsilon$-dependent timestep criterion for non-integer $\epsilon < \nicefrac{3}{2}$:
\begin{equation}
    \frac{\Delta a}{a} < \frac{2 \epsilon}{\frac{3}{2} - \epsilon}.
\end{equation}
The \textsc{LPTFrog} case of $\epsilon = 1$ is an exception: the timestep criterion would imply that $\Delta a < 4 a$; however, since $\epsilon$ is integer in this case, $p(\Delta a, a)$ remains real (but \emph{negative}) even for $\Delta a > 4a$. As discussed in more detail in \ref{sec:contact}, this can be understood by considering the first term on the right-hand side in Eq.~\eqref{eq:d2D_X_LCDM_computed}, which describes a friction force linear in the modified momentum $\vecb{\mPi}$ due to the expansion of the universe. Since the proportionality factor is given by $\mathfrak{f}(a) = 3 / (2 a)$ (for EdS), this term becomes significant at early times.
The value $\epsilon = 1$ is the only integer in the interval $(0, \nicefrac{3}{2})$; therefore, \textsc{LPTFrog} is the only $\epsilon$-integrator that allows negative values of $p_\epsilon$.

It is also interesting to consider the limiting behaviour of this ansatz as $\epsilon \to \infty$,
\begin{equation}
    \lim_{\epsilon \to \infty} p_\epsilon(\Delta a, a) = \exp\left(- \frac{3 \Delta a}{2 a + \Delta a}\right),
    \label{eq:limit_stepper}
\end{equation}
which means that when evaluated at the beginning of the timestep $a_n$, one has $p_\epsilon(\Delta a, a_n) \to \exp\left(-\nicefrac{3}{2} (\Delta a / a_{n+\nicefrac{1}{2}})\right)$, defining another possible $\mPi$-integrator.
\end{example}

We have already seen that the standard Symplectic~2 integrator is not Zel'dovich consistent because it does not satisfy Eq.~\eqref{eq:zeldovich_consistency_condition} (and, even more fundamentally, because the drift does not have the right form). \textsc{FastPM} remedies this in the kick by modifying the factor $q(\Delta D, D)$. Another way of obtaining a Zel'dovich-consistent integrator is to apply the \emph{reverse} procedure and to take the same coefficient function $q(\Delta D, D)$ as for the Symplectic~2 integrator, but to modify instead the factor $p(\Delta D, D)$ such that Eq.~\eqref{eq:zeldovich_consistency_condition} holds. This yields the following integrator, which we call \textsc{TsafPM} (fast, but with reversed logic):

\begin{example}[\textsc{TsafPM}]
\label{example:tsafpm}
We define the \textsc{TsafPM} integrator via
\begin{subequations}
\begin{align}
    p_{\textsc{TsafPM}}(\Delta D, D_n) &= 1 - \frac{3}{2} \Omega_{\mathrm{m}} \frac{D_{n+\nicefrac{1}{2}}}{F(D_{n+1})} \int_{a(D_n)}^{a(D_{n+1})} \frac{\dd a}{H(a) \, a^2} \stackrel{\mathrm{EdS}}{\asymp} 1 - \frac{3 a_{n+\nicefrac{1}{2}} \left(\sqrt{a_{n+1}}-\sqrt{a_n}\right)}{(a_{n+1})^{\nicefrac{3}{2}}}, \\
    q_{\textsc{TsafPM}}(\Delta D, D_n) &= q_{\text{Symplectic~2}}(\Delta D, D_n) = \frac{\int_{a(D_n)}^{a(D_{n+1})} \frac{\dd a}{H(a) \, a^2}}{F(D_{n+1})} \stackrel{\mathrm{EdS}}{\asymp} \frac{2 \left(\sqrt{a_{n+1}} - \sqrt{a_n}\right)}{a_{n+1}^{\nicefrac{3}{2}}}.   
\end{align}
\end{subequations}
Clearly, \textsc{TsafPM} is Zel'dovich consistent by construction, and similarly to \textsc{LPTFrog}, $p_{\textsc{TsafPM}}(\Delta D, D_n)$ becomes negative for $\Delta D / D_n$ sufficiently large. 
\end{example}

\begin{figure*}
\centering
  \noindent
   \resizebox{1\textwidth}{!}{
    \includegraphics{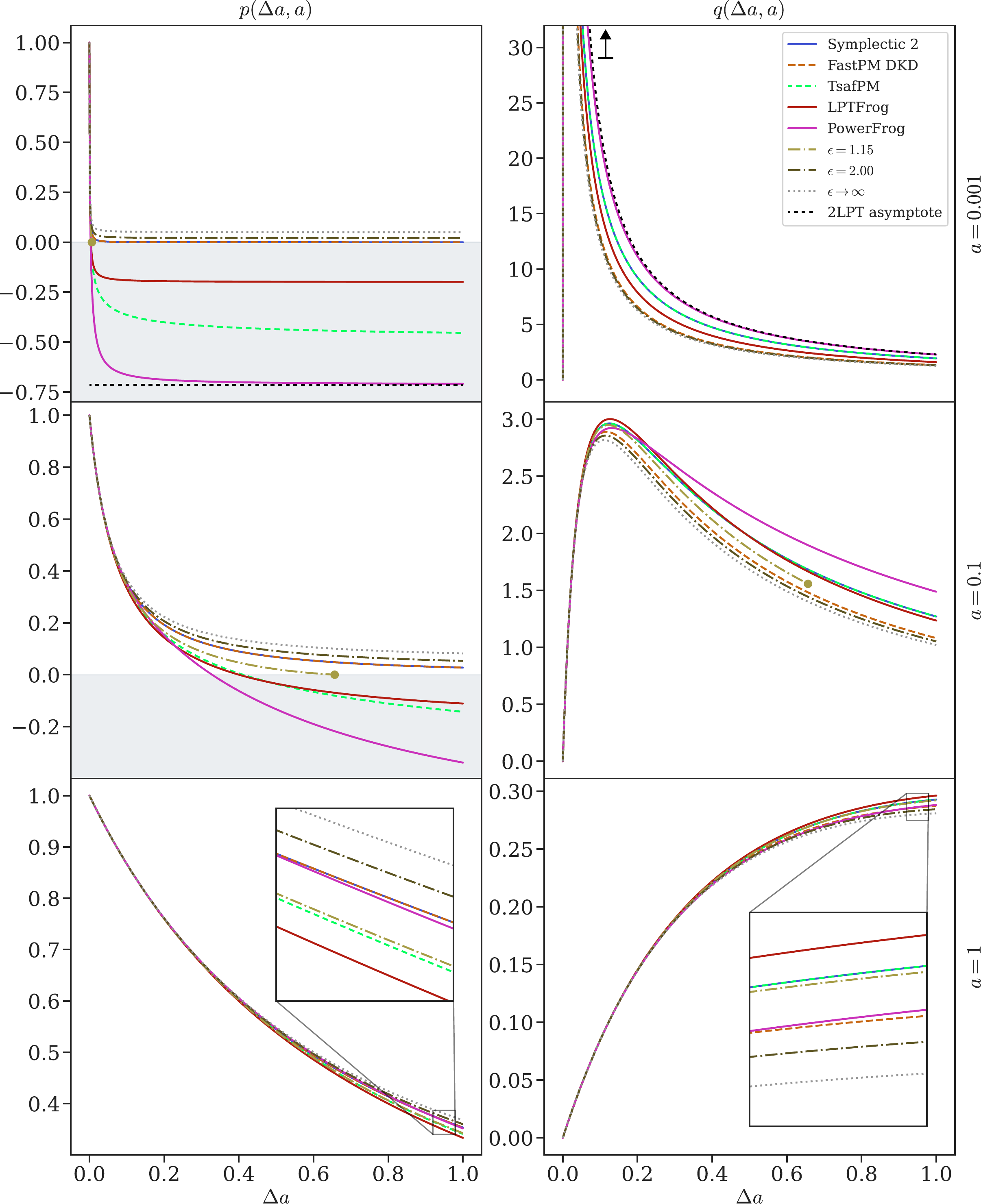}
    }
    \caption{Coefficient functions $p(\Delta a, a)$ and $q(\Delta a, a)$ defining the kick for the DKD integrators that we consider in this work for EdS cosmology, see Eq.~\eqref{eq:kick_ansatz_pi}. By construction, $p_{\mathrm{FastPM \ DKD}} = p_{\text{Symplectic \ 2}}$ and $q_{\text{\textsc{TsafPM}}} = q_{\mathrm{Symplectic \ 2}}$. For non-integer values of $\epsilon < \nicefrac{3}{2}$ (such as $\epsilon = 1.15$ shown here), $\Delta a$ needs to remain small enough such that $p_\epsilon(\Delta a, a) \geq 0$ with the $\epsilon$-integrator; the maximum value of $\Delta a$ for which this is the case is marked with a filled dot. For the smallest value $a = 0.001$, the 2LPT asymptote as $a \to 0$ is indicated with a dashed black line, which is closely traced by the \textsc{PowerFrog} integrator. The region where $p(\Delta a, a) < 0$ is shaded grey. While the differences among the integrators grow with increasing timestep-to-scale factor ratio $\Delta a / a$, all integrators agree in the limit $\Delta a / a \to 0$ up to second order. The (standard) Symplectic~2 integrator is the only scheme plotted that does not satisfy the Zel'dovich consistency condition in Eq.~\eqref{eq:zeldovich_consistency_condition} (in addition to the fact that its drift does not have the form that is required for a $\mPi$-integrator, see Eqs.~\eqref{eq:drift_ansatz_pi_1} \& \eqref{eq:drift_ansatz_pi_2}).}
    \label{fig:integrators_p_and_q}
\end{figure*}

Although \textsc{TsafPM} has not been explicitly constructed with 2LPT in mind, we will see that it performs quite similarly to \textsc{LPTFrog} in practice (and better than \textsc{FastPM}), which is presumably due to the fact that $p_{\textsc{TsafPM}}$ is closer to the coefficient function $p$ of the most powerful integrator presented herein, which we will derive in what follows. Before that, let us briefly give an intuitive argument for why it should be possible to further improve on \textsc{LPTFrog} in terms of matching the 2LPT trajectory. When constructing \textsc{LPTFrog}, we considered a quadratic trajectory within each individual timestep as a starting point, see Eq.~\eqref{eq:LCDM_lptfrog}. However, the resulting \emph{discrete} trajectory travelled by each particle is the concatenation of two straight lines, just as for any other DKD integrator. In particular, a particle located on the exact 2LPT trajectory at $D_n$ will move on a straight line with the constant 2LPT velocity at $D_n$ in the time interval $[D_n, D_{n+\nicefrac{1}{2}})$, thus diverging from the exact 2LPT trajectory that continues on a parabola. The momentum is then updated at $D_{n+\nicefrac{1}{2}}$ using the acceleration $\vecb{A}(\vecb{X}(D_{n+\nicefrac{1}{2}}))$, but the coefficient functions $p_{\text{\textsc{LPTFrog}}}$ and $q_{\text{\textsc{LPTFrog}}}$ do not draw any connection between this acceleration and the evolution of the 2LPT term. As we will see later, this causes the asymptotic behaviour of \textsc{LPTFrog} for $D \to 0$ to differ from that of 2LPT (see Fig.~\ref{fig:integrators_p_and_q}).

In order to improve on \textsc{LPTFrog}, we will now take a closer look at the acceleration and link it to 2LPT at early times. In the single-stream regime where the Jacobian matrix $\mathbf{J} := \bnabla_{\vecb{q}} \vecb{\mPsi}$ is invertible, we can write Poisson's equation as
\begin{equation}
\begin{aligned}
\nabla_{\vecb{x}}^2 \varphi_N(\vecb{X}) = \frac{3}{2} \Omega_{\mathrm{m}} \delta(\vecb{X}) = \frac{3}{2} \Omega_{\mathrm{m}} \left[\left(\det \left( \bnabla_{\vecb{q}} \vecb{X} \right)\right)^{-1} - 1\right],
\end{aligned}
\label{eq:Poisson_with_det}
\end{equation}
with
\begin{equation}
    \det \left( \bnabla_{\vecb{q}} \vecb{X} \right) = \det \left( \mathbb{I} + \bnabla_{\vecb{q}}\vecb{\mPsi} \right) = \det(\mathbb{I} + \mathbf{J}) = 1 + \mu_1 + \mu_2 + \mu_3,
\label{eq:det_in_terms_of_invariants}
\end{equation}
where the invariants $\mu$ are given in terms of $\mathbf{J}$ as
\begin{subequations}
\begin{align}
    &\mu_{1}=\operatorname{tr}(\mathbf{J})= \mPsi_{l, l}, \label{eq:invariants_of_J_1} \\
    &\mu_{2}=\frac{1}{2}\left[\operatorname{tr}(\mathbf{J})^{2}-\operatorname{tr}\left(\mathbf{J}^{2}\right)\right]=\frac{1}{2}\left(\mPsi_{i, i}  \mPsi_{j, j} - \mPsi_{i, j} \mPsi_{j, i}\right), \\
    &\mu_{3}=\operatorname{det}(\mathbf{J})=\frac{1}{6} \varepsilon_{i k l} \varepsilon_{j m n} \mPsi_{j, i}  \mPsi_{m, k} \mPsi_{n, l}.
\label{eq:invariants_of_J_3}
\end{align}
\end{subequations}
Here, we used Einstein's sum convention and the short-hand notation $\mPsi_{i,j} := \partial_{q_j} \mPsi_i$; moreover, $\varepsilon_{ikl}$ is the Levi--Civita symbol. Inserting Eq.~\eqref{eq:det_in_terms_of_invariants} into Eq.~\eqref{eq:Poisson_with_det}, expanding the result in a Taylor series, and plugging in the LPT series for $\vecb{\mPsi}$ (see Eqs.~\eqref{eq:XP_lpt}$-$\eqref{eq:lpt_displacements}) yields to second order:
\begin{equation}
    \vecb{A}(\vecb{X}_i^{n+\nicefrac{1}{2}}) = -\bnabla_{\vecb{x}} \varphi_N(\vecb{X}_i^{n+\nicefrac{1}{2}}) = \frac{3}{2} \Omega_{\mathrm{m}} \left(D_{n+\nicefrac{1}{2}} \vecb{\psi}_i^{(1)} + \frac{7}{3} D_{n+\nicefrac{1}{2}}^2 \vecb{\psi}_i^{(2)} + \text{h.o.t.}\right),
    \label{eq:force_expanded}
\end{equation}
where we ignore the $\Omega_\Lambda$-dependent higher-order contributions to the second-order term for $\Lambda$CDM cosmology derived in Ref.~\cite{Rampf:2022}.

Note that the gradient in Eq.~\eqref{eq:force_expanded} is w.r.t.\ Eulerian coordinates, \emph{not} Lagrangian coordinates. We can, however, exploit the fact that in the limit $D \to 0$, Eulerian and Lagrangian coordinates coincide, and asymptotically match the modified momentum variable $\vecb{\mPi}$ for $D_n \to 0$ between the $\mPi$-integrator and 2LPT. Equation~\eqref{eq:kick_ansatz_pi} gives for $D_n = 0$ 
\begin{equation}
\begin{aligned}
  \vecb{\psi}_i^{(1)} + 2 \Delta D \vecb{\psi}_i^{(2)} \stackrel{!}{=} \vecb{\mPi}^{n+1}_i &= p(\Delta D, 0) \, \vecb{\mPi}_i^{n} + q(\Delta D, 0) \, \vecb{A}(\vecb{X}_i^{n+\nicefrac{1}{2}}) \\
  &\approx p(\Delta D, 0) \, \vecb{\psi}_i^{(1)} + \frac{3}{2} \Omega_{\mathrm{m}} q(\Delta D, 0) \left(\frac{\Delta D}{2} \vecb{\psi}_i^{(1)} + \frac{7}{3} \left(\frac{\Delta D}{2}\right)^2 \vecb{\psi}_i^{(2)} \right).
\end{aligned}
\end{equation}
Requiring that this equation be satisfied for arbitrary $\vecb{\psi}^{(1)}$ and $\vecb{\psi}^{(2)}$ yields two equations. The equation stemming from the $\vecb{\psi}^{(1)}$ term turns out to be the same as the Zel'dovich consistency condition in Eq.~\eqref{eq:zeldovich_consistency_condition} evaluated for $D_n = 0$. Together with the equation for $\vecb{\psi}^{(2)}$, we obtain for the limiting behaviour of $p(\Delta D, D_n)$ as $D_n \to 0$: 
\begin{equation}
    p(\Delta D, 0) = - \frac{5}{7}, \qquad q(\Delta D, 0) = \frac{16}{7 \, \Omega_{\mathrm{m}} \, \Delta D}.
    \label{eq:p_q_2LPT_limiting_behaviour}
\end{equation}
Based on these considerations, we will construct the \textsc{PowerFrog} integrator, which has the correct 2LPT asymptote for $D_n \to 0$.

\begin{example}[\textsc{PowerFrog}]
\label{example:powerfrog}
We need to make an ansatz for $p$ that has sufficient degrees of freedom to match $p_{\text{Symplectic~2}}$ to second order in $\Delta D$ while also yielding the 2LPT limit behaviour as $D_n \to 0$. Let us first consider the EdS case.

A parameterisation that serves our purpose is given by
\begin{equation}
    p_{\textsc{PowerFrog}}^{\text{EdS}}(\Delta a, a_n) = \frac{\alpha a_n^{\epsilon} + \beta a_{n+1}^{\epsilon}}{a_{n+\nicefrac{1}{2}}^{\epsilon}} + \gamma
\label{eq:p_powerfrog}
\end{equation}
with coefficients $\alpha, \beta, \gamma \in \mathbb{R}$ and $\epsilon > 0$ yet to be determined. Enforcing second-order consistency with $p_{\text{Symplectic~2}}$ w.r.t.\ $\Delta a$ (similarly to the $\epsilon$-integrator in Example~\ref{example:epsilon_integrator}) gives rise to the following relations:
\begin{equation}
    \gamma = 1 - \alpha - \beta \quad \text{($0^\text{th}$ order)}, \qquad
    \epsilon = \frac{3}{\alpha - \beta} \quad \text{($1^\text{st}$ order)}, \qquad
    \alpha = \frac{1}{8} \left(3 + 6 \beta \pm \sqrt{9 + 84 \beta + 4 \beta^2} \right) 
    \quad \text{($2^\text{nd}$ order)}.
    \label{eq:powerfrog_coefficient_conditions}
\end{equation}
The solution for $\alpha$ with the `$-$'-sign will lead to divergence as $a_n \to 0$, for which reason we choose the branch with the `$+$'-sign for $\alpha$.

Now, we determine the remaining coefficient $\beta$ such that the integrator matches the 2LPT asymptote for $a_n \to 0$. Computing the limit of $p_{\textsc{PowerFrog}}^{\text{EdS}}(\Delta a, a_n)$ for $a_n \to 0$ yields the $\Delta a$-independent expression
\begin{equation}
    \lim_{a_n \to 0} p_{\textsc{PowerFrog}}^{\text{EdS}}(\Delta a, a_n) = 2^\epsilon \beta + \gamma \stackrel{!}{=} -\frac{5}{7}
    \label{eq:p_limit_an_to_0_powerfrog}
\end{equation}
provided that $\epsilon > 0$, leaving us with a transcendental equation for $\beta$ (after plugging in the expressions for $\alpha$, $\gamma$, and $\epsilon$)
\begin{equation}
    \left(8^{\frac{8}{3 - 2 \beta +\sqrt{9 + 84 \beta + 4 \beta^2}}}-\frac{7}{4}\right) \beta +\frac{1}{8} \left(5-\sqrt{9 + 84 \beta + 4 \beta^2}\right) = -\frac{5}{7}.
\end{equation}
By solving this equation numerically, we obtain the following coefficients for $p_{\textsc{PowerFrog}}^{\text{EdS}}(\Delta a, a_n)$ in Eq.~\eqref{eq:p_powerfrog}, which we provide here with 14-digit precision:
\begin{equation}
\begin{aligned}
    \alpha &= 0.61224056624108, \qquad \beta = -0.049810198916707, \qquad \gamma = 0.43756963267562, \\
    \epsilon &= 4.5313745680590.
\end{aligned}
\label{eq:powerfrog_coeffs_EdS}
\end{equation}
By construction, \textsc{PowerFrog} is
\begin{enumerate}
    \item Zel'dovich consistent (because we define $q_{\textsc{PowerFrog}}$ according to Eq.~\eqref{eq:zeldovich_consistency_condition}), 
    \item \emph{asymptotically consistent} with 2LPT in the limit $a_n \to 0$ (in the sense of Eq.~\eqref{eq:p_limit_an_to_0_powerfrog}), and
    \item second-order consistent with the standard Symplectic~2 integrator w.r.t.\ $\Delta a$.
\end{enumerate}

What now remains is the generalisation of \textsc{PowerFrog} to the $\Lambda$CDM case. Recall that the EdS cosmology is \emph{scale-free}, which is reflected in the fact that $p(\Delta a, a_n)$ only depends on the ratio $\Delta a / a_n$. This is, however, not the case for $\Lambda$CDM cosmology, where the transition from the matter-dominated to the $\Lambda$-dominated era introduces a characteristic timescale. In order to accommodate this complication, we now allow the coefficients $\alpha$, $\beta$, $\gamma$, and $\epsilon$ to evolve as functions of growth-factor time, i.e.\ $\alpha = \alpha(D_n)$ and similarly for the other coefficients. Our ansatz in Eq.~\eqref{eq:p_powerfrog} thus becomes
\begin{equation}
    p_{\textsc{PowerFrog}}(\Delta D, D_n) = \frac{\alpha(D_n) D_n^{\epsilon(D_n)} + \beta(D_n) D_{n+1}^{\epsilon(D_n)}}{D_{n+\nicefrac{1}{2}}^{\epsilon(D_n)}} + \gamma(D_n).
\end{equation}
Second-order consistency with $p_{\text{Symplectic~2}}$ w.r.t.\ $\Delta D$ now requires that
\begin{equation}
    \gamma(D_n) = 1 - \alpha(D_n) - \beta(D_n) \quad \text{($0^\text{th}$ order)}, \qquad
    \epsilon(D_n) = \frac{2 D_n \, (\dd_D F)(D_n)}{\left(\alpha(D_n) - \beta(D_n)\right) F(D_n)} \quad \text{($1^\text{st}$ order)}.
\end{equation}
The equation for $\alpha(D_n)$, which stems from the second-order matching w.r.t.\ $\Delta D$, is lengthy and therefore provided in~\ref{sec:powerfrog_lcdm_material}.
As in the EdS case, we now determine the remaining coefficient $\beta = \beta(D_n)$ such that Eqs.~\eqref{eq:p_q_2LPT_limiting_behaviour} are satisfied. Since the growth factor for $\Lambda$CDM cosmologies (here written in the usual way as a function of the scale factor, rather than the other way around) evolves as
\begin{equation}
    D(a) = a - \frac{2 \Omega_\Lambda}{11 \Omega_{\mathrm{m}}} a^4 + \mathscr{O}(a^7),
\end{equation}
one has in particular that $D(a) / a \to 1$ for $a \to 0$. Therefore, in the limit $D_n \to 0$ where Eqs.~\eqref{eq:p_q_2LPT_limiting_behaviour} are valid (as it is only in this limit that Eulerian and Lagrangian coordinates agree), $\Lambda$CDM cosmologies converge to EdS, and we can again use Eq.~\eqref{eq:p_limit_an_to_0_powerfrog} to determine the remaining coefficient $\beta$. The difference towards the EdS case is that the coefficients need to keep their dependence on $D_n$ despite the fact that Eq.~\eqref{eq:p_limit_an_to_0_powerfrog} has been derived in the limit $a_n \to 0$, which would now correspond to $D_n \to 0$. This is necessary in order to ensure that the method is second-order convergent for arbitrary $D_n > 0$ as $\Delta D \to 0$. Thus, we now solve the following equation numerically,
\begin{equation}
    2^{\epsilon(D_n)} \beta(D_n) + \gamma(D_n) \stackrel{!}{=} -\frac{5}{7},
    \label{eq:powerfrog_LCDM}
\end{equation}
for each given $D_n$. If the timesteps used in the simulation are already known beforehand, the values of the functions $\alpha$, $\beta$, $\gamma$, and $\epsilon$ can be pre-computed. Also, note that the computational cost of numerically solving Eq.~\eqref{eq:powerfrog_LCDM} is usually negligible in comparison with the remaining operations involved in the simulation such as the force evaluation, especially when the spatial resolution is high. An intriguing avenue for future research would be to replace the asymptotic condition in Eq.~\eqref{eq:p_limit_an_to_0_powerfrog} by a 2LPT matching condition \textit{for each individual timestep}, such that the second-order term agrees exactly with 2LPT prior to shell-crossing. This would, however, require a detailed perturbative analysis, which is beyond the scope of this work.
\end{example}

\begin{figure*}[t]
\centering
  \noindent
   \resizebox{1\textwidth}{!}{
    \includegraphics{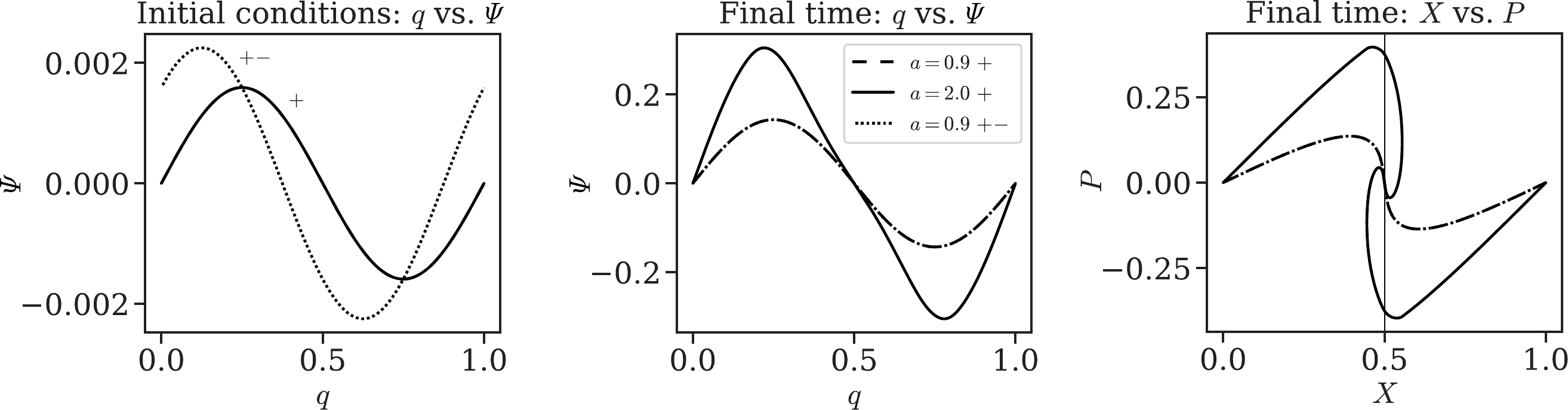}
    }
    \caption{Initial conditions and reference solutions for our numerical experiments in 1D. \textit{Left}: Initial displacement $\mPsi$ at $a_\text{ini} = 0.01$ as a function of the Lagrangian coordinate $q$. `$+$': Growing-mode-only case considered in Sections~\ref{sec:1D_results_pre} and \ref{sec:1D_results_post}, `$+-$': Mixed growing/decaying-mode case considered in Section~\ref{sec:1D_results_decaying}. \textit{Centre}: Displacement at the final time $a_\text{end} = 0.9 < 1.0 = a_\text{cross}$ (pre-shell-crossing, for the examples in Sections~\ref{sec:1D_results_pre} and \ref{sec:1D_results_decaying}) and $a_\text{end} = 2.0$ (post-shell-crossing, for the example in Section~\ref{sec:1D_results_post}). The decaying mode has virtually vanished, and the dashed (growing-mode-only) and dotted (mixed growing/decaying mode) lines overlap. Whereas the displacement described by the Zel'dovich solution prior to shell-crossing is still described by a sine wave, the displacement field at $a_\text{end} = 2.0$ has formed steeper peaks and a wavy shape around $q = 0.5$. \textit{Right}: Eulerian position $X$ vs. momentum $P$. After the first shell-crossing, three fluid streams are present around the centre of the box (which are indicated for the point $X = 0.5$ by the vertical line; for the $a = 2.0$ case this vertical line intersects the curve three times.}
    \label{fig:1D_solutions}
\end{figure*}

Figure~\ref{fig:integrators_p_and_q} shows $p(\Delta a, a)$ and $q(\Delta a, a)$, which uniquely characterise the kick of $\mPi$-integrators (see Eq.~\eqref{eq:kick_ansatz_pi}) as a function of $\Delta a$ for three different values of $a$ in EdS cosmology. For the standard Symplectic~2 integrator, the DKD version of \textsc{FastPM}, and more generally for the $\epsilon$-integrator whenever $\epsilon \geq \nicefrac{3}{2}$, one has $p(\Delta a, a) \geq 0$. In contrast, negative values of $p(\Delta a, a)$ occur with \textsc{LPTFrog}, \textsc{TsafPM}, and \textsc{PowerFrog} when $\Delta a / a$ is sufficiently large. For the $\epsilon$-integrator with $\epsilon \in (0, \nicefrac{3}{2})$, a timestep restriction must be imposed such that $p(\Delta a, a)$ remains positive (or alternatively, one could modify the $\epsilon$-integrator in the regime where the timestep restriction is not satisfied, which we do not pursue herein). The convergence of the \textsc{PowerFrog} integrator towards the 2LPT asymptotes in Eq.~\eqref{eq:p_q_2LPT_limiting_behaviour} for $a \to 0$ can be clearly seen in the $a = 0.001$ panels. The asymptotes of \textsc{LPTFrog} for $a \to 0$ are different, specifically $p_{\text{\textsc{LPTFrog}}} \to -\nicefrac{1}{5}$ and $q_{\text{\textsc{LPTFrog}}} \to 8 / (5 \Delta a)$. For $a = 0.001$, the coefficient functions $q(\Delta a, a)$ rise to a maximum of roughly $300$ (outside the plot range), before decreasing again for larger timestep sizes $\Delta a$. Thus, the maximum of $q(\Delta a, a)$ can be seen to approximate scale as $a^{-1}$. At fixed $a$, the asymptotic behaviour of the functions for $\Delta a \to 0$ is given by $p(\Delta a, a_n) \to 1$ and $q(\Delta a, a_n) \to 0$, just as is the case for consistent canonical DKD integrators with kick as defined in Eq.~\eqref{eq:dkd_kick}. As a matter of fact, $p$ and $q$ for all shown methods match up to second order in $\Delta a$ and are hence second-order accurate time integrators.
In the limit $a \to 0$, the 2LPT asymptotes behave as $p(\Delta a, a) \to -\nicefrac{5}{7}$ and $q(\Delta a, a) \to 16 / (7 \Delta a)$, implying that the former remains constant at a value $\neq 1$ and the latter diverges for $\Delta a \to 0$.

\section{Numerical tests in one dimension}
\label{sec:results_1D}
Since the PT-inspired integrators reproduce the pre-shell-crossing solutions in one dimension exactly, we focus in this section on the performance and convergence of different integrators for a one-dimensional collapse problem in three cases:
\begin{enumerate}
    \item Growing-mode-only solution prior to shell-crossing
    \item Growing-mode-only solution from shell-crossing to the deeply non-linear regime
    \item Mixed growing-/decaying-mode solution prior to shell-crossing
\end{enumerate}

Figure~\ref{fig:1D_solutions} shows the initial conditions and the solutions at final time for each of the three cases. Shell-crossing occurs in box centre at $a_{\text{cross}} = 1.0$, and we take the end times to be $a_{\text{end}} = 0.9$ and $a_{\text{end}} = 2.0$ for the pre-shell-crossing and post-shell crossing experiments, respectively.
The reference solutions prior to shell-crossing have been determined with (first-order) LPT, while the post-shell-crossing solution for the experiment in Section~\ref{sec:1D_results_post} has been computed with a high-resolution $N$-body simulation. The decaying mode is clearly visible in the initial conditions for the experiment in Section~\ref{sec:1D_results_decaying}, but has almost disappeared by the end time $a_\text{end} = 0.9$. From the rightmost panel where we plot the Eulerian position $X$ vs. momentum $P$, it becomes apparent that the flow is still in the single-stream regime at $a = 0.9 < 1.0 = a_\text{cross}$, while three fluid streams have developed around the centre of the box at $X = L/2$ by $a = 2.0$. More details regarding the setup for each case will be provided in the dedicated sections below.

Also, we will study how well the energy of the $N$-body system follows the cosmic energy equation for the different integrators. In this section, we restrict our analysis to an EdS universe (i.e.\ $\Omega_{\mathrm{m}} = 1$, $\Omega_\Lambda = 0$) for simplicity, but the findings in our numerical experiments immediately carry over to the case of $\Lambda$CDM. \\[0.0cm]

\noindent Throughout this section, we will consider the following integrators:
\begin{itemize}
    \item Standard symplectic integrator of 2$^\text{nd}$ and 4$^\text{th}$ order (`Symplectic~2' and `Symplectic~4', respectively)
    \item Runge--Kutta method of 3$^\text{rd}$ order (`RK3')
    \item \emph{Zel'dovich-consistent integrators:} \textsc{FastPM} (in the original KDK formulation), \textsc{LPTFrog}, \textsc{PowerFrog}    
\end{itemize}

The Symplectic~2 integrator is ubiquitous in cosmological simulation codes due to its simplicity, its suitability for adaptive, hierarchical timesteps, its robustness, and low memory requirements, e.g.\ \cite{AnguloHahn:2022}. The Symplectic~4 integrator has been described in Section~\ref{sec:higher_order}. While Runge--Kutta methods are extensively used in the sciences and in engineering, they lack symplecticity (at least in their standard form), which may lead to a systematically growing energy error of the numerical solution, making them less suitable for Hamiltonian systems (see e.g.\ Ref.~\cite[Fig.~4]{Springel:2005}). However, for \emph{approximate} large-scale simulations with few timesteps that do not aim to resolve particle orbits within dark matter haloes, it is much less clear if non-symplectic integrators such as RK3 produce significantly worse results than symplectic schemes.

Another important choice apart from the integrator is given by the spacing of the timesteps. This is particularly relevant for fast approximate simulations that typically evolve the simulation box from the linear regime to redshift $z = 0$ in as few as $N_{\text{timesteps}} < 100$ steps, which makes a clever trade-off between the timestep density at early and late times imperative. At early times, the growth of perturbations is close to linear, and capturing this growth accurately is the main concern in this regime. Logarithmically spacing the timesteps in $a$ rather than linearly leads to a higher timestep density at early times, and might therefore improve the accuracy of the power spectrum on large scales (e.g.\ \cite{Tassev2013}). This comes at the cost of fewer timesteps in the deeply non-linear regime at late times, potentially leading to overly spread-out haloes (as the velocity updates of particles orbiting in haloes are not frequent enough) and often affecting the power spectrum even more significantly than the early-time errors (see Ref.~\cite[Fig.~5]{Izard2016}). Another possible choice is given by uniform steps w.r.t.\ a power of $a$, i.e.\ $a^r$ with $r \in (0, 1)$ as considered by Ref.~\cite{Izard2016}, which interpolates between linear and logarithmic spacing in $a$. Elaborate hybrid strategies for improving the time sampling have been proposed; for example, in the time-stepping scheme by Ref.~\cite{Carlson2009}, two hyperparameters govern the transition from the timestep density at early to late times. Ref.~\cite{Klypin2018} use gradually increasing timesteps at early times $a < a_\text{limit}$ and switch to uniform-in-$a$ timesteps at $a = a_\text{limit}$. 

In the $\Lambda$CDM case, a natural choice for the timestep spacing with $\mPi$-integrators is uniform steps w.r.t.\ the growth factor $D$. We will restrict ourselves to uniform steps w.r.t.\ $D$ (for $\Lambda$CDM), $a$, $\log a$, and superconformal time $\tilde{t}$ in this work. For completeness, let us also mention that, in practice, the force resolution should be refined as the number of timesteps decreases (see e.g.\ Ref.~\cite[Appendix A3]{Klypin2018}); however, since the focus of this work is the comparison of the different integrators relative to one another, we will keep the force resolution fixed in our numerical experiments.

\subsection{Growing-mode-only solution prior to shell-crossing}
\label{sec:1D_results_pre}
\begin{figure}
\centering
  \noindent
   \resizebox{0.5\columnwidth}{!}{
    \includegraphics{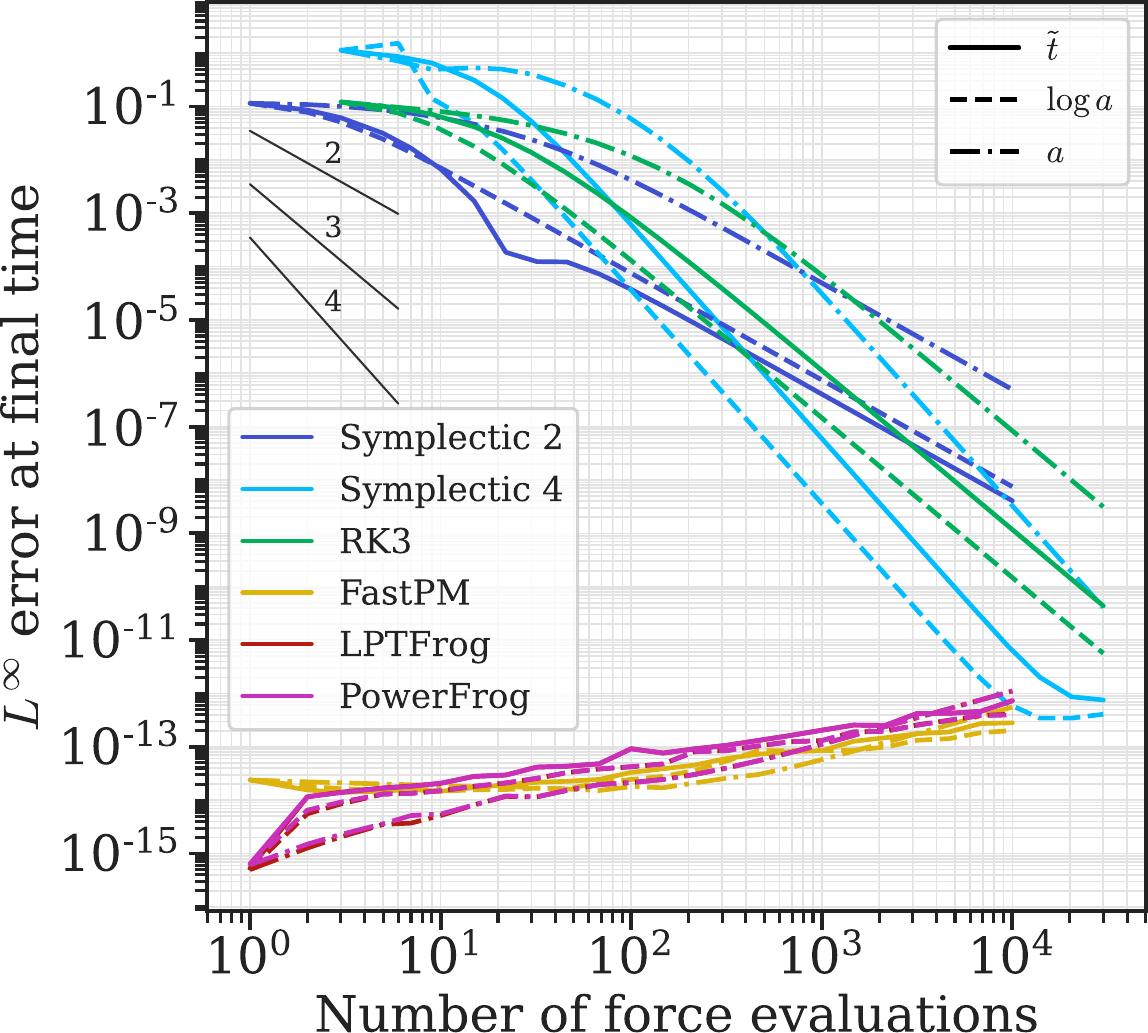}
    }
    \caption{Convergence of the numerically computed displacement field $\mPsi$ for the 1D growing-mode-only solution \textbf{prior to shell-crossing} at $a_\mathrm{end} = 0.9 < 1.0 = a_\mathrm{cross}$. Different colours correspond to different integrators, and different line styles indicate the time variable with respect to which the timesteps were taken to be uniform. To ensure a fair comparison between the methods, the $x$-axis shows the number of total force evaluations, which is given by the number of timesteps times the number of force evaluations per step required by each integrator \textbf{Zel'dovich-consistent methods reproduce the Zel'dovich solution exactly} up to round-off errors, whereas the standard symplectic schemes and the Runge--Kutta method converge at their designated orders.}
    \label{fig:convergence_1D_pre_sc_force_eval}
\end{figure}
As our first numerical test case, we consider a sine wave evolving in a one-dimensional periodic box of length $L = 1$. Specifically, we take the initial displacement field to be
\begin{equation}
     \mPsi _{q}(a_{\text{ini}}) = \frac{a_\text{ini}}{a_\text{cross}} \frac{\sin(2 \pi q)}{2 \pi},
\label{eq:cosine_potential_growing}
\end{equation}
where $a_\text{cross}$ determines the scale factor when shell-crossing occurs in the centre of the box, i.e.\ at $x = L/2 = 0.5$. In our numerical experiments, we choose $a_\text{cross} = 1.0$ and initialise the particles at scale factor $a_\text{ini} = 0.01$, see Fig.~\ref{fig:1D_solutions}.
Unlike in higher dimensions, the Poisson equation in Eq.~\eqref{eq:poisson} can be analytically integrated in one dimension, yielding for the right-hand side of the momentum equation in Eq.~\eqref{eq:eqs_of_motion}
\begin{equation}
    \partial_x \varphi_N(X_{(i)}) = \frac{3}{2} \left(\frac{i - \nicefrac{1}{2}}{N} - (X_{(i)} - \bar{X}) - \frac{1}{2}\right),
\label{eq:exact_potential_gradient}
\end{equation}
where $X_{(i)}$ is the $i$-th sorted particle position in increasing order, and $\bar{X}$ is the arithmetic mean of all particle positions. We take $N = $ 10,000 particles in our one-dimensional experiments, and we do not apply any gravitational softening. In our first experiment, we simulate the evolution of the single-mode perturbation described by the initial displacement in Eq.~\eqref{eq:cosine_potential_growing} under gravity in EdS cosmology until $a_\text{end} = 0.9$, i.e.\ shortly before shell-crossing occurs in the centre of the simulation box.

We study the convergence of different integrators in Fig.~\ref{fig:convergence_1D_pre_sc_force_eval}. Specifically, we plot the $L^\infty$ error of the displacement field $\mPsi$ at $a_{\text{end}} = 0.9 < 1.0 = a_{\text{cross}}$ as a function of the total required number of force evaluations $N_{\text{force}}$ for each integrator. Since the force evaluation is typically the most computationally expensive part of each integration step, $N_{\text{force}}$ can be viewed as a proxy for the computation time required to achieve the desired accuracy, related to the total number of timesteps as $N_{\text{force}} = \kappa N_{\text{timesteps}}$, where $\kappa = 1$ for Symplectic~2, \textsc{FastPM}\footnote{While KDK schemes such as \textsc{FastPM} perform two momentum updates in each timestep and hence in principle require two force evaluations, the second update can be combined with the first update of the subsequent step, effectively reducing the number of required force evaluations per timestep to $\kappa = 1$.}, and the $\mPi$-integrators in DKD form, and $\kappa = 3$ for Symplectic~4 and RK3.

By construction, \textsc{FastPM}, \textsc{LPTFrog}, and \textsc{PowerFrog} agree exactly (up to rounding errors) with the analytic solution, which in the one-dimensional case is given by the Zel'dovich solution prior to shell-crossing, irrespective of the number of timesteps (in fact, the error grows as the number of timesteps increases because the rounding errors accumulate). The lines for the two DKD schemes \textsc{LPTFrog} and \textsc{PowerFrog} are barely distinguishable. The standard symplectic integrators converge as expected in view of their respective order, and the (non-symplectic) RK3 integrator exhibits 3$^\text{rd}$-order convergence. For the Symplectic~2 stepper, uniform timesteps w.r.t.\ superconformal time $\tilde{t}$ yield the smallest error, closely followed by $\log a$-steps, whereas $a$-steps lead to an error that is roughly two orders of magnitude larger for a given number of force evaluations, indicating that a higher time sampling density at early times is preferable in this case. For Symplectic~4 and RK3, logarithmic timesteps in $a$ produce the smallest error.

\subsection{Growing-mode-only solution from shell-crossing to the deeply non-linear regime}
\label{sec:1D_results_post}
\begin{figure*}
\centering
  \noindent
   \resizebox{1\textwidth}{!}{
    \includegraphics{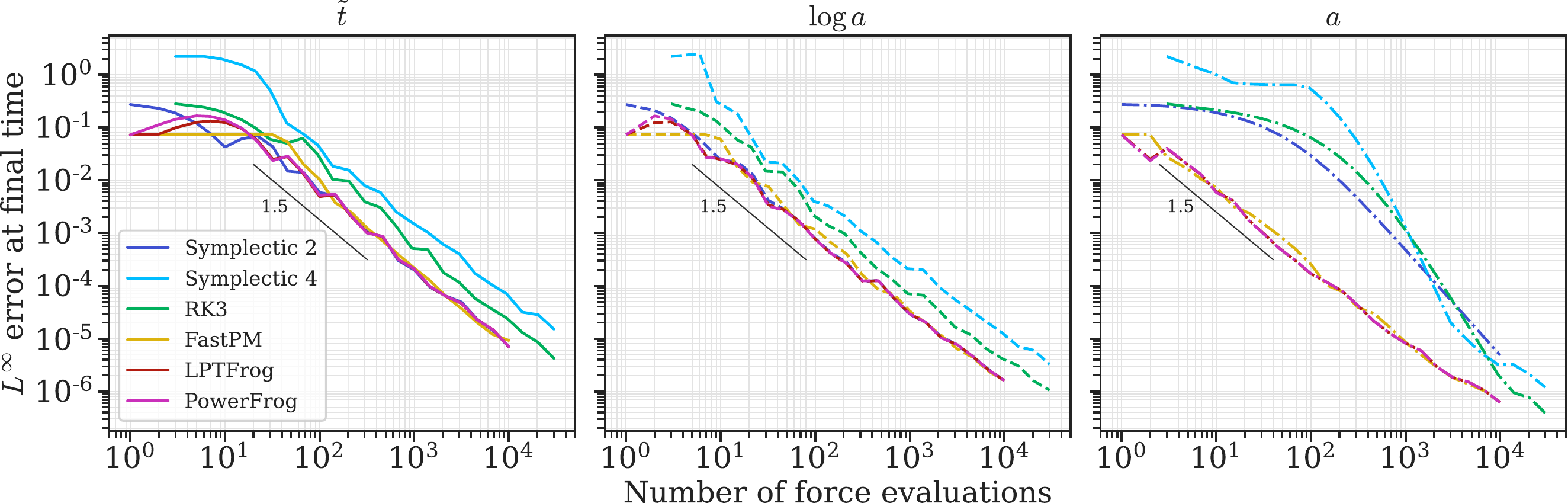}
    }
    \caption{Convergence of the numerically computed displacement field $\mPsi$ for the 1D growing-mode-only solution in the \textbf{post-shell-crossing} regime at $a_\mathrm{end} = 2.0 > 1.0 = a_\mathrm{cross}$. Now, the Zel'dovich-consistent integrators \textsc{FastPM}, \textsc{LPTFrog}, and \textsc{PowerFrog} no longer produce the exact solution and are on par with Symplectic~2 when suitable timesteps are chosen. Importantly, note that \textbf{all methods converge at order} $\nicefrac{\mathbf{3}}{\mathbf{2}}$, irrespective of their nominal order of convergence, which assumes sufficient regularity. Thus, higher-order integrators such as RK3 and Symplectic~4 entail an unnecessary overhead in the post-shell-crossing regime. For uniform steps in $a$, the $L^\infty$ error for Symplectic~2 \& 4 and RK3 comes from particles that are still in the single-stream region, for which reason these integrators exhibit their nominal order of convergence $> \nicefrac{3}{2}$ in this case. Also, note that when considering the $L^p$ error for $p < \infty$ rather than $L^\infty$, the decrease in convergence order would be expected to be less severe because the error then takes \emph{all} particles into consideration, and particularly also those that lie outside the shell-crossed region. We do not use any gravitational softening here.} 
    \label{fig:convergence_1D_post_sc_force_eval}
\end{figure*}

When shell-crossing occurs at $a_\text{cross} = 1.0$, particles originating from the left and right domain halves cross at the centre of the simulation box at $x = L/2 = 0.5$, causing the velocity field to become multivalued when viewed as a function of (Eulerian) position $x$ for $a > a_\text{cross}$. Specifically, three fluid streams develop around $x = L/2$ after the first shell-crossing (see the right panel in Fig.~\ref{fig:1D_solutions}). At the very instant of shell-crossing, the density at $x = 0.5$ becomes infinite, and standard perturbation techniques break down (e.g.\ \cite{ Rampf:2021, Taruya2017, Pietroni2018}). In particular, under the Zel'dovich approximation, crossing particles continue their ballistic motion at constant velocity unaffected by the encounter. Shell-crossing therefore has two important consequences for numerical integrators: 
\begin{enumerate}
\item Since the Zel'dovich approximation does not correctly describe the particle dynamics in the post-shell-crossing regime, building it into the integrator as done in $\mPi$-integrators can no longer be expected to provide more accurate solutions than standard integrators.
\item The sudden decrease in regularity of the (Lagrangian) acceleration field should limit the order of convergence of higher-order integrators, which require a sufficiently regular right-hand side in order to achieve their designated order of convergence. Based on Proposition~\ref{prop:convergence_loss}, we expect the global post-shell-crossing order of convergence to be $\nicefrac{3}{2}$ in the planar-wave collapse scenario, see Eq.~\eqref{eq:psi_singularity_behaviour}.
\end{enumerate}

In what follows, we will demonstrate that these effects occur in practice by means of the same sine-wave example as above, but this time evolved until scale factor $a_\text{end} = 2.0 > 1.0 = a_\text{cross}$. Figure~\ref{fig:convergence_1D_post_sc_force_eval} shows again the $L^\infty$ error of $\mPsi$, now evaluated at $a_\text{end} = 2.0$ using a numerical reference simulation with 100,000 timesteps.
As expected from the lacking smoothness of the Lagrangian acceleration field, all methods converge at order $\nicefrac{3}{2}$. In particular, the higher-order methods RK3 and Symplectic~4 now perform worse than the second-order methods because the additional force evaluations in each step do not improve the order of convergence in this case. Interestingly, the errors of Symplectic~2, \textsc{FastPM}, \textsc{LPTFrog}, and \textsc{PowerFrog} are very similar for uniform steps in $\tilde{t}$ and $\log a$, which indicates that incorporating low-order LPT information into an integrator does not harm its performance even in the post-shell-crossing regime when the LPT series no longer provides the correct solution (or even diverges).
The reason for Symplectic~2 \& 4 and RK3 being able to maintain a higher order of convergence with uniform-in-$a$ steps is that the maximum error at $a_\text{end} = 2.0$ occurs for particles that still lie \emph{outside} the shell-crossed region. Also, note that the convergence w.r.t.\ $L^p$ norms for $p < \infty$ would, in general, be better than the $(\Delta a)^{\nicefrac{3}{2}}$ convergence shown in Fig.~\ref{fig:convergence_1D_post_sc_force_eval} as \emph{all} particles (in single-stream and multi-stream regions) would contribute to the error, not only the single particle whose displacement error is maximal.

\subsection{Mixed growing/decaying mode-solution prior to shell-crossing}
\label{sec:1D_results_decaying}
\begin{figure*}
\centering
  \noindent
   \resizebox{1\textwidth}{!}{
    \includegraphics{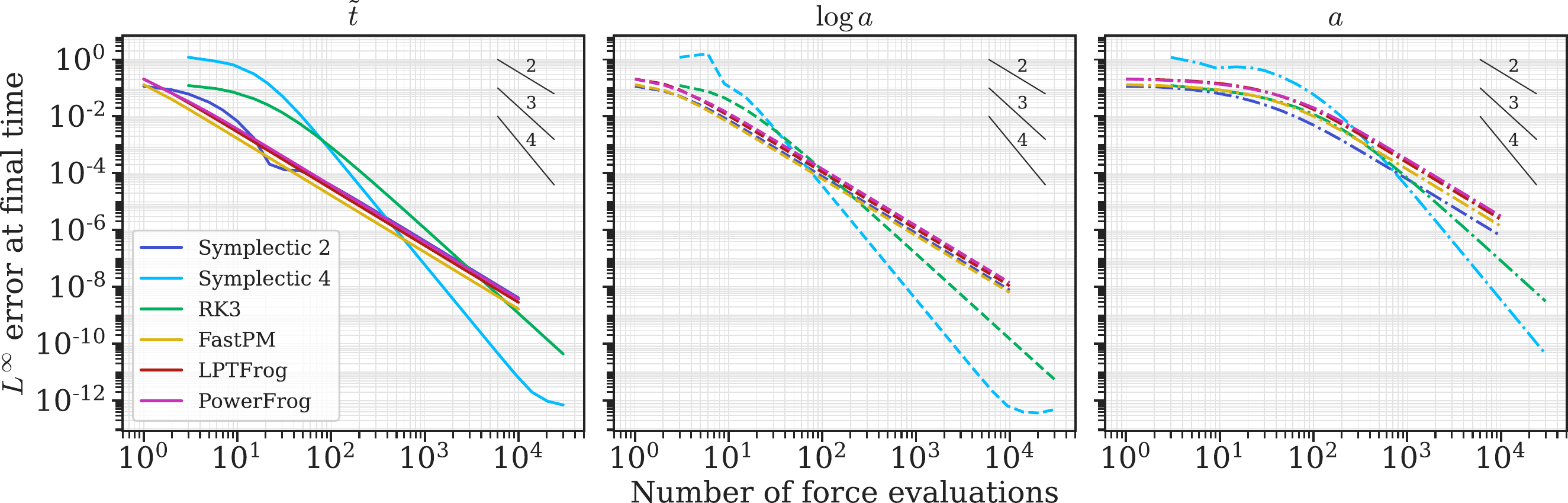}
    }
    \caption{Same as Fig.~\ref{fig:convergence_1D_pre_sc_force_eval} (i.e., prior to shell-crossing), but for the \textbf{mixed growing/decaying-mode} case. Since \textsc{FastPM} and our new Zel'dovich-consistent $\mPi$-integrators are constructed based on the growing mode solution only, they no longer yield the exact solution as was the case for the growing-mode-only case prior to shell-crossing. In particular, the performance of all second-order integrators (standard and $\mPi$-integrators) is comparable.}
    \label{fig:convergence_1D_decaying_force_eval}
\end{figure*}
In our final one-dimensional example, we consider the case of a mixed growing/decaying-mode solution. When setting up initial conditions for cosmological simulations, a `sweet spot' between two incompatible desiderata needs to be found: on one hand, one would like to start the simulation as early as possible in order to ensure the perturbations are still linear and perturbation theory provides an accurate solution; on the other hand, the perturbations at early times are very small, and undesired numerical discreteness effects play a larger role (see e.g.\ \cite{Hahn:2013, Melott1997, Scoccimarro1998, Crocce2006}). 

Attempts to remedy discreteness effects have been presented in the literature, for example particle linear theory \cite{Garrison:2016}. Importantly, initialising $N$-body simulations with 2LPT instead of the Zel'dovich approximation causes transients to decay much more rapidly, significantly improving the higher moments of the density field \cite{Scoccimarro1998}. Recently, Ref.~\cite{Michaux:2021} showed that when going to 3LPT, even later starting times are possible, thus enabling a further reduction of discreteness effects. Since \textsc{FastPM} and our new $\mPi$-integrators are purely based on the growing mode solution and neglect the decaying modes, it is interesting to compare the different integrators in the presence of a decaying mode. To this aim, we consider the following initial displacement field, which describes a mixed growing/decaying-mode scenario:
\begin{equation}
         \mPsi _{q}(a_{\text{ini}}) = \frac{a_\text{ini}}{a_\text{cross}} \frac{\sin(2 \pi q)}{2 \pi} + A_\text{decay} D^{-}(a_\text{ini}) \frac{\cos(2 \pi q)}{{2 \pi}},
\end{equation}
where $D^{-}(a_\text{ini}) = a_\text{ini}^{-\nicefrac{3}{2}}$, and $A_\text{decay}$ sets the amplitude of the decaying mode. We choose $A_\text{decay} = 10^{-5}$ here such that the growing and the decaying mode have equal amplitudes at $a_\text{ini} = 0.01$ and a phase difference of $\nicefrac{\pi}{2}$.
We evaluate the solution in the pre-shell-crossing regime at $a_\text{end} = 0.9 < 1.0 = a_{\text{cross}}$.

The $L^\infty$ errors of the displacement field towards the exact solution at scale factor $a_\text{end}$ are shown in Fig.~\ref{fig:convergence_1D_decaying_force_eval}. Just as in the growing-mode-only case, $\sim$100 $-$ 10,000 force evaluations are required for the higher-order methods Symplectic~4 and RK3 to overtake the standard Symplectic~2 integrator in terms of accuracy. Note that since the acceleration field is still smooth at $a_\text{end} $, the higher-order integrators still exhibit the expected convergence orders despite the presence of the decaying mode. Unlike in the growing-mode-only case, however, the LPT-inspired integrators \textsc{FastPM}, \textsc{LPTFrog}, and \textsc{PowerFrog} no longer reproduce the exact solution. This is of course not surprising as both integrators only consider the fastest growing limit (see Eq.~\eqref{eq:2lpt_fastest_growing}) where the decaying modes have been neglected. Hence, the performance of these methods is very similar to that of the standard Symplectic~2 integrator.

\subsection{Energy balance}
In view of our new $\mPi$-integrators not being symplectic, one might worry whether this might lead to systematic energy errors. In this section, we will show that for the plane-wave example this is not the case and in practice, all integrators that we study perform very similarly in this aspect. In an expanding universe, energy is in fact not conserved, but rather obeys the cosmic energy equation (also known as the Layzer--Irvine equation \cite{Layzer1963, Irvine1965}) for the kinetic and potential energy $T$ and $U$, respectively, which reads
\begin{equation}
    \dd_t (T + U) + H (2 T + U) = 0 \qquad \Leftrightarrow \qquad \dd_a [a (T + U)] = - T.
\label{eq:layzer_irvine}
\end{equation}
We will study energy conservation for different numerical integrators using the latter formulation in Eq.~\eqref{eq:layzer_irvine}.
The kinetic energy $T$ and potential energy $U$ are defined based on the Hamiltonian in cosmic time in Eq.~\eqref{eq:Hamiltonian_t} as
\begin{equation}
    T = \int_{\mathscr{Q}} \dd^3 q \, \frac{\|\vecb{P}\|^2}{2 a^2}  = \frac{1}{N}\sum_{i=1}^N \frac{\|\vecb{P}_i\|^2}{2 a^2} \qquad \text{and} \qquad U = \frac{1}{2 a} \int_{\mathscr{C}} \dd^3x \, \varphi_N \delta.
\label{eq:layzer_irvine_T_and_U}
\end{equation}

Following Ref.~\cite{winther2013layzer}, we integrate the potential energy by parts and use Poisson's equation in Eq.~\eqref{eq:poisson} to obtain
\begin{equation}
    U = \frac{1}{3 a \Omega_{\mathrm{m}}} \int_{\mathscr{C}} \dd^3x \, \varphi_N \nabla_{\vecb{x}}^2 \varphi_N = - \frac{1}{3 a \Omega_{\mathrm{m}}} \int_{\mathscr{C}} \dd^3x \, \|\bnabla_{\vecb{x}} \varphi_N \|^2 \approx -\frac{1}{3 a \Omega_{\mathrm{m}} N_c} \sum_{i=1}^{N_c} \|(\bnabla_{\vecb{x}} \varphi_N)_i\|^2,
\label{eq:U_in_layzer_irvine}
\end{equation}
where we evaluate the forces on a regular grid consisting of $N_c$  cells.

We will measure the deviation of the energy in the $N$-body system from the Layzer--Irvine equation as a function of the scale factor $a$ for the growing-mode-only sine wave potential defined in Eq.~\eqref{eq:cosine_potential_growing} for $a \in [0.01, 2]$, where shell-crossing again occurs at $a_\text{cross} = 1.0$. We consider the diagnostic metric
\begin{equation}
    \Delta E = - \left(1 + \frac{T}{\dd_{a} [a (T + U)]}\right),
\end{equation}
which vanishes when the Layzer--Irvine equation is exactly satisfied. 

\begin{figure*}
\centering
  \noindent
   \resizebox{1\textwidth}{!}{
    \includegraphics{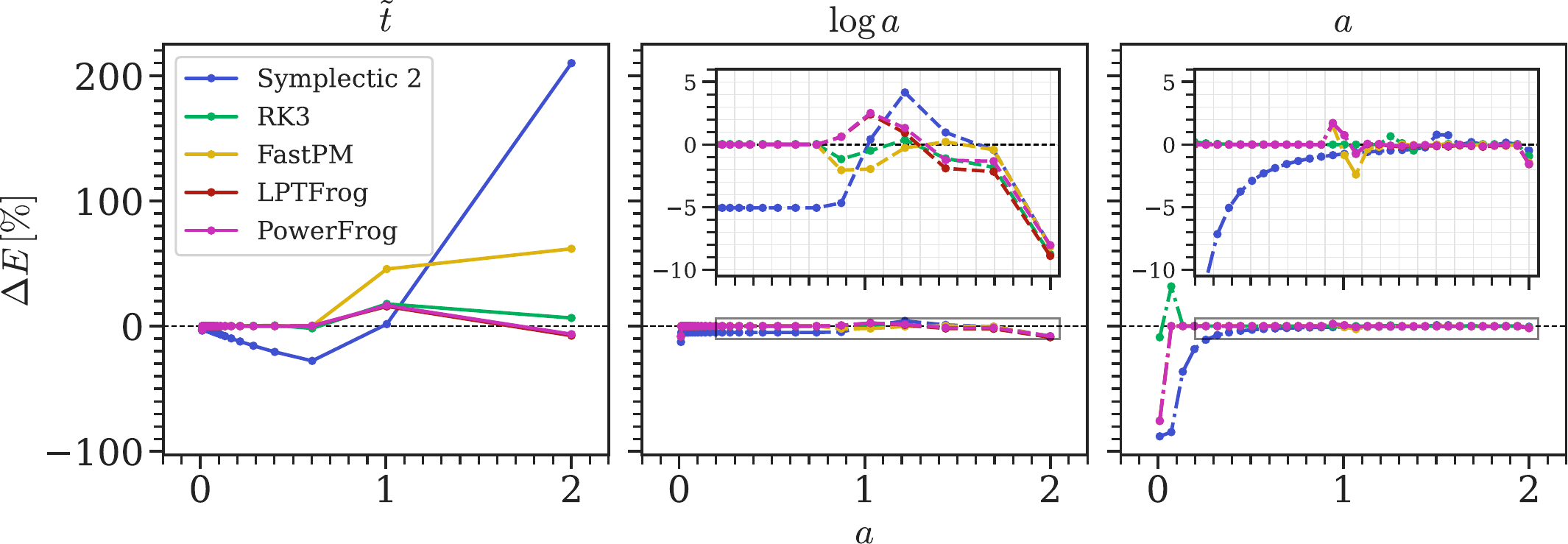}
    }
    \caption{Deviation of the energy in the $N$-body simulation from the cosmic energy (Layzer--Irvine) equation for a 1D growing-mode-only simulation that shell-crosses at $a_\text{cross} = 1.0$, for uniform timesteps w.r.t.\ $\tilde{t}$, $\log a$, and $a$. Clearly, the RK3 integrator and \textsc{LPTFrog} perform similarly w.r.t.\ this metric to the symplectic methods. The major source of energy errors is given by overly large timesteps.}
    \label{fig:1D_layzer_irvine}
\end{figure*}

Figure~\ref{fig:1D_layzer_irvine} shows $\Delta E$ for $N_{\text{timesteps}} = 32$ uniform timesteps in $\tilde{t}$, $\log a$, and $a$, for  $N = $ 10,000 particles and different integrators. Evidently, extremely large timesteps cause the most severe deviations from the cosmic energy equation: for constant steps w.r.t.\ superconformal time $\tilde{t}$, the timesteps are strongly clustered at low values of $a$, and only a single timestep lies between $a_{\text{cross}} = 1.0$ and $a_{\text{end}} = 2.0$, causing a large energy error at $a_{\text{end}} = 2.0$. In the pre-shell-crossing regime, the Zel'dovich-consistent integrators are exact, for which reason also the cosmic energy equation is fulfilled until (shortly before) shell-crossing.\footnote{The non-zero values of $\Delta E$ at the initial scale factor $a_\text{ini}$ are a numerical artefact arising from the finite-difference approximation of the derivative $\dd_a$ at the interval boundary.} The energy balance for the RK3 integrator prior to shell-crossing is also excellent, despite its lack of symplecticity. In contrast, the standard leapfrog Symplectic~2 integrator shows a drift in $\Delta E$ for $\tilde{t}$-steps, which reverses its sign at $a = a_\text{cross}$. Interestingly, the non-symplectic \textsc{LPTFrog}, \textsc{PowerFrog} (for which the lines overlap in the $\tilde{t}$ case), and RK3 exhibit smaller errors than the Symplectic~2 and \textsc{FastPM} schemes after shell-crossing.

For uniform steps w.r.t.\ $\log a$ and $a$, the deviation from the cosmic energy balance is much smaller. For all integrators, fluctuations in $\Delta E$ occur around shell-crossing; afterwards, the values of $\Delta E$ settle again within an error band of $\pm 1\%$ for $a$-steps, whereas the sparser time sampling for  $\log a$-steps causes larger errors at late times. The energy error of the non-symplectic RK3 method is very similar to that of Symplectic~2 and the other integrators for all cases. Our main conclusion from this experiment is that sufficiently small timesteps are much more important than symplecticity for obtaining the correct evolution of the energy in simulations that only resolve mildly non-linear scales. In view of the fact that the Hubble flow effectively introduces a drag term (see \ref{sec:contact}
), this is perhaps not surprising.
(However, we caution that symplecticity becomes crucial when simulating orbits within bound structures that have decoupled from the Hubble flow such as dark matter haloes, which approximate methods such as \textsc{FastPM} do not aim to resolve.) In this experiment, the fact that \textsc{LPTFrog} is connected to contact geometry did not lead to improved results in terms of the $\Delta E$ metric in comparison to e.g.\ \textsc{PowerFrog}, and their difference is miniscule.
We verified that our new $\mPi$-integrators perform similarly to \textsc{FastPM} also in two and three dimensions in terms of the energy balance; however, the evaluation of the potential energy by cloud-in-cell interpolation becomes significantly noisier.

\section{Numerical test in two dimensions}
\label{sec:results_2D}
\begin{figure*}
\centering
  \noindent
   \resizebox{1\textwidth}{!}{
    \includegraphics{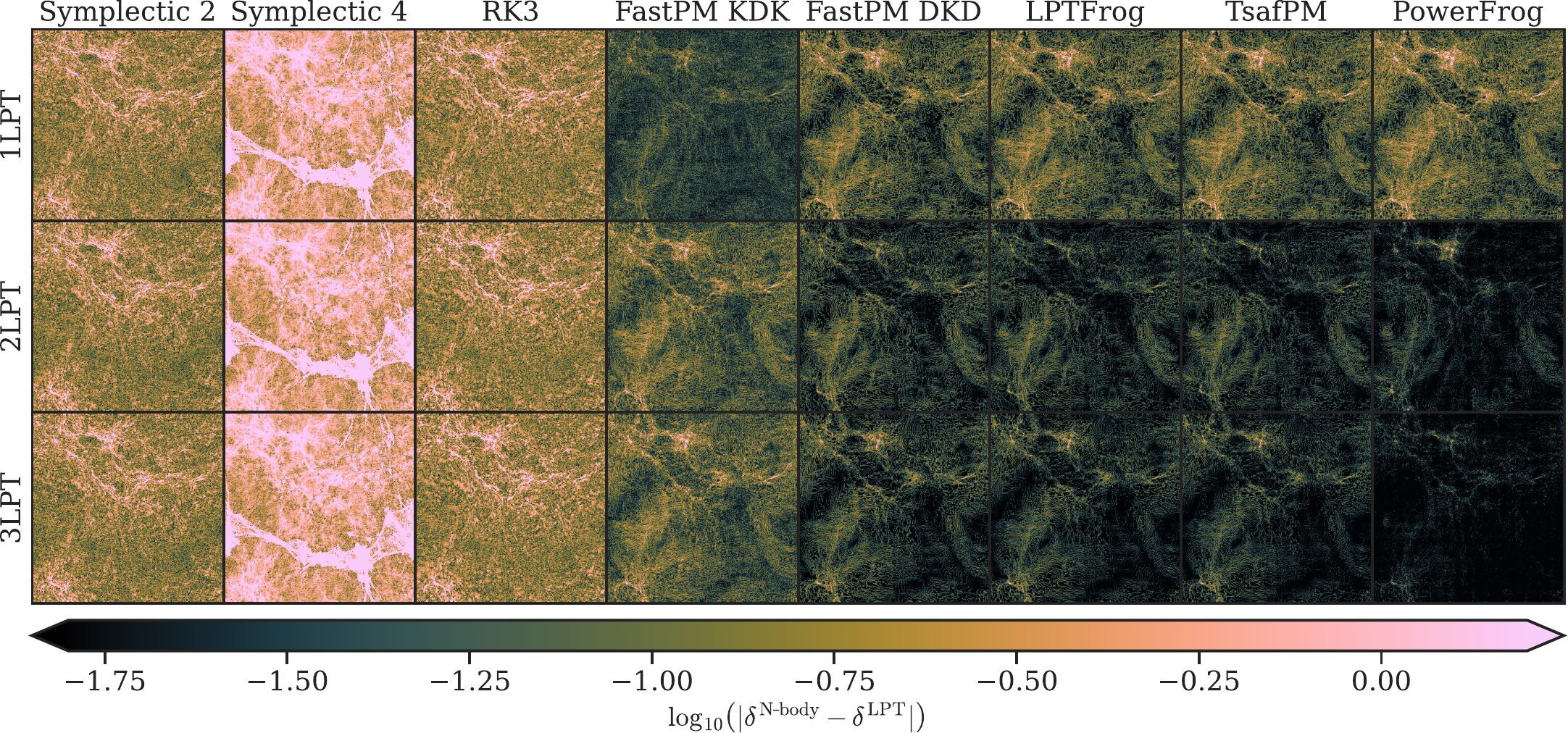}
    }
    \caption{Absolute difference between the $N$-body and LPT density contrast after a \emph{single} timestep in two dimensions. Whereas the original \textsc{FastPM} KDK scheme is closest to the 1LPT (Zel'dovich) solution, it does not capture the corrections by higher LPT orders. On the other hand, the DKD variant of \textsc{FastPM}, \textsc{LPTFrog}, and \textsc{TsafPM} have a much smaller error w.r.t.\ 2LPT. \textsc{PowerFrog} achieves the best performance and produces a solution that actually lies closer to 3LPT than to 2LPT. The corresponding error distributions are shown in Fig.~\ref{fig:2d_kdes}.}
    \label{fig:2d_delta}
\end{figure*}
In two dimensions, the Zel'dovich approximation no longer provides the correct growing-mode solution, and one has instead a recursive hierarchy of perturbative corrections prior to shell-crossing \cite{Rampf:2012,Zheligovsky:2014,Matsubara:2015}. In particular, this implies that Zel'dovich-consistent $\mPi$-integrators do \textit{not} exactly yield the correct solution in this regime, and differences between the different PT-inspired methods will be visible. Here, we will test how well the integrators match the LPT solutions before shell-crossing when performing a single timestep.

For this experiment, we take $N = 512^2$ particles in a simulation box with edge length $L = 1$ (in arbitrary units). We consider EdS cosmology, i.e.\ $\Omega_{\mathrm{m}} = 1$, and the initial power spectrum is taken to scale as $P(k) \propto k^{-2}$, normalised in such a way that the standard deviation of the density field at the initial time $a_\text{ini}$ is $\sigma(\delta(a_{\text{ini}})) = 0.03$. In order to initialise the particle positions and velocities as accurately as possible, we compute them at $a_\text{ini}$ using 9$^\text{th}$-order LPT. We evaluate the gravitational force using a tree-PM scheme \cite{Bode2003}, where the gravitational force is split up into a short-range part (whose evaluation has $\mathscr{O}(N \log N)$ complexity when lumping together the gravitational force exerted by multiple adjacent particles on another far-away particle by means of a tree structure) and a long-range part, which is evaluated using the Fast Fourier Transform in Fourier space (with complexity $\mathscr{O}(N_c \log N_c + N)$, where $N_c$ is the number of grid cells, which we take to be $N_c = N$ here). We use the following differentiation kernels in the PM computation:
\begin{subequations}
\begin{align}
    \nabla_{\vecb{x}}^{-2} &\to -k^{-2} \quad (\text{exact}), \\
    \partial_{x_\ell} &\to \frac{i}{6} \left(8 \sin(k_\ell) - \sin(2 k_\ell)\right) \quad (4^{\text{th}}\text{-order finite-difference kernel}),
\end{align}
\label{eq:differentiation_kernels}
\end{subequations}
where $k_\ell$ is the wave number along the $\ell$-th dimension in units of the Nyquist mode $k_{\text{Nyquist}, \ell} = \pi N_{c}^{\nicefrac{1}{d}} / L$, and $k^2 = \|\vecb{k}\|^2 = \sum_{\ell=1}^d k_\ell^2$ (where the spatial dimension $d = 2$ in this example). We compute the resulting density contrast $\delta^{N\!\text{-body}}$ with cloud-in-cell interpolation.

We perform a single step with different integrators, where we take the ratio between the timestep and initial time to be large, namely $\Delta a / a_\text{ini} = 14$, in order to highlight the differences between the integrators. This leads to a standard deviation of the final density field of $\sigma(\delta(a_\text{end})) = 0.55$. The choice of the intermediate time has a substantial effect when performing only one timestep, for which reason we use the arithmetic mean of the initial and final scale factor also for \textsc{FastPM} here, i.e.\ $(a_\text{ini} + a_\text{end}) / 2$, rather than the geometric mean as in the original \textsc{FastPM} scheme. Since the standard symplectic integrators and RK3 do not use the growth factor ($\Lambda$CDM) / scale factor (EdS) as the time variable and we use the Hamiltonian in superconformal time in our implementation, the intermediate time for these methods is taken as the arithmetic mean between the superconformal time at the beginning and the end of the timestep, i.e.\ $(\tilde{t}_\text{ini} + \tilde{t}_\text{end}) / 2$.

The resulting density residuals between the $N$-body simulation and the LPT solution are shown in Fig.~\ref{fig:2d_delta} for different integrators (columns) and LPT orders (rows). For the standard symplectic integrator as well as RK3, the density error between the $N$-body simulation and LPT is substantially larger than the subtle differences between different LPT orders and is strongly correlated with the density field itself, see Fig.~\ref{fig:data_2d_psi_and_delta_lpt} in \ref{sec:2D_appendix}. The Symplectic~4 integrator exhibits the largest error w.r.t.\ the LPT solution, which is unsurprising given our results for the 1D case prior to shell-crossing, which already indicated that many timesteps are necessary in order for the extra force evaluations to pay off. The original KDK \textsc{FastPM} integrator performs significantly better than the standard integrators, and its error towards the 1LPT solution is much smaller; however, a single timestep is not enough to capture the 2LPT and 3LPT corrections. In contrast, switching to the DKD variant of the \textsc{FastPM} integrator brings the $N$-body solution closer to the 2LPT and 3LPT solution.

\begin{figure}
\centering
  \noindent
   \resizebox{0.82\textwidth}{!}{
    \includegraphics{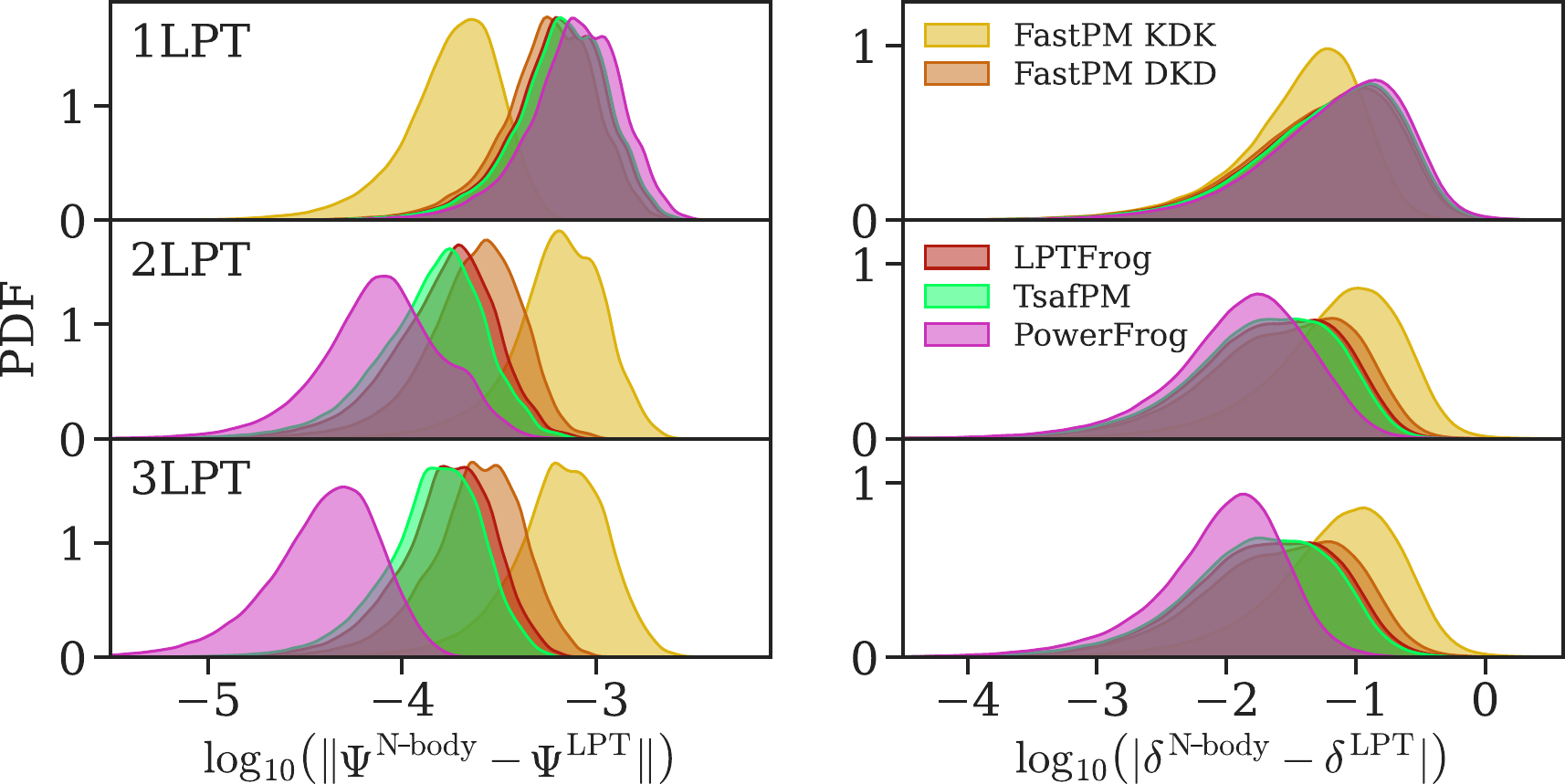}
    }
    \caption{Kernel density estimates of the error distributions between the $N$-body simulation and different LPT orders after a single step in two dimensions, for the displacement (left) and the density field (right). As expected in view of its construction, the \textsc{FastPM} integrator produces the smallest residual w.r.t.\ to the 1LPT solution, but has a much larger error towards the 2LPT and 3LPT solutions. In contrast, the solutions with our new integrators lie significantly closer to 2LPT and 3LPT than to 1LPT. Although we derived \textsc{LPTFrog} and \textsc{PowerFrog} by considering the 2LPT trajectory, the displacement after a DKD partially incorporates the 3LPT contribution, which is excited when computing the momentum update in the kick at the intermediate time.}
    \label{fig:2d_kdes}
\end{figure}

\textsc{PowerFrog} produces the most accurate solution: in fact, the $N$-body displacement field lies even closer to the 3LPT solution than the 2LPT solution in many regions of the 2D simulation box (see e.g.\ the cluster-like structure in the upper centre of the $N$-body vs. LPT residual). Whilst we constructed \textsc{PowerFrog} such that it matches (only) the 2LPT asymptote for $D_n \to 0$, this result is not entirely unexpected: the computation of the acceleration after the first half of the drift also excites higher-order LPT terms, see Eq.~\eqref{eq:force_expanded}. Although these higher-order terms enter the \textsc{PowerFrog} kick with different factors from those that would be required for completely cancelling out higher-order errors, these terms still move the \textsc{PowerFrog} solution towards higher LPT orders, at least partially. It is also interesting to note that although \textsc{TsafPM} -- unlike \textsc{LPTFrog} and \textsc{PowerFrog} -- is not inspired by 2LPT, it performs slightly better than \textsc{LPTFrog}. This can be understood from Fig.~\ref{fig:integrators_p_and_q}, which shows that the coefficient functions $p$ and $q$ defining \textsc{TsafPM} are closer to \textsc{PowerFrog} (and thus to the 2LPT asymptote) than their counterparts for \textsc{LPTFrog}. 

In Fig.~\ref{fig:2d_kdes}, we plot the error distributions of the displacement and density fields for the Zel'dovich-consistent integrators. The errors for Symplectic~2 \& 4 and RK3 are much larger and therefore not shown here, which can already be guessed based on Fig.~\ref{fig:2d_delta}. The error distributions quantitatively confirm the superiority of our new $\mPi$-integrators: while the original \textsc{FastPM} KDK scheme exhibits the smallest error towards the 1LPT solution, the average displacement error of \textsc{PowerFrog} towards 3LPT is more than an order of magnitude smaller than with \textsc{FastPM} KDK. Switching from the KDK to the DKD variant of \textsc{FastPM} improves its match towards 2LPT and 3LPT in this experiment, but our new integrators \textsc{LPTFrog}, \textsc{TsafPM}, and \textsc{PowerFrog} still produce more accurate displacement and density fields.

\section{Numerical tests in three dimensions}
\label{sec:results_3D}
\subsection{\textsc{Quijote} simulation suite: $(1 \ h^{-1} \ \mathrm{Gpc})^3$ box}
Now, we study the efficiency of our new integrators for the case of a $\Lambda$CDM universe in $d = 3$ dimensions. To this aim, we take data from the \textsc{Quijote} gravity-only $N$-body simulation suite \cite{quijote}; specifically, we consider 20 simulations that share the same `fiducial' $\Lambda$CDM cosmology, $\Omega_{\mathrm{m}} = 0.3175$ (density parameter for total matter), $\Omega_b = 0.049$ (density parameter for baryonic matter), $\Omega_\Lambda = 0.6825$ (density parameter for dark energy), $h = 0.6711$ (dimensionless Hubble constant), $n_s = 0.9624$ (spectral index of initial density perturbations), $\sigma_8 = 0.834$ (present-day root-mean-square of matter density averaged over a sphere of radius $8 \ h^{-1} \ \text{Mpc}$), massless neutrinos. These simulations track the evolution of $512^3$ particles in an $L^3 = (1 \ h^{-1} \ \text{Gpc})^3$ box from $z = 127$ down to $z = 0$. 

We evolve the \textsc{Quijote} initial conditions for each of the 20 simulations until $z = 0$ with different integrators, using 2, 4, and 8 timesteps. We also run a reference simulation with $128$ \textsc{FastPM} KDK timesteps, to which we compare the other results. Taking the reference to be a PM simulation performed with the exact same code as the other simulations (rather than comparing to the $z=0$ \textsc{Quijote} results that were obtained with the simulation code \textsc{Gadget-3}, an improved version of \textsc{Gadget-2} \cite{Springel:2005}) allows us to distill the effect of the time-stepping while eliminating all other factors that impact the resulting statistics, most importantly the coarser force resolution of our PM-only code as compared to the tree-PM code \textsc{Gadget-3}.

For our simulations, we take the timesteps to be uniformly spaced in growth-factor time $D$, which is the most natural choice for $\mPi$-integrators and ensures that the time sampling at late times is sufficiently dense when using $\lesssim 10$ timesteps. We compute the gravitational forces in Fourier space using the PM method with a grid of resolution $N_c = 1024^3$ using the differentiation kernels in Eq.~\eqref{eq:differentiation_kernels} (where now $d = 3$), and we dealiase the forces by means of an interlaced grid \cite{Chen1974}.

We will analyse the resulting density fields via the normalised cross-spectrum (also known as the correlation function) between the reference solution at $z = 0$ (denoted with the superscript ${}^\text{ref}$) and the fast solutions with much fewer timesteps, as well as the transfer functions of the power spectra and the equilateral bispectra. As customary, we define the power spectrum $P(k)$ and bispectrum $B(k_1, k_2, k_3)$, and the cross-power spectrum $P^\text{cross}(k)$ between two fields $A$ and $B$ as
\begin{subequations}
\begin{align}
    \langle \tilde{\delta}(\vecb{k}_1) \tilde{\delta}(\vecb{k}_2) \rangle &= (2 \pi)^3 \delta_\textrm{D}^{(3)}(\vecb{k}_1 + \vecb{k}_2) P(k_1), \\
    \langle \tilde{\delta}(\vecb{k}_1) \tilde{\delta}(\vecb{k}_2) \tilde{\delta}(\vecb{k}_3) \rangle &= (2 \pi)^3 \delta_\textrm{D}^{(3)}(\vecb{k}_1 + \vecb{k}_2 + \vecb{k}_3) B(k_1, k_2, k_3), \\
    \langle \tilde{\delta}_A(\vecb{k}_1) \tilde{\delta}_B(\vecb{k}_2) \rangle &= (2 \pi)^3 \delta_\textrm{D}^{(3)}(\vecb{k}_1 + \vecb{k}_2) P^\text{cross}(k_1),
\end{align}
\end{subequations}
where $k = \|\vecb{k}\|$ and $\tilde{\delta}(\vecb{k})$ is the Fourier transform of $\delta(\vecb{x})$. In this work, we will restrict ourselves to analysing the equilateral configuration of the bispectrum, i.e.\ $k_1 = k_2 = k_3 = k$. Further, we define the numerical transfer functions for the power spectrum and equilateral bispectrum $B_{\text{equi}}(k)$ as
\begin{equation}
    T_P(k) = \left(\frac{P(k)}{P^{\text{ref}}(k)}\right)^{\nicefrac{1}{2}}, \qquad T_{B_\text{equi}}(k) = \left(\frac{B_\text{equi}(k)}{B_\text{equi}^{\text{ref}}(k)}\right)^{\nicefrac{1}{3}}\;.
\end{equation}
\begin{figure*}[t]
\centering
  \noindent
   \resizebox{1\textwidth}{!}{
    \includegraphics{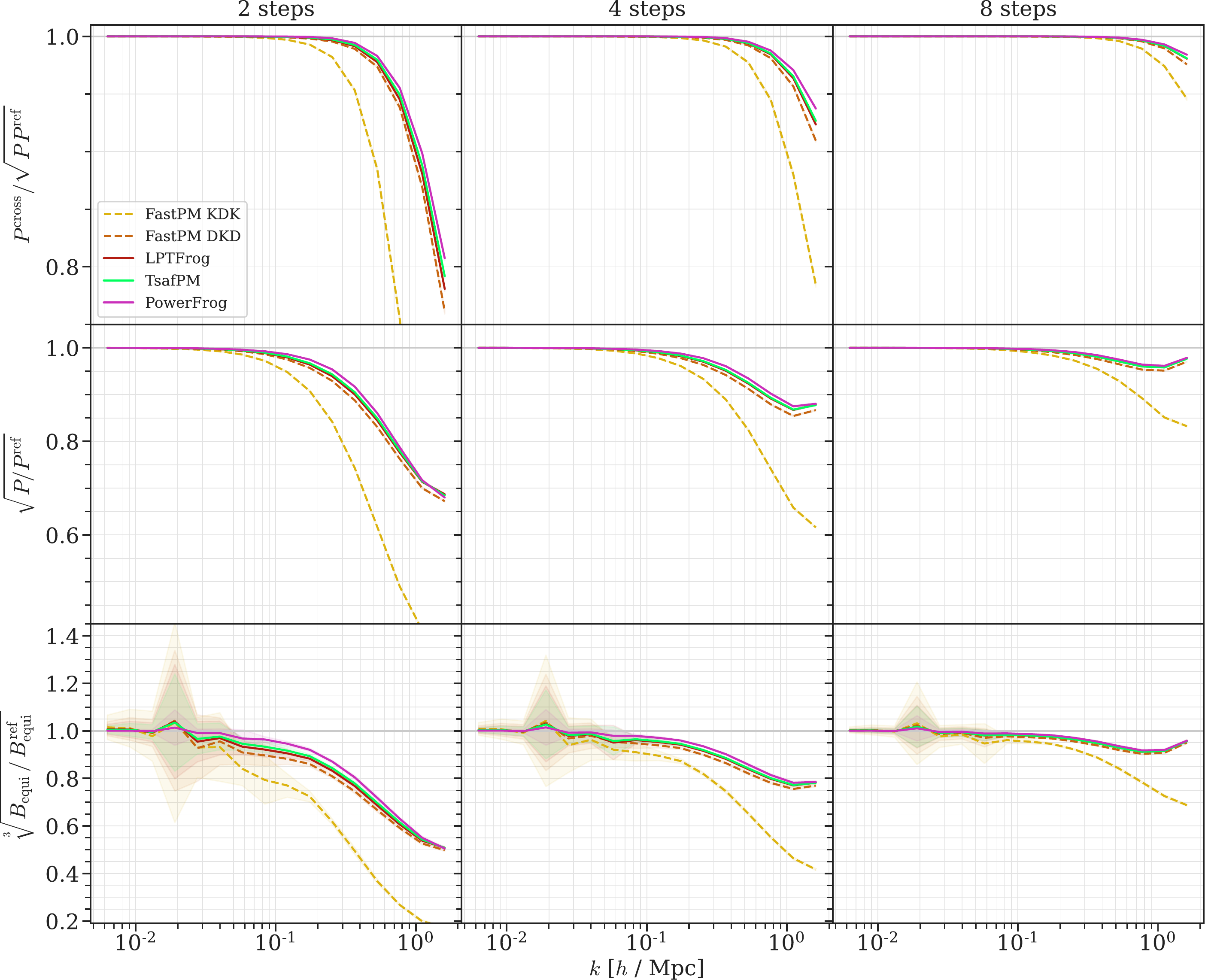}
    }
    \caption{\textsc{Quijote} simulations: normalised cross-power spectra (top), and transfer functions of the power spectrum (middle) and equilateral bispectrum (bottom) between the fast simulations and our reference simulation that used 128 timesteps, for different integrators at $z = 0$. The number of timesteps increases from left to right (uniformly spaced in $D$). Shaded regions indicate the standard deviation computed over 20 realisations. The statistics for the standard Symplectic~2 integrator are significantly worse and are not shown here, which allows us to consider much smaller ranges on the $y$-axis for an easier comparison of the different $\mPi$-integrators.}
    \label{fig:XPk_Pk_Bk_quijote}
\end{figure*}

\begin{figure*}[t]
\centering
  \noindent
   \resizebox{1\textwidth}{!}{
    \includegraphics{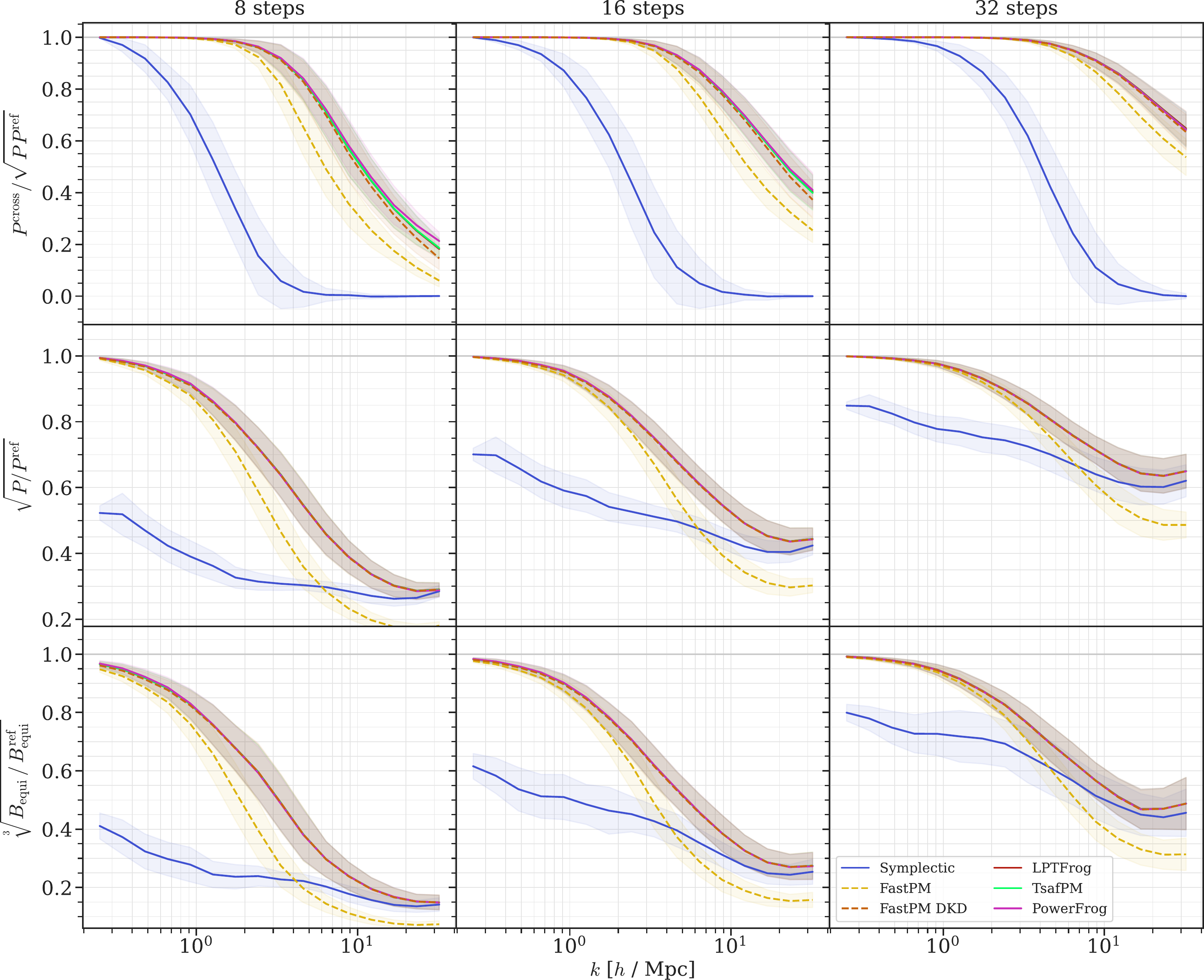}
    }
    \caption{\textsc{Camels} simulations: normalised cross-power spectra (top), and transfer functions of the power spectrum (middle) and equilateral bispectrum (bottom) between the fast simulations and our reference simulation that used 256 timesteps for different integrators at $z = 0$. The number of timesteps increases from left to right (uniformly spaced in $D$). Shaded regions indicate the standard deviation computed over the 27 realisations in the `CV' simulation suite.}
    \label{fig:XPk_Pk_Bk_camels}
\end{figure*}

Figure~\ref{fig:XPk_Pk_Bk_quijote} shows a comparison of the normalised density cross-power spectrum and the transfer functions of the power spectrum and equilateral bispectrum, for 2, 4, and 8 timesteps between $z_{\text{ini}} = 127$ and $z_\text{end} = 0$. For all considered timesteps, our new integrators produce the most accurate results in terms of all three summary statistics. \textsc{PowerFrog} achieves the best results, followed by \textsc{TsafPM} and \textsc{LPTFrog}, whose performance is very similar in our experiment.
The DKD variant of \textsc{FastPM} is far superior to its KDK counterpart for $\leq 8$ steps in terms of all statistics, but the statistics with both variants drop somewhat earlier than with our new integrators. 

For instance, the mean cross power spectra on a scale of $k = 1.1 \ h \ \text{Mpc}^{-1}$ are 57.8\% $/$ 88.0\% $/$  97.4\% (\textsc{FastPM KDK}), 86.9\% $/$ 95.7\% $/$  99.0\% (\textsc{FastPM DKD}), 88.1\% $/$ 96.4\% $/$ 99.2\% (\textsc{LPTFrog}), 88.8\% $/$ 96.5\% $/$ 99.2\% (\textsc{TsafPM}), and 89.8\% $/$ 97.1\% $/$ 99.3\% (\textsc{PowerFrog}) for
2 $/$ 4 $/$ 8 timesteps. The values of the mean transfer function of the power spectrum for $k = 1.1 \ h \ \text{Mpc}^{-1}$ are 39.6\% $/$ 65.8\% $/$ 85.1\% (\textsc{FastPM KDK}), 70.0\% $/$ 85.4\% $/$ 95.1\% (\textsc{FastPM DKD}), 71.4\% $/$ 86.7\% $/$ 95.8\% (\textsc{LPTFrog}), 71.5\% $/$ 86.8\% $/$ 95.8\% (\textsc{TsafPM}), and 71.7\% $/$ 87.5\% $/$ 96.1\% (\textsc{PowerFrog}). All $\mPi$-integrators in DKD form achieve an error $\leq 5\%$ in terms of cross-power spectrum and power spectrum transfer function down to the Nyquist scale $k_\text{Nyquist} = \pi N^{\nicefrac{1}{3}} / L = 1.6 \ h \ \text{Mpc}^{-1}$ with only 8 timesteps. (For clarity, we reiterate that the reference solution has also been computed with the PM method here, for which reason all deviations of the statistics from the reference are solely due to the small number of timesteps and not due to a potential lack of force resolution on small scales.)
 
This is in stark contrast to the standard second-order symplectic integrator (not shown in the plot), whose convergence is very slow: for 8 steps, the normalised cross-power spectrum still lies below $1\%$ on scales $k \geq 1.1 \ h \ \text{Mpc}^{-1}$, implying that the solution is entirely uncorrelated with the reference on these scales. Also, the transfer functions of the power spectrum and bispectrum averaged over the 20 realisations are $< 50\%$ on all scales.
The results of this experiment highlight that our new $\mPi$-integrators produce excellent results in realistic 3D PM simulations with $\Lambda$CDM cosmology.

\subsection{\textsc{Camels} simulation suite: $(25 \ h^{-1} \ \mathrm{Mpc})^3$ box}
Next, we compare the different integrators for the scenario of a small simulation box where the aim is to resolve non-linear scales $k \approx (1 - 30) \ h \ \text{Mpc}^{-1}$. For this purpose, we consider the gravity-only `CV set' of the \textsc{Camels} simulation suite \cite{Villaescusa-Navarro2020} (where `CV' stands for cosmic variance). This set consists of 27 simulations with identical cosmology ($\Omega_{\mathrm{m}} = 0.3$, $\Omega_b = 0.049$, $\Omega_\Lambda = 0.7$, $h = 0.6711$, $n_s = 0.9624$, $\sigma_8 = 0.8$, massless neutrinos) and different random seeds for the initial conditions. The \textsc{Camels} simulations were run with $256^3$ particles in an $L^3 = (25 \ h^{-1} \ \text{Mpc})^3$ box from $z_\text{ini} = 127$ to $z_\text{end} = 0$. Just as we did for the \textsc{Quijote} simulations, we evolve the \textsc{Camels} initial conditions with our PM code (using an $N_c = 512^3$ PM mesh) and compare our solutions at $z = 0$ to a reference run with 256 \textsc{FastPM KDK} timesteps.

Figure~\ref{fig:XPk_Pk_Bk_camels} shows the same statistics that we analysed for the \textsc{Quijote} simulations, namely the normalised cross-power spectrum and the transfer functions of the power spectrum and the equilateral bispectrum. Since accurately reproducing the statistics on the much smaller (and therefore more non-linear) scales in the \textsc{Camels} simulations requires more timesteps, we now consider 8, 16, and 32 timesteps between $z = 127$ and $z = 0$. 

Evidently, the LPT-inspired $\mPi$-integrators outperform the standard second-order symplectic leapfrog integrator again w.r.t.\ all considered summary statistics. For 32 timesteps, the normalised cross-power spectrum for the DKD $\mPi$-integrators is $> 90\%$ on scales $k \approx 9 \ h \ \text{Mpc}^{-1}$, whereas the standard symplectic integrator has a normalised cross-power of $11\%$ on that scale, showing that LPT-inspired integrators achieve significant improvements also on highly non-linear scales (although significantly more steps would be necessary to reach the same accuracy as in the \textsc{Quijote} experiment). On large and intermediate scales, the statistics with only 8 ${\mPi}$-integrator steps are superior to those with 32 standard symplectic steps.

Our new integrators \textsc{PowerFrog}, \textsc{TsafPM}, and \textsc{LPTFrog} slightly improve the cross-spectrum over both variants of \textsc{FastPM}. The \textsc{FastPM} KDK integrator again produces the worst results in terms of all statistics among the LPT-inspired integrators, whereas the DKD variant achieves a very similar performance to \textsc{TsafPM} and \textsc{LPTFrog} in terms of the power spectrum and equilateral bispectrum.
This suggests that -- as long as Zel'dovich consistency is satisfied -- the exact choice of the integrator now plays a smaller role than in the case of the larger simulation box in the \textsc{Quijote} experiment, as the non-linear scales in the \textsc{Camels} box are no longer well described by higher-order LPT corrections.

\subsection{Convergence study}
\label{sec:3D_convergence_study}
\begin{figure*}[htb]
\centering
  \noindent
   \resizebox{.74\textwidth}{!}{
    \includegraphics{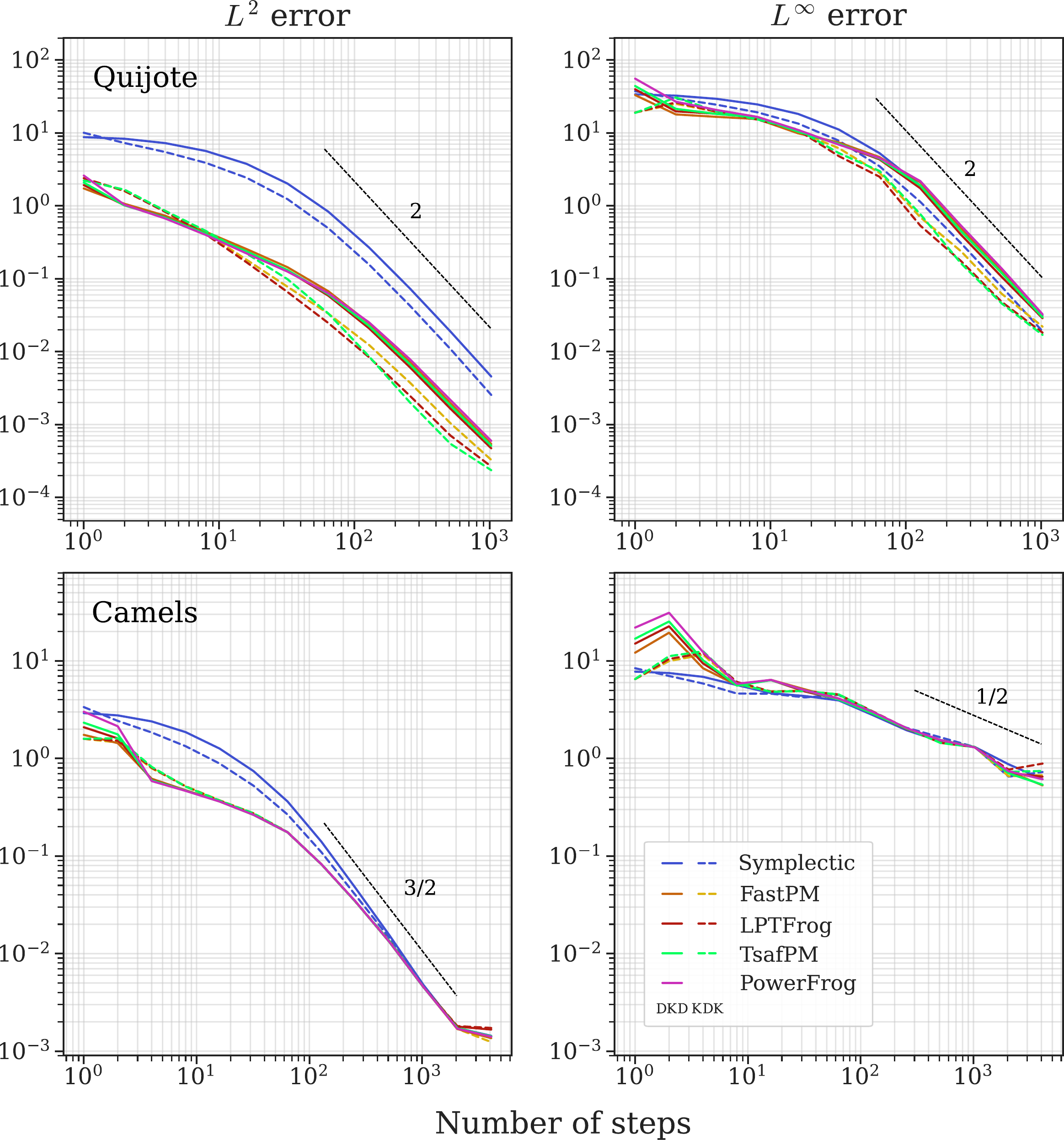}
    }
    \caption{Displacement errors (in $h^{-1} \, \text{Mpc}$) for the \textsc{Quijote} (top) and \textsc{Camels} (bottom) experiments with different integrators as a function of the number of timesteps, w.r.t.\ the $L^2$ norm (left) and $L^\infty$ norm (right). Solid (dashed) lines correspond to the DKD (KDK) variant of the integrator. For \textsc{PowerFrog}, we only consider the DKD case here. The indicated convergence orders are for orientation only and do not necessarily agree with the theoretically expected convergence orders.}
    \label{fig:3D_convergence}
\end{figure*}

Finally, we study the convergence of the different integrators for the realistic 3D $\Lambda$CDM case, both for the large \textsc{Quijote} and the small \textsc{Camels} simulation box. For this experiment, we only consider a single initial white noise field (realisation `10000' for \textsc{Quijote} and `CV0' for \textsc{Camels}), not the entire sets of 20 (\textsc{Quijote}) and 27 (\textsc{Camels}) different ones as in the previous sections. Again, to isolate the effect of the timestepping from other numerical differences, we perform reference simulations with our own PM code -- the same that we also use for all the other simulations. However, since we will now study the field-level convergence of the displacement (at $z = 0$), rather than considering summary statistics as above, we significantly increase the number of timesteps for the reference simulations, to 8,192 (\textsc{Quijote}) and 32,768 (\textsc{Camels}).

The results for both cases are shown in Fig.~\ref{fig:3D_convergence}, for the standard symplectic second-order leapfrog integrator and four $\mPi$-integrators. For the \textsc{Quijote} box, both the $L^\infty$ and the $L^2$ errors start decreasing rather slowly as the number of timesteps increases to $\sim 100$ steps, but then drop off more steeply, exhibiting a numerical convergence order of $\sim 2$ in the $\sim$ 1,000 step regime. Although it is typical for particle trajectories to have crossed by $z = 0$, the reduction of the convergence order observed for the planar-wave collapse in the post-shell-crossing regime does not occur here. We postulate that this might be due to the fact that approximating the true continuous system with $N = 512^3$ particles in a PM-only $N$-body simulation produces a spuriously smoothed displacement field. Since our reference solution is converged in terms of the time discretisation, however not in terms of the spatial discretisation which is the same as for the other simulations, the time convergence of the discrete PM-based solution towards that reference need not be the same as for the underlying continuous system.
Note that probing the time convergence of our PM simulations towards a reference with higher spatial resolution and/or more accurate force computation would be futile, as a residual would remain regardless the number of timesteps, and thus convergence would not be achieved.
Instead, an in-depth time-convergence study with high spatial resolution and very accurate forces (e.g.\ using the \textsc{Abacus} code, \cite{garrison2021abacus}) would be an interesting follow-up of this work.

In order to gather additional evidence for the hypothesis that the lack of spatial convergence affects the convergence order, we revisit the 1D example in \ref{sec:1D_exact_vs_pm} with a much lower spatial resolution, and for both PM and exact forces. In that case, we find that when the reference simulation uses a low resolution and PM forces, the convergence order is in fact limited to $2$, rather than $\nicefrac{3}{2}$ as expected from theory (see Proposition~\ref{prop:convergence_loss}) and observed with sufficiently high spatial resolution (see Fig.~\ref{fig:convergence_1D_post_sc_force_eval}).

Interestingly, the KDK version of \textsc{FastPM} achieves a smaller error for $\gtrsim 10$ steps than its DKD counterpart; therefore, we also show the results with the KDK variants of `Symplectic~2`, \textsc{LPTFrog} and \textsc{TsafPM}. Indeed, in this experiment, the KDK version of the $\mPi$-integrators starts off worse, but then overtakes the respective DKD version as the number of timesteps grows.

In the case of the small \textsc{Camels} box, there is no significant difference between the DKD and KDK variants for $\gtrsim 10$ steps and, as already seen above, the choice of the integrator plays a smaller role due to the fact that the scales resolved in the simulation are no longer well described by LPT at late times. Moreover, since the crossing of particles occurs earlier and much more frequently than for the large \textsc{Quijote} box, the displacement field can be expected to be less regular. This manifests itself in the order of convergence, which is now limited to $\sim \nicefrac{3}{2}$ ($L^2$) and $\sim \nicefrac{1}{2}$ ($L^\infty$).
It would be interesting to study the impact of the box size and resolution on the order of convergence in further detail; however, this is beyond the scope of this paper.

\section{Summary}
\label{sec:conclusions}
In this work, we presented a detailed study of numerical integrators for large-scale simulations of the universe. In the first part, we considered general leapfrog/Verlet DKD integrators. Our main conclusions from this part are:

\begin{itemize}
    \item \emph{All} `canonical' DKD integrators (as defined in Definition~\ref{def:canonical_dkd_integrator}) are symplectic, no matter the time variable used.

    \item We derived conditions for the factors arising in the drift and kick operations in order for canonical DKD integrators to be globally second-order accurate, see Eqs.~\eqref{eq:general_dkd_consistency_conditions}. Also, the proof of Proposition~\ref{prop:convergence} revealed that the time dependence of the cosmological $N$-body Hamiltonian leads to (globally) $3^{\text{rd}}$-order error terms, which add to the `usual' $3^{\text{rd}}$-order leapfrog approximation error.

    \item We demonstrated analytically that using higher-order integrators is futile in the post-shell-crossing regime for perfectly cold flows (see Proposition~\ref{prop:convergence_loss}): due to insufficient regularity of the acceleration field, the order of convergence is limited to $\nicefrac{3}{2}$ for single-axis collapse. This behaviour is confirmed by our numerical experiments in one dimension.    
\end{itemize}

In the second part of this work, we turned towards LPT-inspired integrators. Here, our main contributions are:

\begin{itemize}
    \item We introduced a class of integrators, which we named $\mPi$-integrators to emphasise that they use a modified momentum variable $\vecb{\mPi} = \dd_D \vecb{X}$, i.e.\ the momentum w.r.t.\ growth factor time $D$. We then defined the concept of \emph{Zel'dovich consistency} of numerical integrators (see Definition~\ref{def:zeldovich_consistency}), by which we mean that for one-dimensional initial data prior to shell-crossing particles on the exact Zel'dovich solution trajectory remain on this trajectory. We constructed $\mPi$-integrators in such a way that this is automatically satisfied for the drift part (i.e.\ the position update) of the integration step and endowed the kick operation (i.e.\ the momentum update) with two free coefficient functions $p(\Delta D, D_n)$ and $q(\Delta D, D_n)$, which may depend on the timestep $\Delta D$ and the growth factor $D_n$ at the beginning of the timestep. We then derived a simple criterion that $p$ and $q$ need to satisfy in order for the $\mPi$-integrator to be Zel'dovich consistent, see Proposition~\ref{prop:characterisation_of_zeldovich_consistency}.

    \item We briefly discussed the \textsc{FastPM} scheme by Ref.~\cite{Feng:2016} in this context and showed that it is the unique $\mPi$-integrator that is both Zel'dovich consistent and symplectic (see Proposition~\ref{prop:fastpm}).

    \item We then proceeded to construct new $\mPi$-integrators, starting with the \textsc{LPTFrog} integrator (see Example~\ref{example:lptfrog}), which we derived by assuming that particles move on a quadratic trajectory w.r.t.\ the growth factor $D$, just as they do in 2LPT. Perhaps surprisingly, the coefficient function $p(\Delta D, D_n)$, which determines the contribution of the current momenta to the momenta in the next integration step, can become \emph{negative} when the timestep $\Delta D$ is large: this is necessary in order to adequately account for the expansion of the universe.
    The adaptation of some of our integrators to the KDK case is given in \ref{sec:KDK}.

    \item For EdS cosmology, we showed that \textsc{FastPM} and \textsc{LPTFrog} are special cases of a one-parameter family of `$\epsilon$-integrators', see Example~\ref{example:epsilon_integrator}.

    \item Whereas \textsc{FastPM} has the same coefficient function $p(\Delta D, D_n)$ as the standard symplectic DKD integrator and $q(\Delta D, D_n)$ is modified such that \textsc{FastPM} becomes Zel'dovich consistent, we proposed a new integrator \textsc{TsafPM} derived from the \emph{reverse} procedure (see Example~\ref{example:tsafpm}): starting from the $q(\Delta D, D_n)$ of the standard symplectic DKD stepper, we defined $p(\Delta D, D_n)$ such that \textsc{TsafPM} becomes Zel'dovich consistent. In all our experiments, \textsc{TsafPM} had an edge over \textsc{FastPM}.

    \item Finally, we introduced our most powerful integrator, \textsc{PowerFrog} (Example~\ref{example:powerfrog}). To construct this $\mPi$-integrator, we matched the asymptote of its coefficient function $p(\Delta D, D_n)$ in the limit $D_n \to 0$ to 2LPT, see Eq.~\eqref{eq:p_q_2LPT_limiting_behaviour}.

    \item In our numerical experiments in 1D, we demonstrated that Zel'dovich-consistent integrators indeed reproduce the analytic solution prior to shell-crossing (up to round-off errors). After shell-crossing when the Zel'dovich approximation no longer describes the dynamics of the system, the errors of $\mPi$-integrators tend to be similar to those of the standard symplectic DKD integrator. This suggests that informing integrators with LPT is not detrimental to their post-shell-crossing performance, and one might in fact use the symplectic \textsc{FastPM} scheme as the `default' integrator in cosmological simulations, although further analyses (e.g.\ of bound structures at low redshift) would be necessary. Importantly, we observed convergence order $\nicefrac{3}{2}$ with \emph{all} integrators due to the loss of regularity of the displacement field, as predicted by theory (however, cf.\ \ref{sec:1D_exact_vs_pm} for the same experiment with a low spatial resolution). Also, we studied the errors of different integrators when the initial conditions contain a decaying-mode component (which is not incorporated in $\mPi$-integrators). In that case, Zel'dovich-consistent $\mPi$-integrators no longer yield the exact solution; rather, all integrators converge at their nominal order (in particular, the DKD-based $\mPi$-integrators convergence quadratically) before shell-crossing. Lastly, we analysed the cosmic energy balance, where we found that non-symplectic methods such as our new $\mPi$-integrators as well as the standard $3^{\text{rd}}$-order Runge--Kutta method do not yield systematically larger energy errors than symplectic methods in fast approximate simulations (we used 32 timesteps). Instead, the sampling of the timesteps turned out to be crucial in order for the energy error to remain small.

    \item Then, we performed a single (very large) timestep with different integrators in 2D, where the pre-shell-crossing solution is given by the infinite series of LPT terms. Whereas in 1D Zel'dovich-consistent integrators are exact prior to shell-crossing and they all behaved similarly in the post-shell-crossing regime in our experiments, the 2D experiment revealed the differences between the different integrators at the field level: the original \textsc{FastPM} KDK scheme exhibited the smallest error towards 1LPT (i.e.\ the Zel'dovich approximation), but barely captured any information provided by LPT orders $\geq 2$. The displacement fields resulting from the DKD variant of \textsc{FastPM} and, to a larger degree, our new $\mPi$-integrators lead to solutions closer to 2LPT than to 1LPT. Crucially, the \textsc{PowerFrog} integrator produced a displacement field with a $> 2.5 \times$ (and $> 5 \times$) smaller mean error towards the 2LPT (and 3LPT) solution in comparison to \textsc{FastPM} DKD. The fact that the \textsc{PowerFrog} solution is so similar to the 3LPT solution suggests that it could be attractive to initialise cosmological simulations simply with a single \textsc{PowerFrog} step starting from $a = 0$ rather than with 2LPT, thus making the LPT-based initialisation of $N$-body simulations unneccesary. We leave an investigation in this direction to future work.

    \item Finally, we applied our new integrators to realistic scenarios in 3D with $\Lambda$CDM cosmology. We started from the initial conditions of the \textsc{Quijote} and \textsc{Camels} simulation suites and evolved the particles down to redshift $z = 0$. In the case of the large $(1 \ h^{-1} \ \text{Gpc})^3$ \textsc{Quijote} simulation box, our new $\mPi$-integrators \textsc{LPTFrog}, \textsc{TsafPM}, and \textsc{PowerFrog} achieved the best results in terms of (cross-)power spectrum and equilateral bispectrum,  outperforming \textsc{FastPM} (both the DKD and the KDK variant) for small numbers of timesteps. For the smaller $(25 \ h^{-1} \ \text{Mpc})^3$ \textsc{Camels} box, the differences between the $\mPi$-integrators were small; however, our integrators produced the best cross-power spectrum also in that case. Importantly, in the regime of $\mathcal{O}(10-100)$ steps, all $\mPi$-integrators reproduced the statistics of the density field significantly more accurately than the standard leapfrog integrator by virtue of their Zel'dovich consistency.     
    
    \item In a first exploratory convergence study for the realistic 3D case, we found that the expected reduction of the convergence order due to the lacking regularity of the displacement field does occur for the small \textsc{Camels} box. However, analysing the intrinsic singularities of the underlying \emph{continuous} fields is numerically challenging, as the spatial resolution and force accuracy can affect the convergence (see \ref{sec:1D_exact_vs_pm}). In addition, we did not employ gravitational smoothing, the effect of which on the convergence would also merit further investigation.  
\end{itemize}

Our results highlight that in order to construct numerical integrators for cosmological simulations that are as effective as possible \emph{in practice}, it is not sufficient to ensure the correct behaviour in the limit of vanishing timesteps $\Delta \tau \to 0$, but it is also crucial to optimise the integrators at fixed $\Delta \tau > 0$ by incorporating knowledge of the true solution, which in the pre-shell-crossing regime is conveniently provided by the LPT series. We will present further numerical results with our integrators in a follow-up publication. 


\section*{Data Availability}
\noindent All simulation data presented in this article can be made available upon reasonable request to the authors.

\section*{Acknowledgments}
\noindent The authors thank Cornelius Rampf, Romain Teyssier, Raul Angulo, and Jens Stücker for stimulating discussions. Also, we thank Ewoud Wempe for finding a bug in an earlier \textsc{PowerFrog} implementation. We thank the organisers of the MIAPbP workshop `Advances in cosmology through numerical simulations' held from 9~May to 3~June 2022 at the `ORIGINS' Excellence Cluster in Garching, Germany, where this project was initiated. 
OH acknowledges funding from the European Research Council (ERC) under the European Union’s Horizon 2020 research and innovation programme (grant agreement No. 679145, project `COSMO-SIMS') for part of this work. 

This work made use of the following open-source software: {\sc Matplotlib} \citep{Hunter:2007}, {\sc Seaborn} \citep{Waskom2021}, {\sc NumPy} \citep{Harris:2020}, {\sc SciPy} \citep{Virtanen:2020}, {\sc Jax} \citep{jax2018github}, {\sc Pylians} \citep{Pylians}, {\sc Jax-Cosmo}\footnote{\href{https://github.com/DifferentiableUniverseInitiative/jax_cosmo}{https://github.com/DifferentiableUniverseInitiative/jax\_cosmo}}, {\sc cmcrameri}\footnote{\href{https://pypi.org/project/cmcrameri/}{https://pypi.org/project/cmcrameri/}}. Also, we heavily used the arXiv preprint repository.

\bibliographystyle{elsarticle-num}

\begin{thebibliography}{100}
\expandafter\ifx\csname url\endcsname\relax
  \def\url#1{\texttt{#1}}\fi
\expandafter\ifx\csname urlprefix\endcsname\relax\def\urlprefix{URL }\fi
\expandafter\ifx\csname href\endcsname\relax
  \def\href#1#2{#2} \def\path#1{#1}\fi

\bibitem{Feng:2016}
Y.~{Feng}, M.-Y. {Chu}, U.~{Seljak}, P.~{McDonald}, {FastPM: A new scheme for
  fast simulations of dark matter and haloes}, \mnras 463~(3) (2016)
  2273--2286.
\newblock \href {http://arxiv.org/abs/1603.00476} {\path{arXiv:1603.00476}},
  \href {http://dx.doi.org/10.1093/mnras/stw2123}
  {\path{doi:10.1093/mnras/stw2123}}.

\bibitem{AnguloHahn:2022}
R.~E. {Angulo}, O.~{Hahn}, {Large-scale dark matter simulations}, Living
  Reviews in Computational Astrophysics 8~(1) (2022) 1.
\newblock \href {http://arxiv.org/abs/2112.05165} {\path{arXiv:2112.05165}},
  \href {http://dx.doi.org/10.1007/s41115-021-00013-z}
  {\path{doi:10.1007/s41115-021-00013-z}}.

\bibitem{Peebles:1980}
P.~Peebles, The Large-scale Structure of the Universe, Princeton Series in
  Physics, Princeton University Press, 1980.

\bibitem{Quinn:1997}
T.~{Quinn}, N.~{Katz}, J.~{Stadel}, G.~{Lake}, {Time stepping N-body
  simulations}, \preprint\href {http://arxiv.org/abs/astro-ph/9710043}
  {\path{arXiv:astro-ph/9710043}}.

\bibitem{Springel:2005}
V.~{Springel}, {The cosmological simulation code GADGET-2}, \mnras 364~(4)
  (2005) 1105--1134.
\newblock \href {http://arxiv.org/abs/astro-ph/0505010}
  {\path{arXiv:astro-ph/0505010}}, \href
  {http://dx.doi.org/10.1111/j.1365-2966.2005.09655.x}
  {\path{doi:10.1111/j.1365-2966.2005.09655.x}}.

\bibitem{Euclid}
R.~Laureijs, J.~Amiaux, S.~Arduini, J.-L. Augueres, J.~Brinchmann, R.~Cole,
  M.~Cropper, C.~Dabin, L.~Duvet, A.~Ealet, et~al., Euclid definition study
  report, \preprint\href {http://arxiv.org/abs/1110.3193}
  {\path{arXiv:1110.3193}}.

\bibitem{LSST}
{\v{Z}}.~Ivezi{\'c}, S.~M. Kahn, J.~A. Tyson, B.~Abel, E.~Acosta, R.~Allsman,
  D.~Alonso, Y.~AlSayyad, S.~F. Anderson, J.~Andrew, et~al., {LSST: from
  science drivers to reference design and anticipated data products}, \apj
  873~(2) (2019) 111.
\newblock \href {http://arxiv.org/abs/0805.2366} {\path{arXiv:0805.2366}}.

\bibitem{WFIRST}
J.~Green, P.~Schechter, C.~Baltay, R.~Bean, D.~Bennett, R.~Brown, C.~Conselice,
  M.~Donahue, X.~Fan, B.~Gaudi, et~al., {Wide-field infrared survey telescope
  (WFIRST) final report}, \preprint\href {http://arxiv.org/abs/1208.4012}
  {\path{arXiv:1208.4012}}.

\bibitem{potter2017pkdgrav3}
D.~Potter, J.~Stadel, R.~Teyssier, {PKDGRAV3: beyond trillion particle
  cosmological simulations for the next era of galaxy surveys}, Computational
  Astrophysics and Cosmology 4~(1) (2017) 2.
\newblock \href {http://arxiv.org/abs/1609.08621} {\path{arXiv:1609.08621}},
  \href {http://dx.doi.org/10.1186/s40668-017-0021-1}
  {\path{doi:10.1186/s40668-017-0021-1}}.

\bibitem{AbacusSummit}
N.~A. Maksimova, L.~H. Garrison, D.~J. Eisenstein, B.~Hadzhiyska, S.~Bose,
  T.~P. Satterthwaite, {AbacusSummit: a massive set of high-accuracy,
  high-resolution N-body simulations}, \mnras 508~(3) (2021) 4017--4037.
\newblock \href {http://arxiv.org/abs/2110.11398} {\path{arXiv:2110.11398}},
  \href {http://dx.doi.org/10.1093/mnras/stab2484}
  {\path{doi:10.1093/mnras/stab2484}}.

\bibitem{OuterRim1}
S.~Habib, A.~Pope, H.~Finkel, N.~Frontiere, K.~Heitmann, D.~Daniel, P.~Fasel,
  V.~Morozov, G.~Zagaris, T.~Peterka, et~al., {HACC: Simulating sky surveys on
  state-of-the-art supercomputing architectures}, New Astronomy 42 (2016)
  49--65.
\newblock \href {http://arxiv.org/abs/1410.2805} {\path{arXiv:1410.2805}}.

\bibitem{OuterRim2}
K.~Heitmann, T.~D. Uram, H.~Finkel, N.~Frontiere, S.~Habib, A.~Pope, E.~Rangel,
  J.~Hollowed, D.~Korytov, P.~Larsen, et~al., {HACC Cosmological Simulations:
  First Data Release}, \apjs 244~(1) (2019) 17.
\newblock \href {http://arxiv.org/abs/1904.11966} {\path{arXiv:1904.11966}}.

\bibitem{TianNu}
J.~D. Emberson, H.-R. Yu, D.~Inman, T.-J. Zhang, U.-L. Pen,
  J.~Harnois-D{\'{e}}raps, S.~Yuan, H.-Y. Teng, H.-M. Zhu, X.~Chen, Z.-Z. Xing,
  {Cosmological neutrino simulations at extreme scale}, Research in Astronomy
  and Astrophysics 17~(8) (2017) 085.
\newblock \href {http://arxiv.org/abs/1311.6130} {\path{arXiv:1311.6130}},
  \href {http://dx.doi.org/10.1088/1674-4527/17/8/85}
  {\path{doi:10.1088/1674-4527/17/8/85}}.

\bibitem{Uchuu}
T.~Ishiyama, F.~Prada, A.~A. Klypin, M.~Sinha, R.~B. Metcalf, E.~Jullo,
  B.~Altieri, S.~A. Cora, D.~Croton, S.~{De La Torre}, D.~E.
  Mill{\'{a}}n-Calero, T.~Oogi, J.~Ruedas, C.~A. Vega-Mart{\'{i}}nez, {The
  Uchuu simulations: Data Release 1 and dark matter halo concentrations},
  \mnras 506~(3) (2021) 4210--4231.
\newblock \href {http://arxiv.org/abs/2007.14720} {\path{arXiv:2007.14720}},
  \href {http://dx.doi.org/10.1093/mnras/stab1755}
  {\path{doi:10.1093/mnras/stab1755}}.

\bibitem{Bagla:2002}
J.~S. {Bagla}, {TreePM: A Code for Cosmological N-Body Simulations}, Journal of
  Astrophysics and Astronomy 23 (2002) 185--196.
\newblock \href {http://arxiv.org/abs/astro-ph/9911025}
  {\path{arXiv:astro-ph/9911025}}, \href {http://dx.doi.org/10.1007/BF02702282}
  {\path{doi:10.1007/BF02702282}}.

\bibitem{Barnes:1986}
J.~{Barnes}, P.~{Hut}, {A hierarchical O(N log N) force-calculation algorithm},
  \nat 324~(6096) (1986) 446--449.
\newblock \href {http://dx.doi.org/10.1038/324446a0}
  {\path{doi:10.1038/324446a0}}.

\bibitem{Appel:1985}
A.~W. {Appel}, {An Efficient Program for Many-Body Simulation}, SIAM Journal on
  Scientific and Statistical Computing 6~(1) (1985) 85--103.

\bibitem{Darden:1993}
T.~{Darden}, D.~{York}, L.~{Pedersen}, {Particle mesh Ewald: An N~log(N) method
  for Ewald sums in large systems}, \jcp 98~(12) (1993) 10089--10092.
\newblock \href {http://dx.doi.org/10.1063/1.464397}
  {\path{doi:10.1063/1.464397}}.

\bibitem{Greengard:1987}
L.~{Greengard}, V.~{Rokhlin}, {A Fast Algorithm for Particle Simulations}, \jcp
  73~(2) (1987) 325--348.
\newblock \href {http://dx.doi.org/10.1016/0021-9991(87)90140-9}
  {\path{doi:10.1016/0021-9991(87)90140-9}}.

\bibitem{Hockney:1988}
R.~Hockney, J.~Eastwood, Computer Simulation Using Particles, Taylor \&
  Francis, 1988.

\bibitem{Hahn:2013}
O.~{Hahn}, T.~{Abel}, R.~{Kaehler}, {A new approach to simulating collisionless
  dark matter fluids}, \mnras 434~(2) (2013) 1171--1191.
\newblock \href {http://arxiv.org/abs/1210.6652} {\path{arXiv:1210.6652}},
  \href {http://dx.doi.org/10.1093/mnras/stt1061}
  {\path{doi:10.1093/mnras/stt1061}}.

\bibitem{Michaux:2021}
M.~{Michaux}, O.~{Hahn}, C.~{Rampf}, R.~E. {Angulo}, {Accurate initial
  conditions for cosmological N-body simulations: minimizing truncation and
  discreteness errors}, \mnras 500~(1) (2021) 663--683.
\newblock \href {http://arxiv.org/abs/2008.09588} {\path{arXiv:2008.09588}},
  \href {http://dx.doi.org/10.1093/mnras/staa3149}
  {\path{doi:10.1093/mnras/staa3149}}.

\bibitem{Colombi:2021}
S.~{Colombi}, {Phase-space structure of protohalos: Vlasov versus
  particle-mesh}, \aap 647 (2021) A66.
\newblock \href {http://arxiv.org/abs/2012.04409} {\path{arXiv:2012.04409}},
  \href {http://dx.doi.org/10.1051/0004-6361/202039719}
  {\path{doi:10.1051/0004-6361/202039719}}.

\bibitem{Hayli:1974}
A.~{Hayli}, {The method of the doubly individual step for N-body
  computations.}, in: D.~Bettis (Ed.), Proceedings of the Conference on the
  Numerical Solution of Ordinary Differential Equations. Lecture Notes in
  Mathematics, Springer, Berlin, Heidelberg, 1974, pp. 304--312.
\newblock \href {http://dx.doi.org/10.1007/BFb0066598}
  {\path{doi:10.1007/BFb0066598}}.

\bibitem{McMillan:1986}
S.~L.~W. {McMillan}, {The Vectorization of Small-N Integrators}, in: P.~{Hut},
  S.~L.~W. {McMillan} (Eds.), The Use of Supercomputers in Stellar Dynamics,
  Vol. 267, Springer, Berlin, Heidelberg, New York, 1986, p. 156.
\newblock \href {http://dx.doi.org/10.1007/BFb0116406}
  {\path{doi:10.1007/BFb0116406}}.

\bibitem{Zeldovich:1970}
Y.~B. {Zel'dovich}, {Gravitational instability: An approximate theory for large
  density perturbations.}, \aap 5 (1970) 84--89.

\bibitem{Marcos:2006}
B.~{Marcos}, T.~{Baertschiger}, M.~{Joyce}, A.~{Gabrielli}, F.~{Sylos Labini},
  {Linear perturbative theory of the discrete cosmological N-body problem},
  \prd 73~(10) (2006) 103507.
\newblock \href {http://arxiv.org/abs/astro-ph/0601479}
  {\path{arXiv:astro-ph/0601479}}, \href
  {http://dx.doi.org/10.1103/PhysRevD.73.103507}
  {\path{doi:10.1103/PhysRevD.73.103507}}.

\bibitem{Joyce:2009}
M.~{Joyce}, B.~{Marcos}, T.~{Baertschiger}, {Towards quantitative control on
  discreteness error in the non-linear regime of cosmological N-body
  simulations}, \mnras 394~(2) (2009) 751--773.
\newblock \href {http://arxiv.org/abs/0805.1357} {\path{arXiv:0805.1357}},
  \href {http://dx.doi.org/10.1111/j.1365-2966.2008.14290.x}
  {\path{doi:10.1111/j.1365-2966.2008.14290.x}}.

\bibitem{Jasche2019}
J.~Jasche, G.~Lavaux,
  \href{https://www.aanda.org/10.1051/0004-6361/201833710}{{Physical Bayesian
  modelling of the non-linear matter distribution: New insights into the nearby
  universe}}, \aap 625 (2019) A64.
\newblock \href {http://arxiv.org/abs/1806.11117} {\path{arXiv:1806.11117}},
  \href {http://dx.doi.org/10.1051/0004-6361/201833710}
  {\path{doi:10.1051/0004-6361/201833710}}.
\newline\urlprefix\url{https://www.aanda.org/10.1051/0004-6361/201833710}

\bibitem{Cranmer2020}
K.~Cranmer, J.~Brehmer, G.~Louppe,
  \href{https://pnas.org/doi/full/10.1073/pnas.1912789117}{{The frontier of
  simulation-based inference}}, Proceedings of the National Academy of Sciences
  117~(48) (2020) 30055--30062.
\newblock \href {http://arxiv.org/abs/1911.01429} {\path{arXiv:1911.01429}},
  \href {http://dx.doi.org/10.1073/pnas.1912789117}
  {\path{doi:10.1073/pnas.1912789117}}.
\newline\urlprefix\url{https://pnas.org/doi/full/10.1073/pnas.1912789117}

\bibitem{Modi2021}
C.~Modi, F.~Lanusse, U.~Seljak,
  \href{https://linkinghub.elsevier.com/retrieve/pii/S2213133721000597}{{FlowPM:
  Distributed TensorFlow implementation of the FastPM cosmological N-body
  solver}}, Astronomy and Computing 37 (2021) 100505.
\newblock \href {http://arxiv.org/abs/2010.11847} {\path{arXiv:2010.11847}},
  \href {http://dx.doi.org/10.1016/j.ascom.2021.100505}
  {\path{doi:10.1016/j.ascom.2021.100505}}.
\newline\urlprefix\url{https://linkinghub.elsevier.com/retrieve/pii/S2213133721000597}

\bibitem{Li2022}
Y.~Li, C.~Modi, D.~Jamieson, Y.~Zhang, L.~Lu, Y.~Feng, F.~Lanusse,
  L.~Greengard, {Differentiable Cosmological Simulation with Adjoint Method},
  \preprint\href {http://arxiv.org/abs/2211.09815} {\path{arXiv:2211.09815}}.

\bibitem{quijote}
F.~Villaescusa-Navarro, C.~Hahn, E.~Massara, A.~Banerjee, A.~M. Delgado, D.~K.
  Ramanah, T.~Charnock, E.~Giusarma, Y.~Li, E.~Allys, A.~Brochard, C.~Uhlemann,
  C.-T. Chiang, S.~He, A.~Pisani, A.~Obuljen, Y.~Feng, E.~Castorina,
  G.~Contardo, C.~D. Kreisch, A.~Nicola, J.~Alsing, R.~Scoccimarro, L.~Verde,
  M.~Viel, S.~Ho, S.~Mallat, B.~Wandelt, D.~N. Spergel,
  \href{http://dx.doi.org/10.3847/1538-4365/ab9d82}{{The Quijote Simulations}},
  \apjs 250~(1) (2020) 2.
\newblock \href {http://arxiv.org/abs/1909.05273} {\path{arXiv:1909.05273}},
  \href {http://dx.doi.org/10.3847/1538-4365/ab9d82}
  {\path{doi:10.3847/1538-4365/ab9d82}}.
\newline\urlprefix\url{http://dx.doi.org/10.3847/1538-4365/ab9d82}

\bibitem{Villaescusa-Navarro2020}
F.~Villaescusa-Navarro, D.~Angl{\'{e}}s-Alc{\'{a}}zar, S.~Genel, D.~N. Spergel,
  R.~S. Somerville, R.~Dave, A.~Pillepich, L.~Hernquist, D.~Nelson, P.~Torrey,
  D.~Narayanan, Y.~Li, O.~Philcox, V.~{La Torre}, A.~M. Delgado, S.~Ho,
  S.~Hassan, B.~Burkhart, D.~Wadekar, N.~Battaglia, G.~Contardo, {The CAMELS
  project: Cosmology and Astrophysics with MachinE Learning Simulations}, The
  Astrophysical Journal 915~(1) (2021) 71.
\newblock \href {http://arxiv.org/abs/2010.00619} {\path{arXiv:2010.00619}}.

\bibitem{Rampf:2021}
C.~{Rampf}, U.~{Frisch}, O.~{Hahn}, {Unveiling the singular dynamics in the
  cosmic large-scale structure}, \mnras 505~(1) (2021) L90--L94.
\newblock \href {http://arxiv.org/abs/1912.00868} {\path{arXiv:1912.00868}},
  \href {http://dx.doi.org/10.1093/mnrasl/slab053}
  {\path{doi:10.1093/mnrasl/slab053}}.

\bibitem{Arnold:1989}
V.~Arnold, Mathematical methods of classical mechanics, Vol.~60, Springer,
  1989.

\bibitem{Abel:2012}
T.~{Abel}, O.~{Hahn}, R.~{Kaehler}, {Tracing the dark matter sheet in phase
  space}, \mnras 427~(1) (2012) 61--76.
\newblock \href {http://arxiv.org/abs/1111.3944} {\path{arXiv:1111.3944}},
  \href {http://dx.doi.org/10.1111/j.1365-2966.2012.21754.x}
  {\path{doi:10.1111/j.1365-2966.2012.21754.x}}.

\bibitem{Doroshkevich:1973}
A.~G. {Doroshkevich}, V.~S. {Ryaben'kii}, S.~F. {Shandarin}, {Nonlinear theory
  of the development of potential perturbations}, Astrophysics 9~(2) (1973)
  144--153.
\newblock \href {http://dx.doi.org/10.1007/BF01011421}
  {\path{doi:10.1007/BF01011421}}.

\bibitem{Martel:1998}
H.~{Martel}, P.~R. {Shapiro}, {A convenient set of comoving cosmological
  variables and their application}, \mnras 297~(2) (1998) 467--485.
\newblock \href {http://arxiv.org/abs/astro-ph/9710119}
  {\path{arXiv:astro-ph/9710119}}, \href
  {http://dx.doi.org/10.1046/j.1365-8711.1998.01497.x}
  {\path{doi:10.1046/j.1365-8711.1998.01497.x}}.

\bibitem{Buchert:1993}
T.~{Buchert}, J.~{Ehlers}, {Lagrangian theory of gravitational instability of
  Friedman-Lemaitre cosmologies -- second-order approach: an improved model for
  non-linear clustering}, \mnras 264 (1993) 375--387.
\newblock \href {http://dx.doi.org/10.1093/mnras/264.2.375}
  {\path{doi:10.1093/mnras/264.2.375}}.

\bibitem{Bouchet:1995}
F.~R. {Bouchet}, S.~{Colombi}, E.~{Hivon}, R.~{Juszkiewicz}, {Perturbative
  Lagrangian approach to gravitational instability.}, \aap 296 (1995) 575.
\newblock \href {http://arxiv.org/abs/astro-ph/9406013}
  {\path{arXiv:astro-ph/9406013}}.

\bibitem{Rampf:2012}
C.~{Rampf}, {The recursion relation in Lagrangian perturbation theory}, \jcap
  2012~(12) (2012) 004.
\newblock \href {http://arxiv.org/abs/1205.5274} {\path{arXiv:1205.5274}},
  \href {http://dx.doi.org/10.1088/1475-7516/2012/12/004}
  {\path{doi:10.1088/1475-7516/2012/12/004}}.

\bibitem{Rampf:2021a}
C.~{Rampf}, {Cosmological Vlasov-Poisson equations for dark matter}, Reviews of
  Modern Plasma Physics 5~(1) (2021) 10.
\newblock \href {http://dx.doi.org/10.1007/s41614-021-00055-z}
  {\path{doi:10.1007/s41614-021-00055-z}}.

\bibitem{Rampf:2022}
C.~Rampf, S.~O. Schobesberger, O.~Hahn,
  \href{http://arxiv.org/abs/2205.11347}{{Analytical growth functions for
  cosmic structures in a $\Lambda$CDM Universe}}, \mnras 516~(2) (2022)
  2840--2850.
\newblock \href {http://arxiv.org/abs/2205.11347} {\path{arXiv:2205.11347}},
  \href {http://dx.doi.org/10.1093/mnras/stac2406}
  {\path{doi:10.1093/mnras/stac2406}}.
\newline\urlprefix\url{http://arxiv.org/abs/2205.11347}

\bibitem{Chernin:2003}
A.~D. {Chernin}, D.~I. {Nagirner}, S.~V. {Starikova}, {Growth rate of
  cosmological perturbations in standard model: Explicit analytical solution},
  \aap 399 (2003) 19--21.
\newblock \href {http://arxiv.org/abs/astro-ph/0110107}
  {\path{arXiv:astro-ph/0110107}}, \href
  {http://dx.doi.org/10.1051/0004-6361:20021763}
  {\path{doi:10.1051/0004-6361:20021763}}.

\bibitem{Demianski:2005}
M.~{Demianski}, Z.~A. {Golda}, A.~{Woszczyna}, {Evolution of density
  perturbations in a realistic universe}, General Relativity and Gravitation
  37~(12) (2005) 2063--2082.
\newblock \href {http://arxiv.org/abs/gr-qc/0504089}
  {\path{arXiv:gr-qc/0504089}}, \href
  {http://dx.doi.org/10.1007/s10714-005-0180-2}
  {\path{doi:10.1007/s10714-005-0180-2}}.

\bibitem{skeel2001practical}
R.~D. Skeel, D.~J. Hardy, Practical construction of modified hamiltonians, SIAM
  J. Sci. Comput. 23~(4) (2001) 1172--1188.
\newblock \href {http://dx.doi.org/10.1137/S106482750138318X}
  {\path{doi:10.1137/S106482750138318X}}.

\bibitem{hairer2006geometric}
E.~Hairer, M.~Hochbruck, A.~Iserles, C.~Lubich, Geometric numerical
  integration, Oberwolfach Reports 3~(1) (2006) 805--882.
\newblock \href {http://dx.doi.org/10.14760/OWR-2006-14}
  {\path{doi:10.14760/OWR-2006-14}}.

\bibitem{Yoshida:1990}
H.~{Yoshida}, {Construction of higher order symplectic integrators}, Physics
  Letters A 150~(5-7) (1990) 262--268.
\newblock \href {http://dx.doi.org/10.1016/0375-9601(90)90092-3}
  {\path{doi:10.1016/0375-9601(90)90092-3}}.

\bibitem{mclachlan2004nonlinear}
R.~I. McLachlan, M.~Perlmutter, G.~Quispel, On the nonlinear stability of
  symplectic integrators, BIT Numer. Math. 44 (2004) 99--117.
\newblock \href {http://dx.doi.org/10.1023/B:BITN.0000025088.13092.7f}
  {\path{doi:10.1023/B:BITN.0000025088.13092.7f}}.

\bibitem{shang1999kam}
Z.-j. Shang, Kam theorem of symplectic algorithms for hamiltonian systems,
  Numer. Math. 83 (1999) 477--496.
\newblock \href {http://dx.doi.org/10.1007/s002110050460}
  {\path{doi:10.1007/s002110050460}}.

\bibitem{kolmogorov1954conservation}
A.~N. Kolmogorov, On conservation of conditionally periodic motions for a small
  change in hamilton's function, in: Dokl. Akad. Nauk SSSR, Vol.~98, 1954, pp.
  527--530.

\bibitem{moser1962invariant}
J.~Moser, On invariant curves of area-preserving mappings of an annulus, Nachr.
  Akad. Wiss. G{\"o}ttingen, II (1962) 1--20.

\bibitem{Arnold_1963}
V.~I. Arnol'd, \href{https://dx.doi.org/10.1070/RM1963v018n05ABEH004130}{Proof
  of a theorem of a. n. kolmogorov on the invariance of quasi-periodic motions
  under small perturbations of the hamiltonian}, Russ. Math. Surv. 18~(5)
  (1963) 9.
\newblock \href {http://dx.doi.org/10.1070/RM1963v018n05ABEH004130}
  {\path{doi:10.1070/RM1963v018n05ABEH004130}}.
\newline\urlprefix\url{https://dx.doi.org/10.1070/RM1963v018n05ABEH004130}

\bibitem{Magnus1954}
W.~Magnus, {On the exponential solution of differential equations for a linear
  operator}, Communications on Pure and Applied Mathematics 7~(4) (1954)
  649--673.
\newblock \href {http://dx.doi.org/10.1002/cpa.3160070404}
  {\path{doi:10.1002/cpa.3160070404}}.

\bibitem{Blanes2009}
S.~Blanes, F.~Casas, J.~A. Oteo, J.~Ros, {The Magnus expansion and some of its
  applications}, Physics Reports 470~(5-6) (2009) 151--238.
\newblock \href {http://arxiv.org/abs/0810.5488} {\path{arXiv:0810.5488}},
  \href {http://dx.doi.org/10.1016/j.physrep.2008.11.001}
  {\path{doi:10.1016/j.physrep.2008.11.001}}.

\bibitem{kolmogorov1991mean}
A.~Kolmogorov, On the notion of mean, Selected works of AN Kolmogorov 25 (1991)
  144--146.

\bibitem{Teyssier2002}
R.~Teyssier, {Cosmological hydrodynamics with adaptive mesh refinement: A new
  high resolution code called RAMSES}, \aap 385~(1) (2002) 337--364.
\newblock \href {http://arxiv.org/abs/0111367} {\path{arXiv:0111367}}, \href
  {http://dx.doi.org/10.1051/0004-6361:20011817}
  {\path{doi:10.1051/0004-6361:20011817}}.

\bibitem{campbell1897}
J.~E. Campbell, On a law of combination of operators, Proc. London Math. Soc.
  29 (1897) 14--32.

\bibitem{baker1905}
H.~F. Baker, Alternant and continuous groups, Proc. London Math. Soc. 3 (1905)
  24--47.

\bibitem{hausdorff1906}
F.~Hausdorff, {Die symbolische Exponentialformel in der Gruppentheorie},
  Leipziger Ber. 58 (1906) 19--48.

\bibitem{Chin:2001}
S.~A. Chin, C.-R. Chen, {Fourth order gradient symplectic integrator methods
  for solving the time-dependent Schr{\"o}dinger equation}, The Journal of
  Chemical Physics 114~(17) (2001) 7338--7341.

\bibitem{ForestRuth:1990}
E.~{Forest}, R.~D. {Ruth}, {Fourth-order symplectic integration}, Physica D
  Nonlinear Phenomena 43~(1) (1990) 105--117.
\newblock \href {http://dx.doi.org/10.1016/0167-2789(90)90019-L}
  {\path{doi:10.1016/0167-2789(90)90019-L}}.

\bibitem{CandyRozmus:1991}
J.~{Candy}, W.~{Rozmus}, {A Symplectic Integration Algorithm for Separable
  Hamiltonian Functions}, Journal of Computational Physics 92~(1) (1991)
  230--256.
\newblock \href {http://dx.doi.org/10.1016/0021-9991(91)90299-Z}
  {\path{doi:10.1016/0021-9991(91)90299-Z}}.

\bibitem{Taruya2017}
A.~Taruya, S.~Colombi, {Post-collapse perturbation theory in 1D cosmology -
  beyond shell-crossing}, \mnras 470~(4) (2017) 4858--4884.
\newblock \href {http://dx.doi.org/10.1093/MNRAS/STX1501}
  {\path{doi:10.1093/MNRAS/STX1501}}.

\bibitem{Melott1989}
A.~L. Melott, S.~F. Shandarin,
  \href{http://adsabs.harvard.edu/doi/10.1086/167681}{{Gravitational
  instability with high resolution}}, \apj 343 (1989) 26.
\newblock \href {http://dx.doi.org/10.1086/167681} {\path{doi:10.1086/167681}}.
\newline\urlprefix\url{http://adsabs.harvard.edu/doi/10.1086/167681}

\bibitem{Tassev2013}
S.~Tassev, M.~Zaldarriaga, D.~J. Eisenstein, {Solving large scale structure in
  ten easy steps with COLA}, \jcap 2013~(6) (2013) 036.
\newblock \href {http://arxiv.org/abs/1301.0322} {\path{arXiv:1301.0322}},
  \href {http://dx.doi.org/10.1088/1475-7516/2013/06/036}
  {\path{doi:10.1088/1475-7516/2013/06/036}}.

\bibitem{Howlett2015}
C.~Howlett, M.~Manera, W.~J. Percival, {L-PICOLA: A parallel code for fast dark
  matter simulation}, Astronomy and Computing 12 (2015) 109--126.
\newblock \href {http://arxiv.org/abs/1506.03737} {\path{arXiv:1506.03737}}.

\bibitem{Tassev2015}
S.~Tassev, D.~J. Eisenstein, B.~D. Wandelt, M.~Zaldarriaga, {sCOLA: The N-body
  COLA Method Extended to the Spatial Domain}, \preprint\href
  {http://arxiv.org/abs/1502.07751} {\path{arXiv:1502.07751}}.

\bibitem{Izard2016}
A.~Izard, M.~Crocce, P.~Fosalba, {ICE-COLA: Towards fast and accurate synthetic
  galaxy catalogues optimizing a quasi-N-body method}, \mnras 459~(3) (2016)
  2327--2341.
\newblock \href {http://arxiv.org/abs/1509.04685} {\path{arXiv:1509.04685}},
  \href {http://dx.doi.org/10.1093/mnras/stw797}
  {\path{doi:10.1093/mnras/stw797}}.

\bibitem{Koda2016}
J.~Koda, C.~Blake, F.~Beutler, E.~Kazin, F.~Marin, {Fast and accurate mock
  catalogue generation for low-mass galaxies}, \mnras 459~(2) (2016)
  2118--2129.
\newblock \href {http://arxiv.org/abs/1507.05329} {\path{arXiv:1507.05329}},
  \href {http://dx.doi.org/10.1093/mnras/stw763}
  {\path{doi:10.1093/mnras/stw763}}.

\bibitem{Rampf2021b}
C.~Rampf, O.~Hahn, {Shell-crossing in a $\Lambda$cDM Universe}, \mnrasl 501~(1)
  (2021) L71--L75.
\newblock \href {http://dx.doi.org/10.1093/mnrasl/slaa198}
  {\path{doi:10.1093/mnrasl/slaa198}}.

\bibitem{bayer2021fast}
A.~E. Bayer, A.~Banerjee, Y.~Feng, A fast particle-mesh simulation of
  non-linear cosmological structure formation with massive neutrinos, \jcap
  2021~(01) (2021) 016.
\newblock \href {http://dx.doi.org/10.1088/1475-7516/2021/01/016}
  {\path{doi:10.1088/1475-7516/2021/01/016}}.

\bibitem{Brenier2003}
Y.~Brenier, U.~Frisch, M.~H{\'{e}}non, G.~Loeper, S.~Matarrese, R.~Mohayaee,
  A.~Sobolevskiǐ, {Reconstruction of the early Universe as a convex
  optimization problem}, \mnras 346~(2) (2003) 501--524.
\newblock \href {http://arxiv.org/abs/0304214} {\path{arXiv:0304214}}, \href
  {http://dx.doi.org/10.1046/j.1365-2966.2003.07106.x}
  {\path{doi:10.1046/j.1365-2966.2003.07106.x}}.

\bibitem{mcquinn2016cosmological}
M.~McQuinn, M.~White, Cosmological perturbation theory in 1+ 1 dimensions,
  \jcap 2016~(01) (2016) 043.
\newblock \href {http://dx.doi.org/10.1088/1475-7516/2016/01/043}
  {\path{doi:10.1088/1475-7516/2016/01/043}}.

\bibitem{Corliss1980}
G.~F. Corliss, {Integrating ODEs in the complex plane—pole vaulting},
  Mathematics of Computation 35~(152) (1980) 1181--1189.

\bibitem{Chambers2003}
J.~E. Chambers, {Symplectic Integrators with Complex Time Steps}, The
  Astronomical Journal 126~(2) (2003) 1119--1126.
\newblock \href {http://dx.doi.org/10.1086/376844} {\path{doi:10.1086/376844}}.

\bibitem{Carlson2009}
J.~Carlson, M.~White, N.~Padmanabhan, {Critical look at cosmological
  perturbation theory techniques}, \prd 80~(4) (2009) 043531.
\newblock \href {http://arxiv.org/abs/0905.0479} {\path{arXiv:0905.0479}},
  \href {http://dx.doi.org/10.1103/PhysRevD.80.043531}
  {\path{doi:10.1103/PhysRevD.80.043531}}.

\bibitem{Klypin2018}
A.~Klypin, F.~Prada, {Dark matter statistics for large galaxy catalogues: Power
  spectra and covariance matrices}, \mnras 478~(4) (2018) 4602--4621.
\newblock \href {http://arxiv.org/abs/1701.05690} {\path{arXiv:1701.05690}},
  \href {http://dx.doi.org/10.1093/mnras/sty1340}
  {\path{doi:10.1093/mnras/sty1340}}.

\bibitem{Pietroni2018}
M.~Pietroni, {Structure formation beyond shell-crossing: Nonperturbative
  expansions and late-time attractors}, \jcap 2018~(6).
\newblock \href {http://arxiv.org/abs/1804.09140} {\path{arXiv:1804.09140}},
  \href {http://dx.doi.org/10.1088/1475-7516/2018/06/028}
  {\path{doi:10.1088/1475-7516/2018/06/028}}.

\bibitem{Melott1997}
A.~L. Melott, S.~F. Shandarin, R.~J. Splinter, Y.~Suto, Demonstrating
  discreteness and collision error in cosmological n-body simulations of dark
  matter gravitational clustering, \apj 479~(2) (1997) L79.

\bibitem{Scoccimarro1998}
R.~Scoccimarro, {Transients from initial conditions: A perturbative analysis},
  \mnras 299~(4) (1998) 1097--1118.
\newblock \href {http://arxiv.org/abs/astro-ph/9711187}
  {\path{arXiv:astro-ph/9711187}}, \href
  {http://dx.doi.org/10.1046/j.1365-8711.1998.01845.x}
  {\path{doi:10.1046/j.1365-8711.1998.01845.x}}.

\bibitem{Crocce2006}
M.~Crocce, S.~Pueblas, R.~Scoccimarro, {Transients from initial conditions in
  cosmological simulations}, \mnras 373~(1) (2006) 369--381.

\bibitem{Garrison:2016}
L.~H. {Garrison}, D.~J. {Eisenstein}, D.~{Ferrer}, M.~V. {Metchnik}, P.~A.
  {Pinto}, {Improving initial conditions for cosmological N-body simulations},
  \mnras 461~(4) (2016) 4125--4145.
\newblock \href {http://arxiv.org/abs/1605.02333} {\path{arXiv:1605.02333}},
  \href {http://dx.doi.org/10.1093/mnras/stw1594}
  {\path{doi:10.1093/mnras/stw1594}}.

\bibitem{Layzer1963}
D.~{Layzer}, {A Preface to Cosmogony. I. The Energy Equation and the Virial
  Theorem for Cosmic Distributions.}, \apj 138 (1963) 174.
\newblock \href {http://dx.doi.org/10.1086/147625} {\path{doi:10.1086/147625}}.

\bibitem{Irvine1965}
W.~M. {Irvine}, {Local Irregularities in a Universe Satisfying the Cosmological
  Principle.}, Ph.D. thesis, Harvard University, Massachusetts (Jan. 1961).

\bibitem{winther2013layzer}
H.~A. Winther, {Layzer-Irvine equation for scalar-tensor theories: A test of
  modified gravity N-body simulations}, \prd 88~(4) (2013) 044057.
\newblock \href {http://arxiv.org/abs/1308.4682} {\path{arXiv:1308.4682}}.

\bibitem{Zheligovsky:2014}
V.~{Zheligovsky}, U.~{Frisch}, {Time-analyticity of Lagrangian particle
  trajectories in ideal fluid flow}, Journal of Fluid Mechanics 749 (2014)
  404--430.
\newblock \href {http://arxiv.org/abs/1312.6320} {\path{arXiv:1312.6320}},
  \href {http://dx.doi.org/10.1017/jfm.2014.221}
  {\path{doi:10.1017/jfm.2014.221}}.

\bibitem{Matsubara:2015}
T.~{Matsubara}, {Recursive solutions of Lagrangian perturbation theory}, \prd
  92~(2) (2015) 023534.
\newblock \href {http://arxiv.org/abs/1505.01481} {\path{arXiv:1505.01481}},
  \href {http://dx.doi.org/10.1103/PhysRevD.92.023534}
  {\path{doi:10.1103/PhysRevD.92.023534}}.

\bibitem{Bode2003}
P.~Bode, J.~P. Ostriker, {Tree Particle‐Mesh: An Adaptive, Efficient, and
  Parallel Code for Collisionless Cosmological Simulation}, \apjs 145~(1)
  (2003) 1--13.
\newblock \href {http://arxiv.org/abs/0302065} {\path{arXiv:0302065}}, \href
  {http://dx.doi.org/10.1086/345538} {\path{doi:10.1086/345538}}.

\bibitem{Chen1974}
L.~Chen, A.~{Bruce Langdon}, C.~K. Birdsall, {Reduction of the grid effects in
  simulation plasmas}, \jcp 14~(2) (1974) 200--222.
\newblock \href {http://dx.doi.org/10.1016/0021-9991(74)90014-X}
  {\path{doi:10.1016/0021-9991(74)90014-X}}.

\bibitem{garrison2021abacus}
L.~H. Garrison, D.~J. Eisenstein, D.~Ferrer, N.~A. Maksimova, P.~A. Pinto, The
  abacus cosmological n-body code, \mnras 508~(1) (2021) 575--596.
\newblock \href {http://dx.doi.org/10.1093/mnras/stab2482}
  {\path{doi:10.1093/mnras/stab2482}}.

\bibitem{Hunter:2007}
J.~D. Hunter, Matplotlib: A 2d graphics environment, Computing in science \&
  engineering 9~(3) (2007) 90--95.

\bibitem{Waskom2021}
M.~L. Waskom, seaborn: statistical data visualization, Journal of Open Source
  Software 6~(60) (2021) 3021.
\newblock \href {http://dx.doi.org/10.21105/joss.03021}
  {\path{doi:10.21105/joss.03021}}.

\bibitem{Harris:2020}
C.~R. Harris, K.~J. Millman, S.~J. van~der Walt, R.~Gommers, P.~Virtanen,
  D.~Cournapeau, E.~Wieser, J.~Taylor, S.~Berg, N.~J. Smith, R.~Kern, M.~Picus,
  S.~Hoyer, M.~H. van Kerkwijk, M.~Brett, A.~Haldane, J.~Fernández~del Río,
  M.~Wiebe, P.~Peterson, P.~Gérard-Marchant, K.~Sheppard, T.~Reddy,
  W.~Weckesser, H.~Abbasi, C.~Gohlke, T.~E. Oliphant, Array programming with
  {NumPy}, Nature 585 (2020) 357–362.
\newblock \href {http://dx.doi.org/10.1038/s41586-020-2649-2}
  {\path{doi:10.1038/s41586-020-2649-2}}.

\bibitem{Virtanen:2020}
P.~Virtanen, R.~Gommers, T.~E. Oliphant, M.~Haberland, T.~Reddy, D.~Cournapeau,
  E.~Burovski, P.~Peterson, W.~Weckesser, J.~Bright, S.~J. {van der Walt},
  M.~Brett, J.~Wilson, K.~J. Millman, N.~Mayorov, A.~R.~J. Nelson, E.~Jones,
  R.~Kern, E.~Larson, C.~J. Carey, {\.I}.~Polat, Y.~Feng, E.~W. Moore,
  J.~{VanderPlas}, D.~Laxalde, J.~Perktold, R.~Cimrman, I.~Henriksen, E.~A.
  Quintero, C.~R. Harris, A.~M. Archibald, A.~H. Ribeiro, F.~Pedregosa, P.~{van
  Mulbregt}, {SciPy 1.0 Contributors}, {{SciPy} 1.0: Fundamental Algorithms for
  Scientific Computing in Python}, Nature Methods 17 (2020) 261--272.
\newblock \href {http://dx.doi.org/10.1038/s41592-019-0686-2}
  {\path{doi:10.1038/s41592-019-0686-2}}.

\bibitem{jax2018github}
J.~Bradbury, R.~Frostig, P.~Hawkins, M.~J. Johnson, C.~Leary, D.~Maclaurin,
  G.~Necula, A.~Paszke, J.~Vander{P}las, S.~Wanderman-{M}ilne, Q.~Zhang,
  \href{http://github.com/google/jax}{{JAX}: composable transformations of
  {P}ython+{N}um{P}y programs} (2018).
\newline\urlprefix\url{http://github.com/google/jax}

\bibitem{Pylians}
F.~{Villaescusa-Navarro}, {Pylians: Python libraries for the analysis of
  numerical simulations}, Astrophysics Source Code Library, record
  ascl:1811.008 (Nov. 2018).
\newblock \href {http://arxiv.org/abs/1811.008} {\path{arXiv:1811.008}}.

\bibitem{dyson1949}
F.~J. Dyson, {The radiation theories of Tomonaga, Schwinger, and Feynman},
  Physical Review 75~(3) (1949) 486.

\bibitem{Blanes2010}
S.~Blanes, F.~Casas, J.~A. Oteo, J.~Ros, {A pedagogical approach to the Magnus
  expansion}, European Journal of Physics 31~(4) (2010) 907--918.
\newblock \href {http://dx.doi.org/10.1088/0143-0807/31/4/020}
  {\path{doi:10.1088/0143-0807/31/4/020}}.

\bibitem{joachain1975quantum}
C.~Joachain, \href{https://books.google.at/books?id=Zs3vAAAAMAAJ}{Quantum
  Collision Theory}, North-Holland Publishing Company, 1975.
\newline\urlprefix\url{https://books.google.at/books?id=Zs3vAAAAMAAJ}

\bibitem{Dragt:1976}
A.~J. {Dragt}, J.~M. {Finn}, {Lie series and invariant functions for analytic
  symplectic maps}, Journal of Mathematical Physics 17~(12) (1976) 2215--2227.
\newblock \href {http://dx.doi.org/10.1063/1.522868}
  {\path{doi:10.1063/1.522868}}.

\bibitem{Cary:1981}
J.~R. {Cary}, {Lie transform perturbation theory for Hamiltonian systems},
  Physics Reports 79~(2) (1981) 129--159.
\newblock \href {http://dx.doi.org/10.1016/0370-1573(81)90175-7}
  {\path{doi:10.1016/0370-1573(81)90175-7}}.

\bibitem{Bravetti2017a}
A.~Bravetti, H.~Cruz, D.~Tapias,
  \href{https://linkinghub.elsevier.com/retrieve/pii/S0003491616302469}{{Contact
  Hamiltonian mechanics}}, Annals of Physics 376 (2017) 17--39.
\newblock \href {http://arxiv.org/abs/1604.08266} {\path{arXiv:1604.08266}},
  \href {http://dx.doi.org/10.1016/j.aop.2016.11.003}
  {\path{doi:10.1016/j.aop.2016.11.003}}.
\newline\urlprefix\url{https://linkinghub.elsevier.com/retrieve/pii/S0003491616302469}

\bibitem{DeLeon2020}
M.~de~Le{\'{o}}n, M.~Lainz, {A review on contact Hamiltonian and Lagrangian
  systems}, \preprint\href {http://arxiv.org/abs/2011.05579}
  {\path{arXiv:2011.05579}}.

\bibitem{Vermeeren2019}
M.~Vermeeren, A.~Bravetti, M.~Seri, {Contact variational integrators}, Journal
  of Physics A: Mathematical and Theoretical 52~(44) (2019) 445206.
\newblock \href {http://arxiv.org/abs/1902.00436} {\path{arXiv:1902.00436}},
  \href {http://dx.doi.org/10.1088/1751-8121/ab4767}
  {\path{doi:10.1088/1751-8121/ab4767}}.

\bibitem{bravetti2020numerical}
A.~Bravetti, M.~Seri, M.~Vermeeren, F.~Zadra, Numerical integration in
  celestial mechanics: a case for contact geometry, Celestial Mechanics and
  Dynamical Astronomy 132~(1) (2020) 7.
\newblock \href {http://dx.doi.org/10.1007/s10569-019-9946-9}
  {\path{doi:10.1007/s10569-019-9946-9}}.

\bibitem{bravetti2021new}
A.~Bravetti, M.~Seri, F.~Zadra, New directions for contact integrators, in:
  International Conference on Geometric Science of Information, Springer, 2021,
  pp. 209--216.

\bibitem{Bravetti2017}
A.~Bravetti, \href{http://www.mdpi.com/1099-4300/19/10/535}{{Contact
  Hamiltonian Dynamics: The Concept and Its Use}}, Entropy 19~(10) (2017) 535.
\newblock \href {http://dx.doi.org/10.3390/e19100535}
  {\path{doi:10.3390/e19100535}}.
\newline\urlprefix\url{http://www.mdpi.com/1099-4300/19/10/535}

\bibitem{Rampf2021TwoFluid}
C.~Rampf, C.~Uhlemann, O.~Hahn, {Cosmological perturbations for two cold fluids
  in $\Lambda$cDM}, Monthly Notices of the Royal Astronomical Society 503~(1)
  (2021) 406--425.
\newblock \href {http://dx.doi.org/10.1093/mnras/staa3605}
  {\path{doi:10.1093/mnras/staa3605}}.

\bibitem{Matarrese2002}
S.~Matarrese, R.~Mohayaee, {The growth of structure in the intergalactic
  medium}, \mnras 329~(1) (2002) 37--60.
\newblock \href {http://arxiv.org/abs/0102220} {\path{arXiv:0102220}}, \href
  {http://dx.doi.org/10.1046/j.1365-8711.2002.04944.x}
  {\path{doi:10.1046/j.1365-8711.2002.04944.x}}.

\bibitem{Short2006}
C.~J. Short, P.~Coles,
  \href{https://iopscience.iop.org/article/10.1088/1475-7516/2006/12/012}{{Gravitational
  instability via the Schr{\"{o}}dinger equation}}, \jcap 2006~(12) (2006)
  012--012.
\newblock \href {http://arxiv.org/abs/0605012} {\path{arXiv:0605012}}, \href
  {http://dx.doi.org/10.1088/1475-7516/2006/12/012}
  {\path{doi:10.1088/1475-7516/2006/12/012}}.
\newline\urlprefix\url{https://iopscience.iop.org/article/10.1088/1475-7516/2006/12/012}

\bibitem{Uhlemann2019}
C.~Uhlemann, C.~Rampf, M.~Gosenca, O.~Hahn,
  \href{https://doi.org/10.1103/PhysRevD.99.083524}{{Semiclassical path to
  cosmic large-scale structure}}, \prd 99~(8) (2019) 83524.
\newblock \href {http://arxiv.org/abs/1812.05633} {\path{arXiv:1812.05633}},
  \href {http://dx.doi.org/10.1103/PhysRevD.99.083524}
  {\path{doi:10.1103/PhysRevD.99.083524}}.
\newline\urlprefix\url{https://doi.org/10.1103/PhysRevD.99.083524}

\bibitem{sloan2019scalar}
D.~Sloan, Scalar fields and the flrw singularity, Classical and Quantum Gravity
  36~(23) (2019) 235004.
\newblock \href {http://dx.doi.org/10.1088/1361-6382/ab4eb4}
  {\path{doi:10.1088/1361-6382/ab4eb4}}.

\bibitem{sloan2023herglotz}
D.~Sloan, Herglotz action for homogeneous cosmologies, Classical and Quantum
  Gravity 40~(11) (2023) 115008.
\newblock \href {http://dx.doi.org/10.1088/1361-6382/accef6}
  {\path{doi:10.1088/1361-6382/accef6}}.

\bibitem{Herglotz1930}
G.~Herglotz, {Vorlesungen über die Theorie der Berührungstransformationen}
  (1930).

\bibitem{Heckmann1955}
O.~Heckmann, E.~Sch{\"{u}}cking, {Bemerkungen zur Newtonschen Kosmologie},
  Zeitschrift f{\"{u}}r Astrophysik 38 (1955) 95--109.

\bibitem{Ehlers1997}
J.~Ehlers, T.~Buchert, {Newtonian cosmology in Lagrangian formulation:
  Foundations and Perturbation theory}, General Relativity and Gravitation
  29~(6) (1997) 733--764.
\newblock \href {http://arxiv.org/abs/9609036} {\path{arXiv:9609036}}, \href
  {http://dx.doi.org/10.1023/A:1018885922682}
  {\path{doi:10.1023/A:1018885922682}}.

\bibitem{saga2018lagrangian}
S.~Saga, A.~Taruya, S.~Colombi, Lagrangian cosmological perturbation theory at
  shell crossing, \prl 121~(24) (2018) 241302.
\newblock \href {http://dx.doi.org/10.1103/PhysRevLett.121.241302}
  {\path{doi:10.1103/PhysRevLett.121.241302}}.

\end{thebibliography}
\def\mnras{Monthly Notices of the Royal Astronomical Society
  }\def\mnrasl{Monthly Notices of the Royal Astronomical Society Letters
  }\def\aap{Astronomy and Astrophysics }\def\prd{Physical Review D
  }\def\prl{Physical Review Letters }\def\jcap{Journal of Cosmology and
  Astroparticle Physics }\def\nat{Nature }\def\jcp{Journal of Computational
  Physics }\def\apj{The Astrophysical Journal }\def\apjs{The Astrophysical
  Journal Supplement Series }\def\preprint{Preprint }

\begin{appendix}

\section{Proof of Proposition~\ref{prop:convergence}}
\label{sec:proof_proposition2}
In order to prove Proposition~\ref{prop:convergence}, we will start in a somewhat more general setting and eventually return to the specific case of the leapfrog/Verlet-scheme in Eqs.~\eqref{eq:leapfrog_scheme}. As in the main body of this paper, we will use $\tau$ to denote a general time variable. A Hamiltonian $\mathcal{H} = \mathcal{H}(\vecb{X}, \vecb{P}, \tau)$ defines the Hamiltonian vector field $\vecb{V}_{\mathcal{H}}: \mathbb{R}\to\mathbb{R}^{3N+3N}$ as
\begin{align}
     \vecb{V}_\mathcal{H}(\tau) = \left(\partial_{P_1} \mathcal{H}(\tau), \ldots, \partial_{P_{3N}} \mathcal{H}(\tau), -\partial_{X_1} \mathcal{H}(\tau), \ldots, -\partial_{X_{3N}} \mathcal{H}(\tau)\right).
\end{align}
In what follows, we will use an index $j$ that runs from $j = 1$ to $3N$ over the three coordinates of all the particles, i.e.\ $\vecb{X} = (X_j)_{j=1}^{3N}$ and $\vecb{P} = (P_j)_{j=1}^{3N}$. Let us also define the vector $\vecb{\xi} = (\vecb{X},  \vecb{P}) \in \mathbb{R}^{3N + 3N}$. 
The Lie derivative of a scalar field $f = f(\vecb{X}, \vecb{P}, \tau)$ along the vector field $\vecb{V}_\mathcal{H}$ can be written as
\begin{align}
    \mathscr{L}_{\vecb{V}_\mathcal{H}} f = \sum_{j=1}^{3N} \left( \partial_{P_j} \mathcal{H} \, \partial_{X_j} f - \partial_{X_j} \mathcal{H} \, \partial_{P_j} f \right) = - \{\mathcal{H}, f\},
\label{eq:magnus_proof_lie_derivative}
\end{align}
where $\{\cdot, \cdot\}$ denotes the Poisson bracket. The equation of motion for $f$ is then given by the convective derivative
\begin{align}
\dd_\tau f = \left(\partial_\tau + \mathscr{L}_{\vecb{V}_\mathcal{H}}\right) f,
\end{align}
which implies in particular that the canonical equations of motion become the linear first-order operator equation
\begin{align}
    \dd_\tau \vecb{\xi} = \mathscr{L}_{\vecb{V}_\mathcal{H}} \vecb{\xi}.
\label{eq:magnus_proof_operator_eq}
\end{align}
Formally, the solution of this equation can be written as
\begin{align}
    \vecb{\xi}(\tau) = \mathscr{T}\left(\exp\left(\int_0^\tau \dd s \, \mathscr{L}_{\vecb{V}_\mathcal{H}} \right) \right) \vecb{\xi}_0,
\end{align}
where $\vecb{\xi}_0 := \vecb{\xi}(0)$ and $\mathscr{T}$ is Dyson's time-ordering operator \cite{dyson1949}. Note that for a scalar mapping $\mathscr{L}_{\vecb{V}_\mathcal{H}}: \mathbb{R} \to \mathbb{R}$, as well as in the vector-valued case under the condition that $\mathscr{L}_{\vecb{V}_\mathcal{H}}$ satisfies the commutator relation $\mathscr{L}_{\vecb{V}_\mathcal{H}}(\tau_1) \mathscr{L}_{\vecb{V}_\mathcal{H}}(\tau_2) = \mathscr{L}_{\vecb{V}_\mathcal{H}}(\tau_2) \mathscr{L}_{\vecb{V}_\mathcal{H}}(\tau_1)$ for all $\tau_1, \tau_2 \in \mathbb{R}$ (e.g.\ in the case that $\mathscr{L}_{\vecb{V}_\mathcal{H}}$ is constant over time), the time-ordering operator can be omitted. In the general case, however, the exponential operator cannot be written as a power series as usual, but rather involves a product integral, see e.g.\ Ref.~\cite{Blanes2010} and \cite[Sec.~13.4]{joachain1975quantum}. Still, one can make the \textit{ansatz} of a proper exponential solution of Eq.~\eqref{eq:magnus_proof_operator_eq}
\begin{align}
    \vecb{\xi}(\tau) = \exp(\Theta(\tau)) \, \vecb{\xi}_0,
    \label{eq:exp_solution_magnus}
\end{align}
where $\Theta$ is expanded in a series as 
\begin{equation}
    \Theta(\tau) = \sum_{k=0}^{\infty} \Theta_k(\tau).
\end{equation}
This idea is known as the \emph{Magnus expansion} \cite{Magnus1954}, named after Wilhelm Magnus, who showed that the solution $\vecb{\xi}$ of Eq.~\eqref{eq:magnus_proof_operator_eq} can indeed be written in this form (under sufficient conditions for convergence), where the operator $\Theta$ satisfies
\begin{align}
\dd_\tau \Theta = \sum_{k=0}^{\infty} \frac{B_k}{k!} \operatorname{ad}_\Theta^k \mathscr{L}_{\vecb{V}_\mathcal{H}}
\label{eq:magnus_proof_eq_for_omega}
\end{align}
Here, $B_k$ is the $k$-th Bernoulli number, and the $k$-th application of the adjoint operator $\operatorname{ad}^k$ is recursively defined as
\begin{align}
   \operatorname{ad}_A^0 B = B, \qquad \operatorname{ad}_A^1 B = [A, B], \qquad \operatorname{ad}_A^k B = [A, \operatorname{ad}_A^{k-1} B],
\end{align}
for $k \in \mathbb{N}$, where $[\cdot, \cdot]$ is the Lie bracket.
We will assume the convergence of the Magnus expansion here and refer the reader to Ref.~\cite{Blanes2009} for results on the convergence of this series.

More generally, we can consider Eq.~\eqref{eq:exp_solution_magnus} for an arbitrary time $\tau > 0$ and a timestep $\Delta \tau > 0$:
\begin{align}
    \vecb{\xi}(\tau + \Delta \tau) = \exp(\Theta(\tau, \tau + \Delta \tau)) \, \vecb{\xi}(\tau).
\end{align}
Then, by iteratively solving Eq.~\eqref{eq:magnus_proof_eq_for_omega}, one obtains for the first three terms of the series:
\begin{subequations}
\begin{align}
    \Theta_0(\tau, \tau + \Delta \tau) &= \int_{\tau}^{\tau + \Delta \tau} \dd \tau_1 \, \mathscr{L}_{\vecb{V}_\mathcal{H}}(\tau_1), \\
    \Theta_1(\tau, \tau + \Delta \tau) &= \frac{1}{2} \int_{\tau}^{\tau + \Delta \tau} \dd \tau_1 \int_t^{\tau_1} \dd \tau_2 \, \mathscr{L}_{[{\vecb{V}_\mathcal{H}}(\tau_1), {\vecb{V}_\mathcal{H}}(\tau_2)]}, \\
    \Theta_2(\tau, \tau + \Delta \tau) &= \frac{1}{6} \int_{\tau}^{\tau + \Delta \tau} \dd \tau_1 \int_t^{\tau_1} \dd \tau_2 \int_t^{\tau_2} \dd \tau_3 \, \left( \mathscr{L}_{[\vecb{V}_\mathcal{H}(\tau_1), [\vecb{V}_\mathcal{H}(\tau_2), \vecb{V}_\mathcal{H}(\tau_3)]]} + \mathscr{L}_{[\vecb{V}_\mathcal{H}(\tau_3), [\vecb{V}_\mathcal{H}(\tau_2), \vecb{V}_\mathcal{H}(\tau_1)]]} \right).
\end{align}
\end{subequations}
While these expressions hold for arbitrary Hamiltonians, we limit ourselves to the situation relevant for cosmological $N$-body simulations, where the Hamiltonian is separable (but non-autonomous) as given by Eq.~\eqref{eq:Hamiltonian_general}, which we repeat here for convenience:
\begin{align}
    \mathcal{H}(\vecb{X}, \vecb{P}, \tau) = \mu(\tau) T(\vecb{P}) + \nu(\tau) U(\vecb{X}),
\end{align}
where $T$ and $U$ are the kinetic and potential energy, respectively, and the explicit time dependence of the Hamiltonian is described by the functions $\mu(\tau)$ and $\nu(\tau)$. Then, the Lie derivative in Eq.~\eqref{eq:magnus_proof_lie_derivative} becomes
\begin{align}
    \mathscr{L}_{\vecb{V}_\mathcal{H}} f = \sum_{j=1}^{3N} \left( \mu \, \partial_{P_j} T \, \partial_{X_j} f - \nu \, \partial_{X_j} U \, \partial_{P_j} f \right) =: \left( \mu \mathscr{L}_{\vecb{V}_T} + \nu \mathscr{L}_{\vecb{V}_U}\right) f,
\end{align}
where we defined the kinetic and potential parts of the Hamiltonian vector field as
\begin{align}
 \vecb{V}_T = \left( \partial_{P_1} T, \ldots, \partial_{P_{3N}} T, 0, \ldots, 0 \right), \qquad 
 \vecb{V}_U = \left( 0, \ldots, 0, -\partial_{X_1} U, \ldots, -\partial_{X_{3N}} U \right).
\end{align}
The drift and kick operators are then defined as the infinitesimal generators of the associated Lie groups \cite{Dragt:1976,Cary:1981}. Recall that the relation between generators of transformations as elements of a Lie algebra and the Lie group is made through the well-known limit of infinitely many applications of the generator as the operator exponential
\begin{equation}
\lim_{k\to\infty}\left({\rm \mathbb{I}} + \frac{1}{k} \mathscr{L}_{\vecb{V}_\mathcal{H}} \right)^k = \sum_{k=0}^\infty \frac{\mathscr{L}_{\vecb{V}_\mathcal{H}}^k}{k!} =  \exp\left(\mathscr{L}_{\vecb{V}_\mathcal{H}}\right),
\end{equation}
which is convergent if the Lie derivative is bounded. We thus have for the drift and kick operators for a timestep $\Delta \tau$:
\begin{align}
\hat{D}_{\Delta \tau} = {\mathbb I} + \Delta \tau \mathscr{L}_{\vecb{V}_T},\qquad \hat{K}_{\Delta \tau} = {\mathbb I} + \Delta \tau \mathscr{L}_{\vecb{V}_U},
\label{eq:drift_and_kick_as_generators}
\end{align}
the effect of which on state vectors is easily seen to be
\begin{align}
\hat{D}_{\Delta \tau} \vecb{\xi} = \vecb{\xi} + \Delta \tau \left( \bnabla_{\vecb{P}} T, 0, \ldots, 0\right) \qquad
\hat{K}_{\Delta \tau} \vecb{\xi} =  \vecb{\xi} - \Delta \tau \left( 0, \ldots, 0, \bnabla_{\vecb{X}} U \right).
\end{align}
We see clearly how these infinitesimal generators decouple for a separable Hamiltonian, so that one operator advances the positions using only the momenta, and the other advances the momenta using only the positions. 
Explicitly computing the first three terms of the Magnus expansion for such a Hamiltonian yields:
\begin{subequations}
\begin{align}
    \Theta_0(\tau, \tau + \Delta \tau) &= C_\mu(\tau, \tau + \Delta \tau) \mathscr{L}_{\vecb{V}_T} + C_\nu(\tau, \tau + \Delta \tau) \mathscr{L}_{\vecb{V}_U}, \\
    \Theta_1(\tau, \tau + \Delta \tau) &= C_{[\mu, \nu]}(\tau, \tau + \Delta \tau) \mathscr{L}_{[\vecb{V}_T, \vecb{V}_U]}, \\
    \Theta_2(\tau, \tau + \Delta \tau) &= C_{[\mu, [\mu, \nu]]}(\tau, \tau + \Delta \tau) \mathscr{L}_{[\vecb{V}_T, [\vecb{V}_T, \vecb{V}_U]]} + C_{[\nu, [\nu, \mu]]}(\tau, \tau + \Delta \tau) \mathscr{L}{[\vecb{V}_U, [\vecb{V}_U, \vecb{V}_T]]},    
\end{align}
\label{eq:magnus_proof_omega_012}
\end{subequations}
with the functionals
\begin{subequations}
\begin{align}
C_f(\tau, \tau + \Delta \tau) &:= \int_{\tau}^{\tau + \Delta \tau} \dd \tau_1\, f(\tau_1) = \Delta \tau f(\tau) + \frac{(\Delta \tau)^2}{2}\dot{f}(\tau) + \frac{(\Delta \tau)^3}{6}\ddot{f}(\tau) + \frac{(\Delta \tau)^4}{24}f^{(3)}(\tau) +\mathscr{O}((\Delta \tau)^5)\\
C_{[f,g]}(\tau, \tau + \Delta \tau) &:= \frac{1}{2}\int_{\tau}^{\tau + \Delta \tau} \dd \tau_1\, \int_{\tau}^{\tau_1} \dd \tau_2\, \bigl( f(\tau_1) g(\tau_2) - f(\tau_2) g(\tau_1) \bigr) \nonumber\\
&\phantom{:}= \frac{(\Delta \tau)^3}{12}\left( \dot{f}(\tau)\,g(\tau)-f(\tau)\,\dot{g}(\tau) \right)+\frac{(\Delta \tau)^4}{24}\left(\ddot{f}(\tau)g(\tau)-f(\tau)\ddot{g}(\tau)\right)+\mathscr{O}((\Delta \tau)^5) \\
C_{[f,[f,g]]}(\tau, \tau + \Delta \tau) &:= \frac{1}{6} \int_{\tau}^{\tau + \Delta \tau} \dd \tau_1\, \int_{\tau}^{\tau_1} \dd \tau_2\, \int_{\tau}^{\tau_2} \dd \tau_3\,  \left( f(\tau_1)f(\tau_2)g(\tau_3)+f(\tau_2)f(\tau_3)g(\tau_1)-2f(\tau_1)f(\tau_3)g(\tau_2)\right) \nonumber\\
&\phantom{:}= \mathscr{O}((\Delta \tau)^5),
\end{align}
\end{subequations}
where a dot denotes the derivative w.r.t.\ $\tau$.
Now, let us consider a leapfrog/Verlet-like operator-split propagator that factors into a triple sequence of kick and drift operators. As in the main body of the paper, we will only consider the DKD case, but note that the KDK case can be treated analogously. Following the notation in Eqs.~\eqref{eq:leapfrog_scheme}, let $\alpha$, $\beta$, $\gamma$, denote the multipliers specifying the lengths of the first drift, kick, and second drift, respectively. From the Baker--Campbell--Hausdorff formula \cite{campbell1897,baker1905,hausdorff1906} generalised to three operators, we obtain the following:
\begin{equation}
\begin{aligned}
    \log\left( e^{\alpha  \mathscr{L}_{\vecb{V}_T}} e^{\beta  \mathscr{L}_{\vecb{V}_U}} e^{\gamma  \mathscr{L}_{\vecb{V}_T}} \right) &= (\alpha + \gamma) \mathscr{L}_{\vecb{V}_T} \;+\; \beta \mathscr{L}_{\vecb{V}_U} 
    \;+\; \frac{(\alpha-\gamma)\beta}{2} \mathscr{L}_{[\vecb{V}_T, \vecb{V}_U]} \\
    &\quad + \frac{(\alpha^2 + \gamma^2)\beta - 4 \alpha \beta \gamma}{12} \mathscr{L}_{[\vecb{V}_T, [\vecb{V}_T, \vecb{V}_U]]} \;+\; \frac{\beta^2 (\alpha + \gamma)}{12} \mathscr{L}_{[\vecb{V}_U, [\vecb{V}_U, \vecb{V}_T]]} + \text{h.o.t.}
\end{aligned}
\end{equation}
Matching the coefficients $\alpha$, $\beta$, and $\gamma$ with the Magnus expansion in Eq.~\eqref{eq:magnus_proof_omega_012} to second order in $\Delta \tau$ gives rise to the following requirements for local third-order accuracy:
\begin{equation}
    \alpha + \gamma  = C_\mu(\tau, \tau + \Delta \tau) + \mathscr{O}((\Delta \tau)^3), \qquad 
    \beta = C_\nu(\tau, \tau + \Delta \tau) + \mathscr{O}((\Delta \tau)^3), \qquad
    (\alpha - \gamma) \beta = \mathscr{O}((\Delta \tau)^3).
\end{equation}

These are the necessary conditions stated in Proposition~\ref{prop:convergence}. Since the total number of timesteps needed to cover a fixed time interval scales as $1 / \Delta \tau$, local third-order accuracy for an individual timestep of length $\Delta \tau$ results in global second-order accuracy as usual, which finishes the proof.

\section{Proof of Proposition~\ref{prop:second_order_convergence}}
\label{sec:proof_second_order}
First, we will show that choosing the coefficient functions $p(\Delta D, D)$ and $q(\Delta D, D)$ such that they agree with $p_{\text{Symplectic~2}}(\Delta D, D)$ and $q_{\text{Symplectic~2}}(\Delta D, D)$, respectively, to second order in $\Delta D$ is necessary and sufficient in order for a $\mPi$-integrator to be a second-order accurate method.
To this aim, let us Taylor expand the result of a single DKD step to second-order in $\Delta D$, which yields an expression of the form
\begin{subequations}
\begin{align}    
    \vecb{X}_i^{n+1} &= \vecb{X}_i^n + \vecb{v}_i^{n,(1)} \Delta D + \vecb{v}_i^{n,(2)} (\Delta D)^2 +\mathscr{O}((\Delta D)^3), \\
    \vecb{P}_i^{n+1} &= \vecb{w}_i^{n,(0)} + \vecb{w}_i^{n,(1)} \Delta D + \vecb{w}_i^{n,(2)} (\Delta D)^2  + \mathscr{O}((\Delta D)^3),
\end{align}
\end{subequations}
where $\vecb{v}_i^{n,(r)}$ and $\vecb{w}_i^{n,(r)}$ collect the terms of $r$-th order in $\Delta D$ for $\vecb{X}_i^{n+1}$ and $\vecb{P}_i^{n+1}$, respectively. We will now show that matching the second-order Taylor coefficients for the positions $\vecb{v}_i^{n,(1)}, \vecb{v}_i^{n,(2)}$ and momenta  $\vecb{w}_i^{n,(0)}, \vecb{w}_i^{n,(1)}, \vecb{w}_i^{n,(2)}$ is equivalent to matching the Taylor coefficients of the coefficient functions for the kick $p(\Delta D, D_n)$ and $q(\Delta D, D_n)$ w.r.t.\ $\Delta D$ up to second-order.

Explicitly computing the coefficients for the Symplectic~2 integrator expressed in $D$-time, one finds
\begin{subequations}
\begin{align}
    \vecb{v}^{n,(1)}_{i, \text{Symplectic~2}} &= \vecb{P}_i^n \frac{{ \dd_D a}}{{a^3 H}} = \frac{{\vecb{P}_i^n}}{F} \stackrel{\mathrm{EdS}}{\asymp} \frac{\vecb{P}_i^n}{a^{\nicefrac{3}{2}}}, \label{eq:v1_sym}\\
    \vecb{v}^{n,(2)}_{i, \text{Symplectic~2}} &= \vecb{A}(\vecb{X}_i^{n}) \frac{a}{2 F^2} - \vecb{P}_i^{n} \frac{\dd_D F}{2 F^2}
    \stackrel{\mathrm{EdS}}{\asymp} \frac{\vecb{A}(\vecb{X}_i^n)}{2 a^2} - \frac{3 \vecb{P}_i^n}{4 a^{\nicefrac{5}{2}}}, \label{eq:v2_sym} \\  
    \vecb{w}^{n,(0)}_{i, \text{Symplectic~2}} &= \vecb{P}_i^n, \\
    \vecb{w}^{n,(1)}_{i, \text{Symplectic~2}} &= \vecb{A}(\vecb{X}_i^n) \frac{\dd_D a}{a^2 H} = \vecb{A}(\vecb{X}_i^n) \frac{a}{F} \stackrel{\mathrm{EdS}}{\asymp} \frac{\vecb{A}(\vecb{X}_i^n)}{\sqrt{a}}, \\
    \vecb{w}^{n,(2)}_{i, \text{Symplectic~2}} &= \bnabla_{\vecb{x}} \vecb{A}(\vecb{X}_i^n) \cdot \vecb{P}_i^n \frac{a}{2 F^2} - \vecb{A}(\vecb{X}_i^n) \,
    \dd_D \left(\frac{a}{2 F}\right)
    \stackrel{\mathrm{EdS}}{\asymp} \frac{\bnabla_{\vecb{x}} \vecb{A}(\vecb{X}_i^n) \cdot \vecb{P}_i^n}{2a^2} - \frac{\vecb{A}(\vecb{X}_i^n)}{4 a^{\nicefrac{3}{2}}},    
    \label{eq:w2_sym}
\end{align}
\end{subequations}
where all time-dependent functions on the right-hand side are evaluated at time $D_n$. 

Similarly, for $\mPi$-integrators, one obtains the first coefficient for the positions as
\begin{equation}
\vecb{v}^{n,(1)}_{i, \mPi} = \vecb{P}_i^n \frac{1 + p(0, D_n)}{2 F} + \vecb{A}(\vecb{X}_i^n) \frac{q(0, D_n)}{2}, \label{eq:v1_pi}
\end{equation}
where $p(0, D_n) := p(\Delta D, D_n)|_{\Delta D = 0}$ and similarly for $q$.
Comparing Eqs.~\eqref{eq:v1_sym} and \eqref{eq:v1_pi} immediately yields the conditions
\begin{equation}
    p(0, D_n) = 1 \qquad \text{and} \qquad q(0, D_n) = 0. \label{eq:condition_1_on_p_q}
\end{equation}

Using this result, the second-order coefficient is given by
\begin{equation}
\vecb{v}^{n,(2)}_{i, \mPi} = \vecb{A}(\vecb{X}_i^n) \frac{\partial_{\Delta D} q(0, D_n)}{2} + \vecb{P}_i^n \frac{\partial_{\Delta D} p(0, D_n)}{2 F}, 
\end{equation}
where $\partial_{\Delta D} p(0, D_n) := \partial_{\Delta D} p(\cdot, D_n)|_{\Delta D = 0}$ and similarly for $q$.
Comparison to Eq.~\eqref{eq:v2_sym} leads to
\begin{equation}
    \partial_{\Delta D} p(0, D_n) 
    = - \frac{\dd_D F}{F} \stackrel{\mathrm{EdS}}{\asymp} -\frac{3}{2a} \qquad \text{and} \qquad
    \partial_{\Delta D} q(0, D_n) =
    \frac{a}{F^2}\stackrel{\mathrm{EdS}}{\asymp} \frac{1}{a^2}. \label{eq:condition_2_on_p_q}
\end{equation}
Note that these are exactly the functions $-\mathfrak{f}$ and $\mathfrak{g}$ defined in Eq.~\eqref{eq:f_and_g}.

Now, we continue with the coefficients for the velocity: computing $\vecb{w}^{n,(0)}_{i, \mPi}$ and $\vecb{w}^{n,(1)}_{i, \mPi}$ and making use of Eqs.~\eqref{eq:condition_1_on_p_q} and \eqref{eq:condition_2_on_p_q}, one finds that $\vecb{w}^{n,(0)}_{i, \mPi} = \vecb{w}^{n,(0)}_{i, \text{Symplectic~2}}$ and $\vecb{w}^{n,(1)}_{i, \mPi} = \vecb{w}^{n,(1)}_{i, \text{Symplectic~2}}$, without imposing any further conditions on $p$ and $q$. 

The final coefficient to be determined $\vecb{w}^{n,(2)}_{i, \mPi}$ is then found to be
\begin{equation}
    \vecb{w}^{n,(2)}_{i, \mPi} = \vecb{A}(\vecb{X}_i^n) \left(\frac{a \dd_D F}{F^2} + \frac{\partial_{\Delta D}^2 q(0, D_n) F}{2} \right) + \bnabla_{\vecb{x}} \vecb{A}(\vecb{X}_i^n) \cdot \vecb{P}_i^n  \frac{a}{2 F^2} + \vecb{P}_i^n \left(\frac{-2 (\dd_D F)^2 + F\left(\dd_D^2 F + F \partial_{\Delta D}^2 p(0, D_n) \right)}{2 F^2} \right),
\end{equation}
from which one infers that
\begin{equation}
    \partial^2_{\Delta D} p(0, D_n) = \frac{2 (\dd_D F)^2 - F \dd_D^2 F}{F^2} \stackrel{\mathrm{EdS}}{\asymp} \frac{15}{4 a^2} \qquad \text{and} \qquad
    \partial^2_{\Delta D} q(0, D_n) = \frac{F \dd_D a - 3 a \dd_D F}{F^3} \stackrel{\mathrm{EdS}}{\asymp} -\frac{7}{2 a^3}.
\end{equation} 
Thus, a $\mPi$-integrator is second-order accurate if and only if the asymptotic behaviour of $p$ and $q$ w.r.t.\ $\Delta D$ matches that of Symplectic~2 up to second order, see Eqs.~\eqref{eq:p_consistency_with_symplectic_integrator} and \eqref{eq:q_consistency_with_symplectic_integrator}. 

Finally, it remains to be shown that when $p$ agrees with $p_{\text{Symplectic~2}}$ to second order and $q$ is chosen according to the Zel'dovich consistency condition in Eq.~\eqref{eq:zeldovich_consistency_condition}, $q$ will also agree with $q_{\text{Symplectic~2}}$ to second order. Using Eq.~\eqref{eq:G_identity}, Eq.~\eqref{eq:zeldovich_consistency_condition} can be expressed as
\begin{equation}
    q(\Delta D, D_n) = \frac{a_{n+\nicefrac{1}{2}} (1 - p(\Delta D, D_n))}{F_{n+\nicefrac{1}{2}} (\dd_D F)_{n+\nicefrac{1}{2}}}, \label{eq:q_of_p_in_proof}
\end{equation}
where quantities with a subscript $_{n+\nicefrac{1}{2}}$ are evaluated at $D_n + \Delta D / 2$, rather than at $D_n$.
Then, given a coefficient function 
\begin{equation}
    p(\Delta D, D_n) = 1 - \frac{\dd_D F}{F} \Delta D + \frac{2 (\dd_D F)^2 - F \dd_D^2 F}{2 F^2} (\Delta D)^2 + \mathscr{O}((\Delta D)^3),
\end{equation}
Taylor expanding Eq.~\eqref{eq:q_of_p_in_proof} around $\Delta D = 0$ leads to
\begin{equation}
    q(\Delta D, D_n) = \frac{a}{F^2} \Delta D + \frac{F \dd_D a - 3 a \dd_D F}{2 F^3} (\Delta D)^2 + \mathscr{O}((\Delta D)^3),
\end{equation}
and hence to agreement with $q_{\text{Symplectic~2}}$ to second order, finishing the proof.

\section{\textsc{LPTFrog} in the context of contact geometry}
\label{sec:contact}
Symplectic geometry is in fact not the only suitable toolkit for the construction of integrators in the cosmological setting: in fact, \textsc{LPTFrog} bears a connection to another closely related framework, namely \emph{contact geometry}. Since the focus of this work lies on incorporating LPT into cosmological integrators, however, we will keep this digression brief and refer the reader to Refs.~\cite{Arnold:1989, Bravetti2017a, DeLeon2020} for in-depth introductions to contact geometry and to Refs.~\cite{Vermeeren2019, bravetti2020numerical, bravetti2021new} for contact integrators, which respect the contact structure of the underlying system.
Contact geometry has been dubbed the `odd-dimensional cousin'  of symplectic geometry \cite{Bravetti2017}. It is typically applied to physical problems with \emph{dissipation}; for instance, the archetypal example of such a system would be the damped harmonic oscillator. In the Hamiltonian framework discussed above, and in particular in the Hamiltonians in Eqs.~\eqref{eq:Hamiltonian_t} and \eqref{eq:Hamiltonian_ttilde}, the dissipative nature of the cosmological Vlasov--Poisson system, i.e.\ the loss of momentum due to the expansion of the universe, is implicitly expressed through the time dependence of the terms. Due to this time dependence, the Hamiltonian is not conserved over time. 

To expose the friction term in $D$-time more clearly, we can perform a mapping from the canonical variables $(\vecb{X}, \vecb{P})$ to the variables $(\vecb{X}, \vecb{\mPi})$. While this transformation is not canonical, contact geometry provides a suitable setting to justify this transformation. As a matter of fact, changing to $(\vecb{X}, \vecb{\mPi})$ remedies the degenerate behaviour of the Vlasov--Poisson system as $D \to 0$ as shown in Ref.~\cite[Appendix~D]{Rampf2021TwoFluid} (see also Refs.~\cite{Matarrese2002, Short2006, Uhlemann2019, sloan2019scalar, sloan2023herglotz} for papers that explicitly or implicitly make use of contact geometry in the context of cosmology). Specifically, the `asymptotic' initial conditions for the Vlasov--Poisson system $\lim_{D \to 0} \delta = 0$ and $\lim_{D \to 0} \vecb{\mPi} = -\bnabla_{\vecb{q}} \varphi_{\text{ini}}$ (which are natural when considering growing modes only) can be imposed in the contact Hamiltonian framework, and the solution remains regular as $D \to 0$, but they are not compatible with the standard Hamiltonian framework (see the discussion in that reference).

In contact geometry, the phase space is augmented with an additional scalar variable $\mathcal{S}$, which in fact turns out to be Hamilton's principal function, i.e.\ the solution to the Hamilton--Jacobi equation (which is simply the action up to an additive integration constant).\footnote{More formally, the contact manifold in our cosmological $N$-body setting in $d = 3$ dimensions is a $(2 \, d \, N + 1)$-dimensional manifold supplemented with a 1-form $\eta$ (known as the `contact form'). By Darboux' theorem, there exists a set of local coordinates $(\vecb{X}, \vecb{\mPi}, \mathcal{S})$ such that the contact form can be expressed in local coordinates as $\eta = \dd \mathcal{S} - \sum_{j=1}^{3N} \mPi_j \, \dd X_j$, where the index $j$ runs over the 3 coordinates for all $N$ particles. Compare this to the canonical coordinates $(\vecb{X}, \vecb{P})$ and the canonical 1-form (also known as Poincar\'{e} or Liouville 1-form) $\eta = \sum_{j=1}^{3N} P_j \, \dd X_j$ in the symplectic case, whose exterior derivative is the usual symplectic form. Here, we wrote $\vecb{P}$ again instead of $\vecb{\mPi}$ to indicate that this is the canonical momentum variable. Note that more general contact transformations that also affect $\vecb{X}$ are of course possible too.} One can then define contact transformations (see Eqs.~(60)$-$(62) in Ref.~\cite{Bravetti2017a}), which are the contact counterpart of canonical transformations. It is easy to show that the transformation $(\vecb{X}, \vecb{P}, \mathcal{S}) \to (\vecb{X}, \vecb{\mPi}, \mathcal{S}/F)$ is contact and leads to the following contact Hamiltonian (using Eq.~(82) in \cite{Bravetti2017a})
\begin{equation}
    \mathscr{H}_c(\vecb{X}, \vecb{\mPi}, \mathcal{S}, D) = \sum_{i=1}^N \frac{\|\vecb{\mPi}_i\|^2}{2} + \mathfrak{g}(D) \sum_{i=1}^N \varphi_N(\vecb{X}_i) + \mathfrak{f}(D) \mathcal{S},
\label{eq:contact_hamiltonian}
\end{equation}
where, recall, $\mathfrak{f}(D) = (\dd_D F)(D) / F(D)$ and $\mathfrak{g}(D) = a(D) / F^2(D)$. The kinetic term w.r.t.\ $D$-time is no longer explicitly time-dependent, but a damping term $\mathfrak{f}(D) \mathcal{S}$ appeared, which accounts for the expanding universe. This contact Hamiltonian gives rise to the following flow equations (see Eqs.~(37)$-$(39) in \cite{Bravetti2017a}):
\begin{subequations}
\begin{alignat}{-1}
    \dd_D \vecb{X}_i &= \bnabla_{\vecb{\mPi}_i} \mathscr{H}_c \ \ &=& \ \vecb{\mPi}_i, \label{eq:contact_position} \\
    \dd_D \vecb{\mPi}_i &= - \vecb{\mPi}_i \, \partial_{\mathcal{S}} \mathscr{H}_c - \bnabla_{\vecb{X}_i} \mathscr{H}_c  \ \ &=& \  -\vecb{\mPi}_i \mathfrak{f}(D) - \mathfrak{g}(D) \bnabla _{\vecb{x}} \varphi_N(\vecb{X}_i), \label{eq:contact_momentum} \\
    \dd_D \mathcal{S} &= \sum_{i=1}^N \vecb{\mPi}_i \cdot \bnabla_{\vecb{\mPi}_i} \mathscr{H}_c - \mathscr{H}_c \ \ &=& \ \sum_{i=1}^N \frac{\|\vecb{\mPi}_i\|^2}{2} - \mathfrak{g}(D) \sum_{i=1}^N \varphi_N(\vecb{X}_i) - \mathfrak{f}(D) \mathcal{S}. \label{eq:contact_action}
\end{alignat}
\end{subequations}
Equations~\eqref{eq:contact_position} and \eqref{eq:contact_momentum} are identical to Eqs.~\eqref{eq:dD_X_LCDM_computed} and Eq.~\eqref{eq:d2D_X_LCDM_computed}, respectively. Observe that the right-hand side of Eq.~\eqref{eq:contact_action} is the Lagrangian $\mathscr{L}$ of the system, related to the contact Hamiltonian via a Legendre transform, i.e.
\begin{equation}
    \mathscr{H}_c = \sum_{i=1}^N \bnabla_{\vecb{\mPi}_i}\mathscr{L} \cdot \vecb{\mPi}_i - \mathscr{L},
\end{equation}
confirming that $\mathcal{S}$ is indeed Hamilton's principal function. Importantly, the evolution equations for $\vecb{X}$ and $\vecb{\mPi}$ are decoupled from the  variable $\mathcal{S}$.

Recall that in Lagrangian mechanics, Hamilton's principle of stationary action states that solution curves are critical points of the action integral, from which it can then be derived that solution curves must equivalently satisfy the Euler--Lagrange equations. A generalisation of this principle to contact geometry is Herglotz' variational principle \cite{Herglotz1930}, which defines the action via a differential equation rather than an integral, namely $\dd_D \mathcal{S}(D) = \mathcal{L}(\vecb{X}(D), \vecb{\mPi}(D), \mathcal{S}(D), D)$, see e.g.\ \cite[Definition~1]{Vermeeren2019}. The authors of that reference then proceed to prove that a \emph{discrete} curve satisfies a discrete counterpart of Herglotz' principle if and only if it solves the discrete Euler--Lagrange equations (Theorem~1 in that work); moreover, these equations give rise to a contact transformation (Theorem~2). Based on this idea, the authors derive different contact integrators for contact Hamiltonians similar to our Eq.~\eqref{eq:contact_hamiltonian}. 

In particular, the kick of the second-order contact method in their Example~2 in position-momentum formulation reads \textit{mutatis mutandis}
\begin{equation}
     \vecb{\mPi}_i^{n+1} = \frac{\vecb{\mPi}_i^n \left(1 - \frac{\Delta D}{2} \mathfrak{f}(D_{n+\nicefrac{1}{2}})\right) + \frac{\Delta D}{2} \left[\mathfrak{g}(D_n) \vecb{A}(\vecb{X}_i(D_{n})) + \mathfrak{g}(D_{n+1}) \vecb{A}(\vecb{X}_i(D_{n+1}))\right]}{1 + \frac{\Delta D}{2} \mathfrak{f}(D_{n+\nicefrac{1}{2}})}.
\end{equation}
Remarkably, this almost coincides with the kick operation of \textsc{LPTFrog} in Eq.~\eqref{eq:kick_LPTFrog}, which we derived by considering a quadratic trajectory w.r.t.\ $D - D_n$ without invoking contact geometry, apart from the minor difference that we approximate the acceleration with the midpoint rule rather than the trapezoidal rule, i.e.\ $\mathfrak{g}(D_{n+\nicefrac{1}{2}}) \vecb{A}(\vecb{X}_i^{n+\nicefrac{1}{2}})$ instead of $(\mathfrak{g}(D_n) \vecb{A}(\vecb{X}_i^{n}) + \mathfrak{g}(D_{n+1})\vecb{A}(\vecb{X}_i^{n+1})) / 2$, and the fact that their constant friction coefficient $\alpha$ becomes $\mathfrak{f}(D_{n+\nicefrac{1}{2}})$ in our case.\footnote{Interestingly, the form of \textsc{LPTFrog} bears a resemblance to the non-canonical integrator discussed already in 1988 by Ref.~\cite{Hockney:1988} (Eq.~(11-59), see also \cite[Appendix~A]{Quinn:1997}), which also accounts for the friction by multiplying the old velocity $\vecb{P}_i^{n}$ with a factor $< 1$ in the kick when computing the new velocity $\vecb{P}_i^{n+1}$. However, since that integrator drifts w.r.t.\ cosmic time $t$ rather than growth-factor time $D$, it is not Zel'dovich consistent.} However, the second-order contact integrator presented in Ref.~\cite{Vermeeren2019} updates the positions and momenta synchronously and is built in such a way that it explicitly conserves the contact structure. In contrast, \textsc{LPTFrog} as a leapfrog scheme is similar to the St\"{o}rmer--Verlet method also considered in that reference (although they formulate it as KDK rather than DKD), which is based on the same discrete generalised Euler--Lagrange equation as their contact method (Eq.~20 in that work), but differs in terms of the initialisation, see their Example~2 for more details. It would be interesting to study contact integrators such as those developed in Ref.~\cite{Vermeeren2019} for the cosmological Vlasov--Poisson system in future work.

More general, in the light of this discussion, we conjecture that symplecticity might after all be a dispensable property for the construction of integrators for the \textit{cosmological} Vlasov--Poisson equations, and their dissipative nature might in fact be better accommodated by frameworks more suited to the description of non-conservative systems such as contact geometry. 

An important follow-up question would be the treatment of bound structures such as dark matter haloes that decouple from the Hubble flow and are therefore no longer subject to the dissipative effect of the expanding universe. Here, a promising approach could be hybrid schemes that (locally) switch from a contact integrator to a standard symplectic integrator in a scale-dependent manner when clusters virialise.

\section{Galilean invariance of $\mPi$-integrators}
\label{sec:galilean_invariance}
We also note that $\mPi$-integrators with $p(\Delta D, D_n) \neq F(D_n)/F(D_{n+1})$ are \emph{not} Galilean invariant. To illustrate this point, let us consider a simulated universe with particles at rest that are distributed in such a way that the density is exactly homogeneous, which implies for the acceleration $\vecb{A}(\vecb{X}) \equiv 0$. If we change to a moving reference frame by adding the same initial velocity $\vecb{V}_0$ to every particle, the particles will keep this velocity with canonical DKD integrators (and in particular for \textsc{FastPM}), regardless of the timestep $\Delta D$. This is, however, in general not the case for $\mPi$-integrators with $p(\Delta D, D_n) \neq F(D_n) / F(D_{n+1})$ such as \textsc{LPTFrog}, which follows immediately from rewriting Eq.~\eqref{eq:kick_ansatz_pi} in terms of $\vecb{P}$ as 
\begin{equation}
\vecb{P}^{n+1}_i = p(\Delta D, D_n) \frac{F(D_{n+1})}{F(D_n)} \vecb{P}^n_i + q(\Delta D, D_n) F(D_{n+1}) \vecb{A}(\vecb{X}_i^{n+\nicefrac{1}{2}}),    
\end{equation}
where the factor $p(\Delta D, D_n) F(D_{n+1}) / F(D_n) \neq 1$ (but converges at third order to 1 for $\Delta D \to 0$ for a second-order scheme). This implies that the particles will not maintain their constant velocity $\vecb{V}_0$. For $p(\Delta D, D_n) < 0$, which occurs with \textsc{LPTFrog} for sufficiently large timesteps (see Fig.~\ref{fig:integrators_p_and_q}), the velocity even changes direction. In view of the friction term $f(D) \mathcal{S}$ in the contact Hamiltonian in Eq.~\eqref{eq:contact_hamiltonian}, which gives rise to a damping force proportional to $\vecb{\mPi}$ in Eq.~\eqref{eq:contact_momentum}, this is not a flaw of the method, but simply a consequence of the dissipative nature of the cosmological Vlasov--Poisson system. For completeness, let us also remark that the cosmological Vlasov--Poisson system is in fact invariant under a non-Galilean transformation discovered by Heckmann \& Sch\"{u}cking \cite{Heckmann1955, Ehlers1997} (see also \cite{Rampf:2021}).

\section{1D experiments: impact of the force computation on the convergence}
\label{sec:1D_exact_vs_pm}
\begin{figure*}[h]
\centering
  \noindent
   \resizebox{0.85\textwidth}{!}{
    \includegraphics{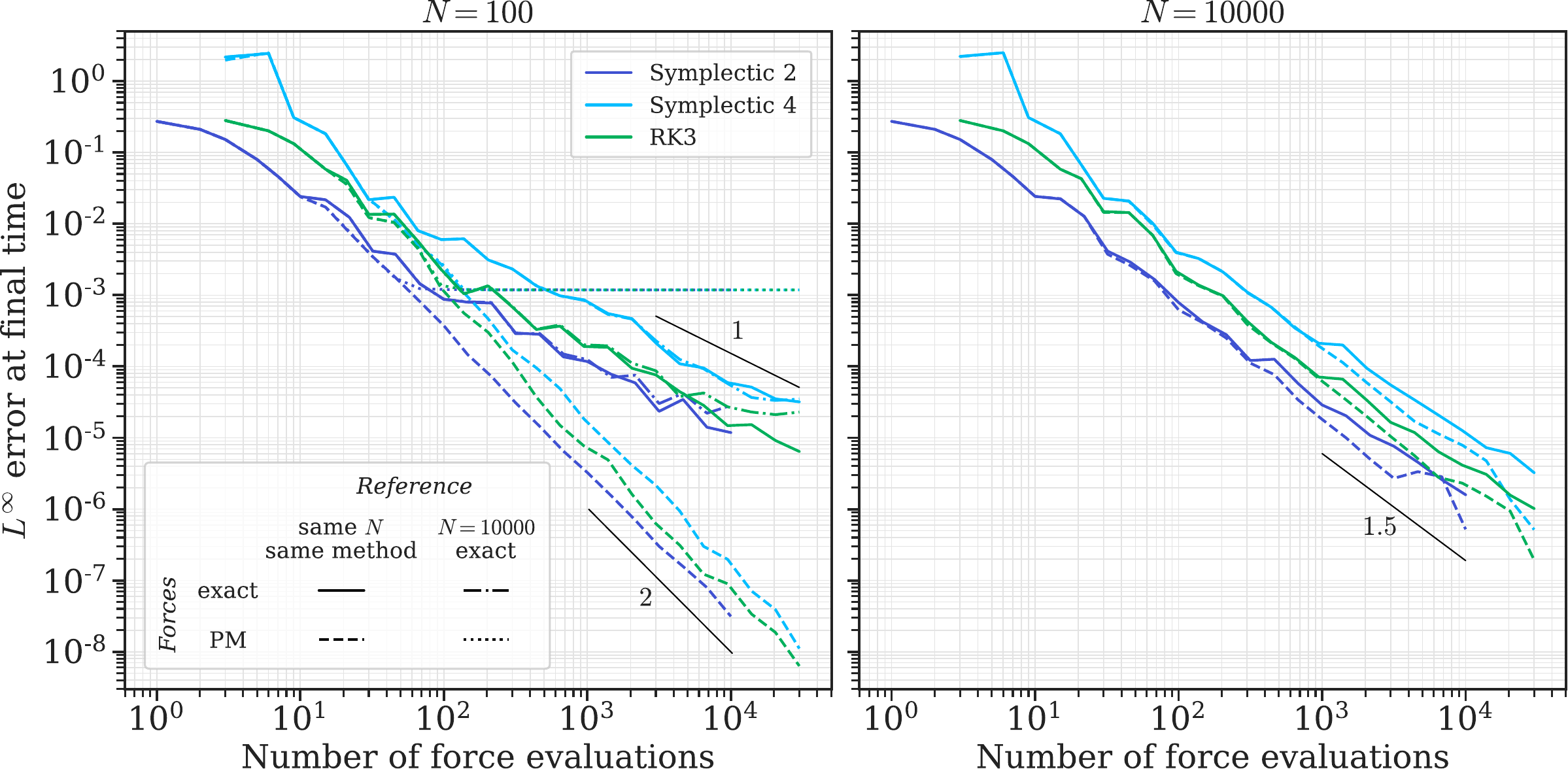}
    }
    \caption{Convergence of the numerically computed displacement field $\mPsi$ for the 1D growing-mode-only solution in the \textbf{post-shell-crossing} regime at $a_\mathrm{end} = 2.0 > 1.0 = a_\mathrm{cross}$, for $N = $ 100 (left) and $N = $ 10,000 (right) particles, for three selected integrators.
    We consider two different force computation methods, exact (Eq.~\ref{eq:exact_potential_gradient}) and PM. For the low-resolution $N = 100$ case, we compute the errors (1) towards a reference simulation that uses the same $N$ and force computation method and (2) towards the high-resolution $N = $ 10,000 reference with exact forces, see the main text for details.
    While convergence order $\nicefrac{3}{2}$ is observed for sufficiently many particles as in Fig.~\ref{fig:convergence_1D_post_sc_force_eval} (where also $N =$ 10,000 particles were used), the force computation method affects the convergence order when the spatial resolution is low.}    \label{fig:convergence_1D_pm_vs_exact}
\end{figure*}
In Section~\ref{sec:1D_results_post}, we showed experimentally that for one-dimensional collapse, the convergence after shell-crossing is limited to order $\nicefrac{3}{2}$, as expected in view of Proposition~\ref{prop:convergence_loss} from the loss of regularity of the displacement field as computed by Ref.~\cite{Rampf:2021}. In our convergence study for the 3D case at $z = 0$ in Section~\ref{sec:3D_convergence_study}, however, we obtained a rather complex picture:
for the large \textsc{Quijote} box ($1 \ h^{-1} \ \text{Gpc}$, average inter-particle spacing $\sim 2 \ h^{-1} \ \text{Mpc}$ in each dimension), convergence was second order w.r.t.\ the $L^2$ and $L^\infty$ norms, whereas for the small \textsc{Camels} box ($25 \ h^{-1} \ \text{Mpc}$, average inter-particle spacing $\sim 100 \ h^{-1} \ \text{kpc}$ in each dimension), we found convergence orders of $\sim \nicefrac{3}{2}$ and $\sim \nicefrac{1}{2}$ w.r.t.\ the $L^2$ and $L^\infty$ norm, respectively. 

Given that one would expect consecutive one-dimensional collapse of structures to be prevalent also in 3D, and simultaneous collapse along more than one axis would rather worsen the regularity (e.g.\ \cite{saga2018lagrangian}), it seems surprising that we still observed second-order convergence in the \textsc{Quijote} experiment. As mentioned in the main text, we hypothesise that when using a rather modest spatial resolution such as $N = 512^3$ in our case, in combination with PM force computation, the time convergence is affected.

In order to test this hypothesis, we revisit the controlled setup in the 1D post-shell-crossing experiment from Section~\ref{sec:1D_results_post}, as the forces can be computed exactly in the 1D case (see Eq.~\eqref{eq:exact_potential_gradient}), and a very high spatial resolution is feasible. Now, we also consider the convergence behaviour when using only $N = 100$ instead of $N = $ 10,000 particles, and when replacing the exact force computation with the PM method.

Figure~\ref{fig:convergence_1D_pm_vs_exact} shows the convergence of the post-shell-crossing displacement error in terms of the $L^\infty$ norm for different cases. We consider three different time integrators, namely the standard second-order `Symplectic~2' stepper, the fourth-order `Symplectic~4' stepper from Section~\ref{sec:higher_order}, and the third-order Runge--Kutta integrator `RK~3'. In the left (right) panel, $N = 100$ ($N = $ 10,000) particles were used. For the reference solution, with respect to which the errors are computed, we always perform 100,000 timesteps just as in Section~\ref{sec:1D_results_post}.

We begin with $N = $ 10,000 particles and consider the following two cases:\vspace{-0.5em}
\begin{itemize}
    \item \textit{simulations:} exact forces, \textit{reference:} exact forces (solid)
    
    This is the same scenario as studied in Section~\ref{sec:1D_results_post}, and we observe convergence of order $\nicefrac{3}{2}$ for all integrators, regardless their nominal convergence order. We will use the reference solution for this case as the `spatially converged' reference also later in the $N = 100$ case.
    
    \item \textit{simulations:} PM forces, 
    \textit{reference:} PM forces (dashed)

    With N = 10,000 particles, using PM forces for both the reference and all the other simulations only makes a minor difference as compared to the exact force computation and also leads to convergence order $\nicefrac{3}{2}$.
\end{itemize}
Thus, we conclude that when the number of particles is very large, the time convergence is not significantly affected by the force computation method, and one obtains the convergence order expected for the underlying continuous solution. Note that we do not incorporate any gravitational softening in this work.

Now, let us turn towards the low-resolution case with $N = 100$. Here, we study four different cases:\vspace{-0.5em}
\begin{itemize}
    \item \textit{simulations:} exact forces, \textit{reference:} exact forces $\&$ $N = 100$ (solid)
    
    Interestingly, for $N = 100$, the convergence occurs more slowly when using exact forces for the simulations and the reference than for $N = $ 10,000, roughly with an order of $1$. This suggests that the coarse spatial approximation of the continuous displacement field with few particles further decreases its regularity when solving the 1D Poisson equation exactly (treating the $N$-body particles as point masses). 

    \item \textit{simulations:} PM forces,
    \textit{reference:} PM forces $\&$ $N = 100$ (dashed)
    
    This is the most relevant case for drawing conclusions on the observed second-order convergence in the 3D \textsc{Quijote} experiment. The reference simulation is converged in terms of the number of timesteps, but not in terms of the number of particles, and the forces have been computed with the PM method. Strikingly, all methods convergence with second order in this scenario, just as what we found in the 3D \textsc{Quijote} case. We therefore suspect that the effect of particle discreteness, together with the approximate force computation via the PM method, might explain the higher-than-expected second-order convergence for our \textsc{Quijote} experiment, just as in this present case.

    \item \textit{simulations}: exact forces,
    \textit{reference:} exact forces $\&$ $N = $ 10,000 (dash-dotted)
    
    This case only differs from the solid lines in that the reference uses $N = $ 10,000 instead of $N = 100$. The results are very similar, and the convergence order is also $\sim 1$.

    \item \textit{simulations}: PM forces,
    \textit{reference:} exact forces $\&$ $N = $ 10,000 (dotted)
    
    When comparing the low-resolution $N = 100$ simulations that use PM forces with the high-resolution $N = $ 10,000 reference, the error decreases when going up to $\sim 100$ timesteps, but then remains constant, revealing the intrinsic error of the PM method at this low spatial resolution.    
\end{itemize}

To summarise, this experiment showed that the observed order of time convergence becomes sensitive to the force computation method for low particle resolutions, and differs from the time convergence order expected based on considerations for the continuous (i.e.\ spatially perfectly resolved) solution. When using PM forces in the simulations and computing the error towards a PM-based reference, we found the numerical convergence to be faster than expected from theory, while the opposite was the case with the exact force computation. A more thorough investigation of the convergence order as a function of resolution, resolved scales, force computation methods, gravitational softening, etc.\ would be an interesting follow-up of this work.

\section{2D experiments: displacement and density fields}
\label{sec:2D_appendix}
\begin{figure*}[htb]
\centering
  \noindent
   \resizebox{0.7\textwidth}{!}{
    \includegraphics{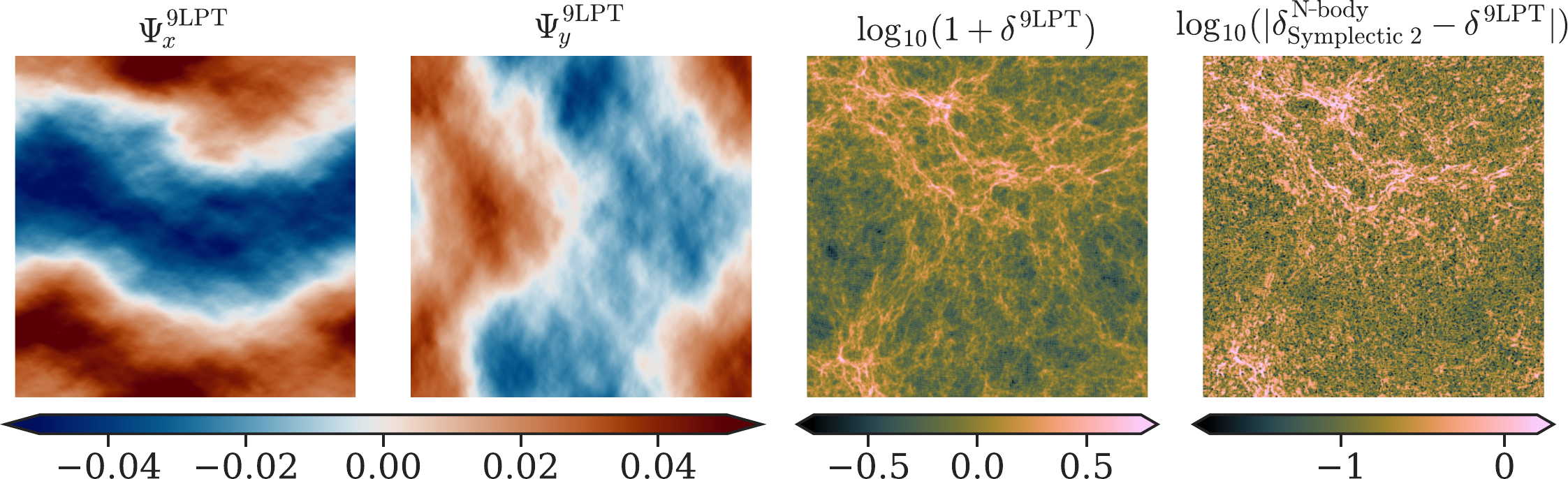}
    }
    \caption{9LPT displacement field (first two panels for the $x$ and $y$ component, respectively), density field (third panel), and density error of the Symplectic~2 integrator towards 9LPT after a single step (fourth panel) for the 2D experiment at the same time as the errors shown in Fig.~\ref{fig:2d_delta}. The correlation between the errors and the density field are clearly visible.}
    \label{fig:data_2d_psi_and_delta_lpt}
\end{figure*}

In Fig.~\ref{fig:data_2d_psi_and_delta_lpt}, we show the displacement and density field at the end of the single timestep after which we evaluated the different integrators for the 2D experiment in Fig.~\ref{fig:2d_delta}. We computed these fields using 9$^\text{th}$-order LPT. The rightmost panel shows the density error of the $N$-body solution after a single step with the Symplectic~2 integrator (see Fig.~\ref{fig:2d_delta} for the errors towards 1LPT, 2LPT, and 3LPT, which look very similar). The strong spatial correlations between the density error and the density field itself can clearly be seen. 

\section{Additional material for the \textsc{PowerFrog} integrator in $\Lambda$CDM cosmology}
\label{sec:powerfrog_lcdm_material}
\begin{figure*}[htb]
\centering
  \noindent
   \resizebox{1\textwidth}{!}{
    \includegraphics{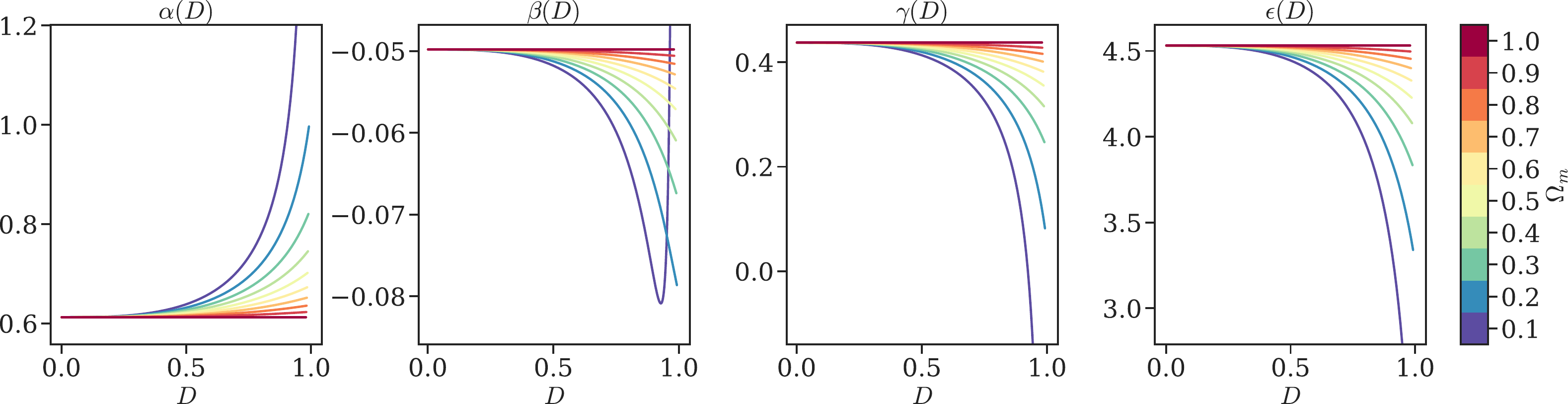}
    }
    \caption{\textsc{PowerFrog} coefficients as a function of the growth factor $D$, for different values of $\Omega_{\mathrm{m}}$ (and $\Omega_\Lambda = 1 - \Omega_{\mathrm{m}}$). For the EdS case of $\Omega_{\mathrm{m}} = 1$, the coefficients are constant and given by the values in Eq.~\eqref{eq:powerfrog_coeffs_EdS}.}
    \label{fig:powerfrog_coeffs_LCDM}
\end{figure*}
Figure~\ref{fig:powerfrog_coeffs_LCDM} shows the coefficients $\alpha(D)$, $\beta(D)$, $\gamma(D)$, and $\epsilon(D)$ as a function of the growth factor $D$ for different values of $\Omega_{\mathrm{m}}$ (and a flat universe such that $\Omega_\Lambda = 1 - \Omega_{\mathrm{m}}$). Notably, for EdS (i.e.\ $\Omega_{\mathrm{m}} = 1$), the coefficients become time-independent, indicative of the fact that the EdS cosmology is scale-free.

For completeness, we also provide the equation for $\alpha(D_n)$ here, which arises from a matching between the 2$^\text{nd}$-order Taylor coefficient of $p_{\text{\textsc{PowerFrog}}}(\Delta D, D_n)$ and $p_{\text{Symplectic~2}}(\Delta D, D_n)$ w.r.t.\ $\Delta D$: 
\begin{equation}
    \alpha(D_n) = \frac{\mathcal{T}_1(D_n) \pm \sqrt{\mathcal{T}_2(D_n)}}{\mathcal{T}_3(D_n)}
    \label{eq:powerfrog_lcdm_alpha}
\end{equation}
with
\begin{equation}
\begin{aligned}
\mathcal{T}_1(D_n) &= (1 + 4 \beta) \, D_n \, (\dd_D F)^2 - 2 \, \beta \, F [\dd_D F + D_n \, \dd^2_D F], \\
\mathcal{T}_2(D_n) &= (\dd_D F)^2 \left[\beta^2 \, F^2 + (1 + 16 \, \beta) \, D_n^2 \, (\dd_D F)^2 - 2 \, \beta \, D_n \, F \left(3 \, \dd_D F + 4 \, D_n \, \dd^2_D F\right)\right], \\
\mathcal{T}_3(D_n) &= 4 \, D_n \, (\dd_D F)^2 - F \, [\dd_D F + 2 \, D_n \, \dd_D^2 F],
\end{aligned}
\end{equation}
where $F$ and its derivatives and $\beta$ are always evaluated at $D_n$, which we suppressed here for brevity. Just as in the EdS case (cf.\ Eq.~\eqref{eq:powerfrog_coefficient_conditions}), we proceed with the solution with the `+'-sign in Eq.~\eqref{eq:powerfrog_lcdm_alpha} from here on. Since $F$ is a function of the scale factor $a$ in our implementation, we convert the derivatives w.r.t.\ $D$ into derivatives w.r.t.\ $a$ in practice.

\section{Taylor series expansions of the integrators for EdS}
\label{sec:taylor_appendix}
For the sake of completeness, we provide the first few terms of the Taylor series expansions of the coefficient functions $p(\Delta a, a)$ and $q(\Delta a, a)$ defining the $\mPi$-integrators considered in this work for EdS cosmology (see Section~\ref{sec:lpt_integrators}). Importantly, all functions $p$ and $q$ agree with the standard second-order symplectic DKD integrator up to second order in $\Delta a$, for which reason all integrators are second-order accurate.

\begin{subequations}
\begin{align}
    p_\text{\textsc{FastPM}}(\Delta a, a) = p_\text{{Symplectic~2}}(\Delta a, a) &= 1 - \frac{3}{2} \frac{\Delta a}{a} + \frac{15}{8} \left(\frac{\Delta a}{a}\right)^2 - \frac{35}{16} \left(\frac{\Delta a}{a}\right)^3 + \frac{315}{128} \left(\frac{\Delta a}{a}\right)^4 + \mathscr{O}\left((\Delta a)^5\right),  \\ 
    p_{\textsc{TsafPM}}(\Delta a, a) &= 1 - \frac{3}{2} \frac{\Delta a}{a} + \frac{15}{8} \left(\frac{\Delta a}{a}\right)^2 - \frac{9}{4} \left(\frac{\Delta a}{a}\right)^3 + \frac{333}{128} \left(\frac{\Delta a}{a}\right)^4 + \mathscr{O}\left((\Delta a)^5\right), \\
    p_\text{\textsc{PowerFrog}}(\Delta a, a) &= 1 - \frac{3}{2} \frac{\Delta a}{a} + \frac{15}{8} \left(\frac{\Delta a}{a}\right)^2 - 2.059 \left(\frac{\Delta a}{a}\right)^3 + 1.960 \left(\frac{\Delta a}{a}\right)^4 + \mathscr{O}\left((\Delta a)^5\right), \\
    p_\epsilon(\Delta a, a) &= 1 - \frac{3}{2} \frac{\Delta a}{a} + \frac{15}{8} \left(\frac{\Delta a}{a}\right)^2 - \frac{3 (3 + 22 \epsilon^2)}{32 \epsilon^2} \left(\frac{\Delta a}{a}\right)^3 + \frac{3 (36 + 89 \epsilon^2)}{128 \epsilon^2} \left(\frac{\Delta a}{a}\right)^4 \nonumber\\ &\quad + \mathscr{O}\left((\Delta a)^5\right),    
\end{align}
\end{subequations}
and
\begin{subequations}
\begin{align}
    q_{\textsc{TsafPM}}(\Delta a, a) = q_\text{{Symplectic~2}}(\Delta a, a) &= \frac{1}{a} \frac{\Delta a}{a} - \frac{7}{4a} \left(\frac{\Delta a}{a}\right)^2 + \frac{19}{8a} \left(\frac{\Delta a}{a}\right)^3 - \frac{187}{64 a} \left(\frac{\Delta a}{a}\right)^4 + \mathscr{O}\left((\Delta a)^5\right), \\
    q_{\textsc{PowerFrog}}(\Delta a, a) &= \frac{1}{a} \frac{\Delta a}{a} - \frac{7}{4a} \left(\frac{\Delta a}{a}\right)^2 + \frac{1.750}{a} \left(\frac{\Delta a}{a}\right)^3 - \frac{2.431}{a} \left(\frac{\Delta a}{a}\right)^4 + \mathscr{O}\left((\Delta a)^5\right), \\
    q_\epsilon(\Delta a, a) &= \frac{1}{a} \frac{\Delta a}{a} - \frac{7}{4a} \left(\frac{\Delta a}{a}\right)^2 + \frac{3 (1 + 12 \epsilon^2)}{16 \epsilon^2 a} \left(\frac{\Delta a}{a}\right)^3 - \frac{7 (6 + 23 \epsilon^2)}{64 \epsilon^2 a} \left(\frac{\Delta a}{a}\right)^4 \nonumber\\ &\quad + \mathscr{O}\left((\Delta a)^5\right).
\end{align}
\end{subequations}
In particular, the values $\epsilon = \nicefrac{3}{2}$ and $\epsilon = 1$ correspond to \textsc{FastPM} and \textsc{LPTFrog}, respectively. Since $p_{\textsc{PowerFrog}}$ and $q_{\textsc{PowerFrog}}$ rely on coefficients computed as the numerical solution of a transcendental equation, the Taylor coefficients have also been determined numerically.

\section{The KDK case}
\label{sec:KDK}
In this section, we briefly comment on the KDK variant of some of the $\mPi$-integrators. In that case, one obtains a scheme of the form
\begin{subequations}
\begin{align}    
    \vecb{\mPi}_i^{n+\nicefrac{1}{2}} &= p^{(1)}(\Delta D, D_n) \vecb{\mPi}_i^{n} + q^{(1)}(\Delta D, D_n) \vecb{A}(\vecb{X}_i^n), \\ 
    \vecb{X}_i^{n+1} &= \vecb{X}_i^{n} + \Delta D \vecb{\mPi}_i^{n+\nicefrac{1}{2}}, \\ 
    \vecb{\mPi}_i^{n+1} &= p^{(2)}(\Delta D, D_n) \vecb{\mPi}_i^{n+\nicefrac{1}{2}} + q^{(2)}(\Delta D, D_n) \vecb{A}(\vecb{X}_i^{n+1}),
\end{align}
\label{eq:pi_integrator_kdk}
\end{subequations}
cf.\ Eqs.~\eqref{eq:pi_integrator} for the DKD case. One has the freedom to introduce two potentially different coefficient functions for each part of the kick now, $p^{(1)}$ and $q^{(1)}$ for the first half, and $p^{(2)}$ and $q^{(2)}$ for the second half.
The Zel'dovich consistency condition in Eq.~\eqref{eq:zeldovich_consistency_condition} for the kick now becomes
\begin{equation}
    \frac{1 - p^{(1)}(\Delta D, D_n)}{q^{(1)}(\Delta D, D_n)} = \frac{3}{2} \Omega_{\mathrm{m}} D_{n} \stackrel{\mathrm{EdS}}{\asymp} \frac{3}{2} a_n \quad \text{and} \quad \frac{1 - p^{(2)}(\Delta D, D_n)}{q^{(2)}(\Delta D, D_n)} = \frac{3}{2} \Omega_{\mathrm{m}} D_{n+1} \stackrel{\mathrm{EdS}}{\asymp} \frac{3}{2} a_{n+1}.
\label{eq:zeldovich_consistency_condition_kdk}
\end{equation}
For example, the original \textsc{FastPM} KDK scheme is given by
\begin{equation}
    p^{(1)}_{\textsc{FastPM}}(\Delta D, D_n) = \frac{F(D_n)}{F(D_{n+\nicefrac{1}{2}})}, \qquad p^{(2)}_{\textsc{FastPM}}(\Delta D, D_n) = \frac{F(D_{n+\nicefrac{1}{2}})}{F(D_{n+1})}, \label{eq:fastpm_in_kdk_form}
\end{equation}
where $D_{n+\nicefrac{1}{2}}$ is an intermediate time, and the coefficients $q^{(1)}_{\textsc{FastPM}}$ and $q^{(2)}_{\textsc{FastPM}}$ are determined by Eq.~\eqref{eq:zeldovich_consistency_condition_kdk}.

Symplecticity then requires that 
\begin{equation}
    p^{(1)}(\Delta D, D_n) \, p^{(2)}(\Delta D, D_n) = \frac{F(D_n)}{F(D_{n+1})},
    \label{eq:symplecticity_condition_kdk}
\end{equation}
which is clearly satisfied for \textsc{FastPM} for any choice of $D_{n+\nicefrac{1}{2}}$. In principle, one could also replace $F(D_{n+\nicefrac{1}{2}})$ in Eq.~\eqref{eq:fastpm_in_kdk_form} by a more general expression (as long as the second-order accuracy of the integrator is unaffected), given that it cancels out in Eq.~\eqref{eq:symplecticity_condition_kdk}.

\subsection{\textsc{LPTFrog}}
Carrying over \textsc{LPTFrog} to the KDK case can be done along the lines of the leapfrog integrator in KDK form for a dissipative oscillator in Ref.~\cite{Vermeeren2019}, which yields
\begin{subequations}
\begin{align}
    p^{(1)}_{\textsc{LPTFrog}}(\Delta D, D_n) &= \left(1 + \frac{\Delta D}{2} \mathfrak{f}(D_n)\right)^{-1}, &\qquad q^{(1)}_{\textsc{LPTFrog}}(\Delta D, D_n) &= \frac{\frac{\Delta D}{2} \mathfrak{g}(D_n)}{1 + \frac{\Delta D}{2} \mathfrak{f}(D_n)}, \\
    p^{(2)}_{\textsc{LPTFrog}}(\Delta D, D_{n+1}) &= \left(1 - \frac{\Delta D}{2} \mathfrak{f}(D_{n+1})\right), &\qquad q^{(2)}_{\textsc{LPTFrog}}(\Delta D, D_n) &= \frac{\Delta D}{2} \mathfrak{g}(D_{n+1}).
\end{align}
\end{subequations}
This is motivated by the ansatz
\begin{subequations}
    \begin{align}
        \vecb{\mPi}_i^{n+\nicefrac{1}{2}} &= \vecb{\mPi}_i^{n} + \frac{\Delta D}{2} \mathfrak{g}(D_n) \vecb{A}(\vecb{X}_i^n) - \frac{\Delta D}{2} \mathfrak{f}(D_n) \vecb{\mPi}_i^{n+\nicefrac{1}{2}}, \\
        \vecb{X}_i^{n+1} &= \vecb{X}_i^{n} + \Delta D \vecb{\mPi}_i^{n+\nicefrac{1}{2}}, \\
        \vecb{\mPi}_i^{n+1} &= \vecb{\mPi}_i^{n+\nicefrac{1}{2}} + \frac{\Delta D}{2} \mathfrak{g}(D_{n+1}) \vecb{A}(\vecb{X}_i^{n+1}) - \frac{\Delta D}{2} \mathfrak{f}(D_{n+1}) \vecb{\mPi}_i^{n+\nicefrac{1}{2}}.
    \end{align}
\end{subequations}

Note that the coefficient functions $p^{(1)}_{\textsc{LPTFrog}}$ and $p^{(2)}_{\textsc{LPTFrog}}$ agree with those of the standard symplectic KDK integrator (which equal those of \textsc{FastPM}) only up to first order, i.e.
\begin{subequations}
\begin{align}
p^{(1)}_{\textsc{LPTFrog}}(\Delta D, D_n) - \frac{F(D_n)}{F(D_{n+\nicefrac{1}{2}})} &= \mathscr{O}((\Delta D)^2), \\
p^{(2)}_{\textsc{LPTFrog}}(\Delta D, D_n) - \frac{F(D_{n+\nicefrac{1}{2}})}{F(D_{n+1})} &= \mathscr{O}((\Delta D)^2).
\end{align}
\end{subequations}
 However, this is not a problem because $p^{(1)}_{\textsc{LPTFrog}}$ is multiplied with another $\Delta D$ in the drift, and the kick is second-order accurate as well, which can be seen by explicitly computing the momentum update
  \begin{equation}
     \vecb{\mPi}_i^{n+1} = p^{(1)}(\Delta D, D_n) \, p^{(2)}(\Delta D, D_n) \vecb{\mPi}_i^{n} + q^{(1)}(\Delta D, D_n) \, p^{(2)}(\Delta D, D_n) \vecb{A}(\vecb{X}_i^n) + q^{(2)}(\Delta D, D_n) \vecb{A}(\vecb{X}_i^{n+1})
 \end{equation}
 and noting that
 \begin{equation}
     p^{(1)}_{\textsc{LPTFrog}}(\Delta D, D_n) \, p^{(2)}_{\textsc{LPTFrog}}(\Delta D, D_n) - \frac{F(D_n)}{F(D_{n+1})} = \mathscr{O}((\Delta D)^3),
 \end{equation} 
 overall resulting in a Zel'dovich-consistent second-order accurate scheme.

 \subsection{\textsc{TsafPM}}
For the \textsc{TsafPM} integrator, one obtains
\begin{subequations}
\begin{alignat}{-1}
    q^{(1)}(\Delta D, D_n) &\,=\,& \frac{\int_{a(D_n)}^{a(D_{n+\nicefrac{1}{2}})} \frac{\dd a}{H(a) a^2}}{F(D_{n+\nicefrac{1}{2}})} &\,\stackrel{\mathrm{EdS}}{\asymp}\,& \frac{2 \left(\sqrt{a_{n+\nicefrac{1}{2}}} - \sqrt{a_n}\right)}{a_{n+\nicefrac{1}{2}}^{\nicefrac{3}{2}}}, \\ 
    q^{(2)}(\Delta D, D_n) &\,=\,& \frac{\int_{a(D_{n+\nicefrac{1}{2}})}^{a(D_{n+1})} \frac{\dd a}{H(a) a^2}}{F(D_{n+1})}  &\,\stackrel{\mathrm{EdS}}{\asymp}\,& \frac{2 \left(\sqrt{a_{n+1}} - \sqrt{a_{n+\nicefrac{1}{2}}}\right)}{a_{n+1}^{\nicefrac{3}{2}}},
\end{alignat}
\end{subequations}
and the coefficients $p^{(1)}$, $p^{(2)}$ follow immediately from Eq.~\eqref{eq:zeldovich_consistency_condition_kdk}. 

\end{appendix}
\label{lastpage}
\end{document}